\documentclass{aa}  

\usepackage{graphicx}
\usepackage{txfonts}
\usepackage{xcolor}
\usepackage{natbib}
\usepackage{float}
\usepackage{makecell}
\usepackage{comment}
\usepackage{subcaption}
\usepackage{soul}
\usepackage{lscape}
\usepackage{placeins}
\usepackage[colorlinks=true, linkcolor=blue, citecolor=blue, urlcolor=blue]{hyperref}

\newcommand{\disperse}{\mbox{{\tt DisPerSE}}\xspace}

\defcitealias{pacheco2026}{PA26}

\begin{document} 
\title{IllustrisTNG50 angular momentum maps: tracing the morpho-kinematic evolution of galaxies}
   
\subtitle{}

\author{Juan Manuel Pacheco-Arias \inst{1}, Benoît Epinat\inst{1,2}, Philippe Amram \inst{1}, Wilfried Mercier \inst{1} \and Katarina Kraljic\inst{3}}

\institute{Aix-Marseille Univ., CNRS, CNES, LAM, 38 Rue Frédéric Joliot Curie, 13338 Marseille, France\\ 
\email{juan.pacheco@lam.fr}
\and French-Chilean Laboratory for Astronomy, IRL 3386, CNRS and Universidad de Concepción, Departamento de Astronomía, Barrio Universitario s/n, Concepción, Chile
\and Observatoire Astronomique de Strasbourg, UMR 7550, CNRS, Université de Strasbourg, F-67000 Strasbourg, France}

\date{Received; accepted}

\abstract
{The first study of the two-dimensional spatial distribution of stellar specific angular momentum (sAM) in observations of galaxies, revealed that there is an unexpected morpho-kinematic diversity within galactic discs.}
{Our goal is to quantify and understand the morpho-kinematic diversity in galaxy simulations using the newly proposed $j_\star$-types classification, originally developed for late-type galaxy observations.}
{We studied the spatial distribution of the stellar sAM surface density (sAMSD) for about 8000 \textsc{TNG50} stellar discs spanning redshifts from 0 to 3.5, across a stellar mass range of $10^{9.5}$ to $10^{11.2}$ $\rm M_\odot$. We built this sample by tracking back in time the \textsc{SubHalos} component of the \textsc{TNG50 MW/M31} catalogue, selecting all galaxies that host a stable, rotationally supported stellar disc. We computed a set of four morpho-kinematic metrics characterizing their $j_\star$-substructures by comparing their sAMSD maps with the expected Freeman sAMSD distribution and by performing a Fourier mode decomposition of their sAMSD. We applied a Gaussian mixture model with four fully covariant components to the parameter space defined by our metrics. We interpreted the probability to belong to each component as the probability for each galaxy to be classified into one of the four $j_\star$-types we identified in this population.}
{Cosmological hydrodynamic simulations of galactic discs exhibit a morpho-kinematic diversity similar to that reported for observations. The \textsc{TNG50} stellar discs store and redistribute their stellar angular momentum predominantly in four $j_\star$-substructures, that evolve, following their expected class fractions with respect to redshift, as follows: $j_\star$-irregular ($\bar{z}=0.91$), $j_\star$-spiral ($\bar{z}=0.76$), $j_\star$-ring ($\bar{z}=0.62$), and $j_\star$-bar ($\bar{z}=0.39$). The gas fraction ($f_{\rm gas}$) and the stellar rotation support ($V/\sigma$) are the main drivers behind this morpho-kinematic evolution. Primarily, $f_{\rm gas}$ separates the $j_\star$-types into two groups, causing gas-rich galaxies to form $j_\star$-irregulars or $j_\star$-spirals, while gas-poor galaxies exhibit $j_\star$-bars or $j_\star$-rings. Once $f_{\rm gas}$ is fixed, high $V/\sigma$ values favours $j_\star$-spirals in the first scenario, and $j_\star$-rings in the second. These results suggest that gas accretion and stellar feedback are the primary physical processes driving the redistribution of angular momentum. In this scenario, galaxy simulations move across the $j_\star-M_\star$ diagram following their morpho-kinematic evolution, clustering in different regions of the diagram depending on their $j_\star$-type. The prevalence of $j_\star$-rings and $j_\star$-bars from the sample selection criteria, and the lack of strong correlations between our morpho-kinematic metrics and galaxy environment at $z=0$, are discussed in this study.}
{There is a canonical pathway for the redistribution of angular momentum within galactic discs undergoing secular evolution in \textsc{TNG50}. This scenario, derived from the morpho-kinematic description of stellar discs, differs from the frameworks established by traditional galaxy morphology and kinematic indicators as $j_\star$-substructures do not always coincide with their morphological counterparts and encode spatial information beyond what standard kinematic indicators capture. The sAMSD analysis links variations in stellar dynamics to their consequences for mass redistribution, enabling the reconstruction of comprehensive galactic evolutionary histories.}


\titlerunning{Morpho-kinematic evolution of galaxies from IllustrisTNG50 angular momentum maps}
\authorrunning{Pacheco-Arias, J. M., et al.}
\maketitle

\section{Introduction}

The origin, transfer, and redistribution of angular momentum are among the fundamental processes in galaxy formation and evolution \citep{fall1980,mo1998}. Understanding the physical mechanisms that govern angular momentum evolution has been essential for reproducing realistic galaxy populations in cosmological volume simulations \citep[see e.g.][]{genel2015,lagos2017,rodriguez2022}. Reaching this point was a lengthy process that involved, in parallel, the development of increasingly sophisticated models for multi-scale baryonic physics and continuous improvements in numerical resolution \citep[see e.g.][]{somerville2015,naab2017,Vogelsberger2020}. As a result, modern simulation suites now produce extensive catalogues of galaxies hosting extended, rotationally supported stellar and gaseous discs \citep[see e.g.][]{schaye2015,springel2018,pillepich2019TNG50,dubois2021}. These structures acquire their angular momentum through exchanges with their dark matter \textsc{SubHaloes} and the cosmic web, driven by gravitational interactions \citep[see e.g.][]{codis2012,welker2014,kang2015,zjupa2017,lagos2018} and tidal torques \citep[see e.g.][]{peebles1969origin,white1984angular,barnes1987,codis2015,cadiou2022}.

The relationship between simulated and observed angular momentum has been explored primarily through comparisons of the specific angular momentum (sAM) measured in both cases. In particular, most studies have focused on the total stellar sAM ($j_\star$) and its relation to the total stellar mass ($M_\star$) for which observations reveal a tight scaling relation, commonly known as the Fall relation \citep[see e.g.][]{fall1983,romanowsky2012,fall2013,posti2018}. The ability of modern cosmological simulations to broadly reproduce this relation demonstrates that the total angular momentum acquired by galactic discs is reasonably well captured as a global and integrated property \citep[see e.g.][]{genel2015,lagos2017,rodriguez2022}. However, the diversity of slopes, normalisations, and even functional forms reported for the $j_\star$--$M_\star$ relation, both in observations \citep[see e.g.][]{obreschkow2014,posti2018,hardwick2022} and simulations \citep[see e.g.][]{rodriguez2022}, suggests that reproducing this scaling relation alone is insufficient to fully understand the coupling between galaxy evolution and angular momentum acquisition. Progress therefore requires understanding how stellar angular momentum is redistributed within galactic discs, together with the physical processes responsible for driving this redistribution.

A first step in this direction was taken by \citet[hereafter \citetalias{pacheco2026}]{pacheco2026}, who defined and quantified the stellar sAM surface density (sAMSD) for galactic discs in a sample of late-type galaxies from the Gassendi H$\alpha$ survey of SPirals \citep[GHASP;][]{epinat2008ghaspb,epinat2008ghaspa,korsaga2019ghasp}. This analysis led to the proposal of a new morpho-kinematic classification system, in which galaxies were grouped into five $j_\star$-types according to the dominant stellar angular momentum substructures responsible for storing the bulk of their sAM ($j_\star$-substructures). Galaxies sharing the same $j_\star$-substructures were found to occupy distinct regions of the $j_\star-M_\star$ plane. In \citetalias{pacheco2026}, this result was interpreted within the context of a morpho-kinematic evolutionary sequence followed by late-type galaxies as they transition between different $j_\star$-types. This framework provides a natural and physically motivated formalism that integrates the morphological and kinematic properties of the stellar component, thereby offering a significant contribution to our understanding of galactic angular momentum. Motivated by these results, in the present study we extend the sAMSD analysis to cosmological hydrodynamic simulations. To this end, we compute the stellar sAMSD for a subsample of disc galaxies from \textsc{IllustrisTNG50} \citep{pillepich2024milky}, with the aim of assessing whether the diversity of observed $j_\star$-substructures is also reproduced in simulations. The main motivation is to propose a scenario for the angular momentum redistribution within secularly evolving discs, which will help us to understand the morpho-kinematic diversity of galaxies, and with it the growth and accumulation of stellar sAM inside numerical simulations.

This paper is organised as follows: We present our sample selection in Sect.~\ref{sec: The sample}. We describe the methodology to compute, quantify and classify the stellar sAMSD for simulations in Sect.~\ref{Sec: Methodology}. The morpho-kinematic diversity and evolution results are presented in Sect.~\ref{sec: Results} while we interpret and discuss these results in Sect.~\ref{sec: Discussion} to finally provide our conclusions in Sect. \ref{sec: Conclusions}.

\section{Sample selection from IllustrisTNG50}\label{sec: The sample}

Resolving individual galaxies with high resolution in cosmological volume simulations is not a simple task. Among the wide range of simulations that attempt to do so \citep[see e.g.][]{pillepich2019TNG50,dave2019,dubois2021,feldmann2023}, \textsc{IllustrisTNG50} stands out for striking a balance between the modelling of realistic physical baryonic processes and its cosmological scale, enabling it to reproduce realistic galaxy populations whilst tracking their formation and evolution. We chose to analyse one of these populations, the \textsc{TNG50 MW/M31} subsample \citep{pillepich2024milky}, from which we selected the galaxies to apply our sAMSD formalism.

\subsection{\textsc{TNG50 MW/M31} galaxies}\label{sec: TNG50_MW/M31}

The \textsc{TNG50} Milky Way and Andromeda sample\footnote{\url{https://www.tng-project.org/data/milkyway+andromeda/}} \citep{pillepich2024milky} is a set of 198 simulated galaxies that resemble the morphological, kinematic, and environmental characteristics of the Milky Way and M31 at $\rm z = 0$. As part of the \textsc{TNG50-1} cosmological magnetohydrodynamical suite \citep{pillepich2019TNG50,nelson2019TNG}, the galaxies in this sample were simulated using the \textsc{arepo} moving-mesh code \citep{springel2010arepo,weinberger2020arepo} and the fiducial \textsc{TNG} galaxy formation model \citep{weinberger2017model,pillepich2018model}, with a target mass for gas cells and stellar particles of $8.5\times10^{4}$ $\rm M_\odot$, and of $4.6\times10^{5}$ $\rm M_\odot$ for dark matter particles. The stellar softening length at $z=0$ ($\epsilon_\star$) was set to $\sim 300$ pc, while the average gas cells spatial resolution in the star-forming regions is 
$\sim 150$ pc.

The selection criteria for this subset, established by \cite{pillepich2024milky}, impose that each chosen galaxy must exhibit a disc-like stellar morphology, a total stellar mass ranging between $10^{10.5}$ and $10^{11.2}$ $\rm M_\odot$, a dark matter halo less massive than $10^{13}$ $\rm M_\odot$, and must not have any massive companions within a range of 500 $\rm kpc$ in the final snapshot of its evolution. \textsc{SubHalos} that met these conditions were tracked from $\rm z=0$ up to $\rm z=7$ in a cubic cutout of 800 comoving kpc, with a separation time between snapshots of approximately 150 Myr. The aforementioned characteristics, combined with a sub-kpc numerical resolution, made this sample ideal to compute the temporal evolution of stellar sAMSD maps in a cosmological context.

\subsection{Large undisturbed discs}\label{Sec: sample_selection}

We applied four additional criteria to select the \textsc{TNG50 MW/M31} galaxies for which we performed their sAMSD analysis. The criteria were defined using only the gravitationally bound stellar particles flagged as \textsc{InSitu} by \cite{pillepich2024milky}. In this way, we ensured that the computed properties reflected the stellar dynamics of their respective \textsc{SubHalo} and not those of their surrounding environment. The coordinates and velocities of these particles were aligned and rotated by \cite{pillepich2024milky} with respect to the centre and galactic plane of the \textsc{SubHalo} in each snapshot, ensuring that the $\mathbf{\hat{z}}$ axis of the coordinate system points in the same direction as the total angular momentum vector. Throughout this paper, we use two coordinate systems to describe the position and velocity of the particles: the Cartesian system ($x,y,z$) and the cylindrical system ($R,\theta,z$). Vector quantities are denoted using the boldface notation. We defined the maximum radius ($R_{\max}$) of each galaxy as the 90th percentile of the distribution of stellar particle radii for every snapshot.

The first sample selection criterion seeks to ensure adequate spatial resolution in the sAMSD maps and Fourier analysis; therefore, we required $R_{\max} \geq 7$ kpc. The second criterion aims to ensure sufficient mass resolution and numerical stability for the analyzed \textsc{SubHalos}; accordingly, we imposed $M_\star \geq 10^{9.5}~\rm M_\odot$ within $R_{\max}$. The final two criteria were adopted to identify galaxies that are not strongly disturbed and that exhibit orderly rotation. Specifically, for the third criterion we required $f_{\rm CR}\leq0.20$, where $f_{\rm CR}=N(v_{\star\theta,k}<0)/N_\star$ is the stellar counter-rotating fraction within $R_{\max}$, with $v_{\star\theta,k}$ the tangential velocity of stellar particle $k$, and $N(v_{\star\theta,k}<0)$ the number of stellar particles with negative $v_{\star\theta,k}$ within $R_{\max}$. For the fourth criterion we required $V/\sigma \geq 0.50$, where $V/\sigma=|\langle v_{\star\theta,k}\rangle_{m_\star}|/(\langle\sigma_{\star R,k}\rangle_{m_\star}^2+\langle\sigma_{\star z,k}\rangle_{m_\star}^2)^{1/2}$ is the ratio of ordered to random stellar motions within $R_{\rm max}$, $\langle v_{\star\theta,k}\rangle_{m_\star}$ being the mass-weighted mean $v_{\star\theta,k}$, and $\langle\sigma_{\star R,k}\rangle_{m_\star}$ and $\langle\sigma_{\star z,k}\rangle_{m_\star}$ the radial and vertical mass-weighted velocity dispersions of the stellar particles within $R_{\max}$, respectively. This definition follows standard stellar-dynamical practice \citep{binney2008}, separating ordered rotation from random motions in disc-dominated systems. Throughout the paper, angle brackets $\langle \cdot \rangle$ denote averaged quantities, while the subscript indicates the weighting scheme adopted in each case.

The application of our selection criteria to all snapshots (from $0 \leq z \leq 7$), reduced the \textsc{TNG50 MW/M31} available galaxies by $\sim$ 54\% (from 17 424 to 7 966) and determined a natural redshift range of $0 \leq z \leq 3.50$ for our studied sample. The size, mass, morphology, and stellar dynamics of this sample, following our selection process, make these \textsc{SubHalos} comparable to the observed galaxies presented in \citetalias{pacheco2026} at any time during their evolution (see Appendix~\ref{app: Obs_Vs_Simu} for a direct comparison between both samples).

\section{Methodology}\label{Sec: Methodology}

In this study, we present a methodology for generating sAMSD maps of galaxy simulations based on their stellar particles properties. This formalism ensures a fair and direct comparison between simulated and observed maps for disc-like galaxies. Furthermore, we have proposed a set of four morpho-kinematic metrics that quantify the prevalence and extent of substructures of variable symmetry in sAMSD space, inspired by the $j_\star$-types proposed in \citetalias{pacheco2026}. 
\subsection{sAMSD maps for simulations}
\label{sec:sAMSDmaps}

We calculated the stellar sAM corresponding to a given stellar particle $k$ as 

\begin{equation}
    j_{\star,k} \equiv \frac{J_{\star,k}}{M_{\star}} = \frac{m_{\star,k} \hspace{0.1cm} \vec{r}_{\star,k}\times\vec{v}_{\star,k}}{M_{\star}} \hspace{0.1cm} \cdot \mathbf{\hat{z}},
    \label{eq: vectorial_j}
\end{equation}

\noindent where $\vec{r}_{\star,k}$ and $\vec{v}_{\star,k}$ are the position (in $\rm kpc$) and velocity (in $\rm km~s^{-1}$) vectors of the stellar particle with respect to the galactic centre. $m_{\star,k}$ is the mass of stellar particle $k$ and $M_\star$ is described in Sect. \ref{Sec: sample_selection}. $j_{\star,k}$ is a scalar quantity defined as the $\mathbf{\hat{z}}$-axis projection of the total stellar sAM of each particle. The $\mathbf{\hat{z}}$ direction is, by construction, where most of the angular momentum is stored in undisturbed rotating discs (see Sect. \ref{Sec: sample_selection}). Furthermore, this component is the only one to which we have access through galaxy observations, and therefore this definition ensures a fair comparison between both approaches. From Eq. (\ref{eq: vectorial_j}), it is straightforward to define $j_\star = \sum_k^{N_\star} j_{\star,k}$ as the total stellar sAM  withing $R_{\max}$.

We constructed the stellar sAMSD ($\iota_\star$) maps for our galaxies by projecting their $dj_\star/dS$ surface density onto their galactic plane over a regular grid, such that

\begin{equation}
\iota_\star(x,y) = \frac{dj_\star}{dS}(x,y) = 
\frac{1}{\Delta x\,\Delta y}
\sum\limits_{k \in \mathcal{P}_{x,y}} j_{\star,k},
\label{eq: iota}
\end{equation}

\noindent with $\mathcal{P}_{x,y}$ being the set of particles whose positions $(x_k,y_k)$ fall inside the pixel centered at $(x,y)$ with size $\Delta x\,\Delta y$ in pc$^2$. All our maps were generated imposing $\Delta x = \Delta y = 500$ pc, as a compromise between $\epsilon_\star$ and an adequate particle sampling. After projecting onto the grid, we smoothed $\iota_\star(x,y)$ by applying a Gaussian filter with $\sigma = \epsilon_\star$ in order to suppress any sub-resolution fluctuations. The grid onto which $dj_\star/dS$ was projected, and the size of the Gaussian smoothing kernel, were kept constant for all maps used throughout this paper, regardless of the property they represent (i.e. any spatially resolved quantity defined as a function of $(x,y)$). Equation (\ref{eq: iota}) is the direct equivalent of the observed $\iota_\star$, defined and calculated for the first time in \citetalias{pacheco2026} for a set of 30 late-type galaxies.

\begin{figure*}[ht!]
\centering
\includegraphics[width=1\textwidth]{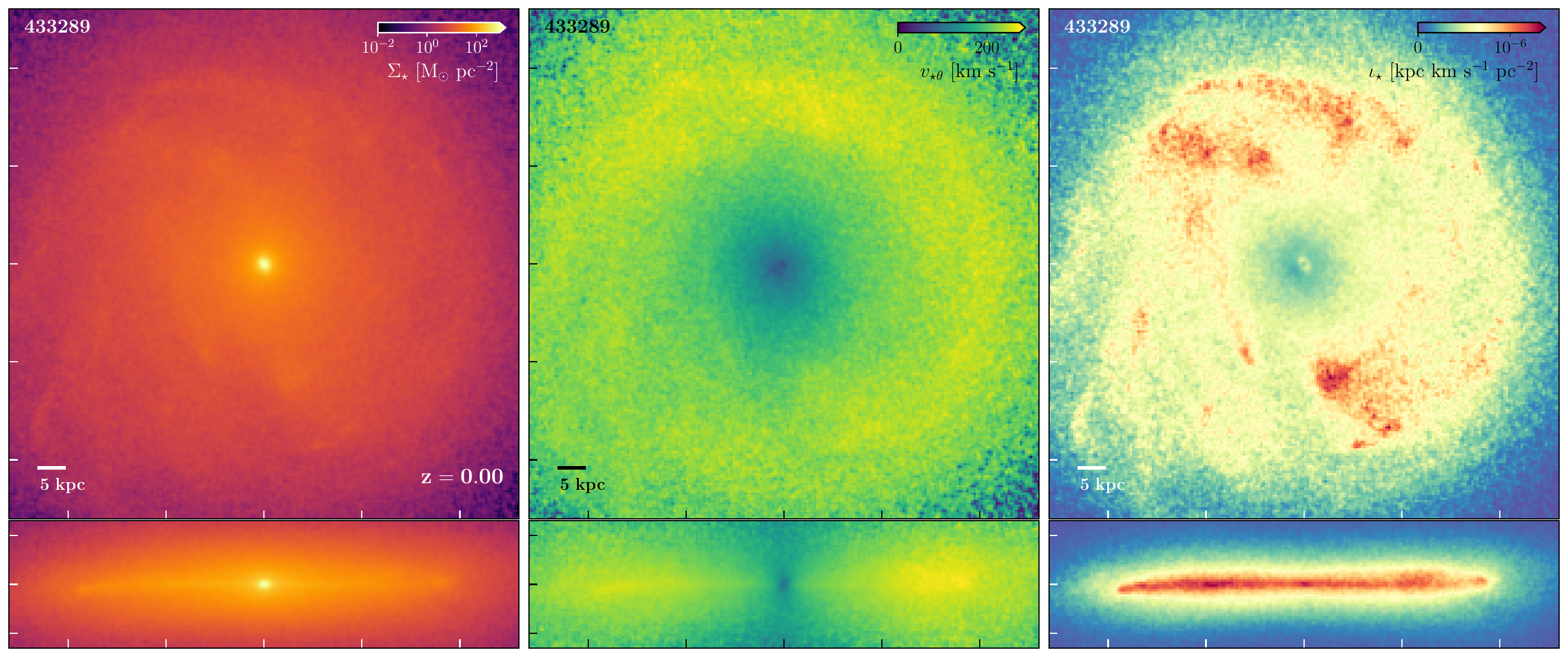}
\caption{Face-on (top) and edge-on (bottom) view of $\Sigma_\star$ (left), $v_{\star\theta}$ (middle), and $\iota_\star$ (right) for the galaxy with the highest $j_\star$ of our sample. The \textsc{SubHaloID} and the 5 kpc length scale are located at the top and bottom left of all face-on maps, respectively, while the redshift corresponding to the snapshot is located at the bottom right of the face-on $\Sigma_\star$.}
\label{fig: density_vel_j_maps}
\end{figure*}

Figure \ref{fig: density_vel_j_maps} shows the spatial distribution of the angular momentum of stellar particles for the galaxy with the highest $j_\star$ in our sample. The panels, from left to right, show the stellar density map $\Sigma_\star(x,y)$, stellar tangential velocity map $v_{\star\theta}(x,y)$, and $\iota_\star(x,y)$. This example illustrates the difference between the morphological and kinematic features of a galaxy when considered separately, compared to those found in the combined morpho-kinematics of its $\iota_\star$ space (for more examples, see Appendix~\ref{app: Morpho_zoo}).

\subsection{Morpho-kinematic metrics}

We quantified the spatial distribution of the stellar sAM in each galaxy based on four morpho-kinematic metrics. The main idea was to estimate how predominant the $j_\star$-substructures are in the sAMSD maps based on their direct comparison with the Freeman disc sAMSD stellar model \citepalias{pacheco2026} and their decomposition into Fourier modes.

\subsubsection{Bi-dimensional correlation}\label{Sec: C_2D}

The first of these metrics seeks to quantify how closely $\iota_\star(x,y)$ resembles the stellar sAMSD distribution of a Freeman disc by means of its bi-dimensional correlation, defined as

\begin{equation}
    \mathcal{C}_{\rm 2D} = \frac{\sum\limits_{x,y}\left(\iota_\star(x,y)\times\iota_{\rm model}(x,y)\right)}{\sqrt{\sum\limits_{x,y}\left(\iota_\star(x,y)^2\right)\times\sum\limits_{x,y}\left(\iota_{\rm model}(x,y)^2\right)}},
    \label{eq: C2D}
\end{equation}

\noindent where

\begin{equation}
    \iota_{\rm model}(x,y) = \frac{\Sigma_{\rm F}(x,y)~v_{\rm RC}(x,y)~R}{\sum\limits_{x,y|R\leq R_{\max}}\Sigma_{\star}(x,y)}.
    \label{eq: iota_model}
\end{equation}

\noindent 

In Eq. (\ref{eq: C2D}), the sums are applied to all pixels $(x,y)$ whose associated radial coordinate $R$ satisfies $3\epsilon_\star \leq R \leq R_{\max}$. This radial interval is adopted throughout the paper for all radial profiles unless explicitly stated otherwise. We include the inner cut to exclude the central region where the gravitational force resolution is softened and the kinematic properties are not numerically reliable. In Eq. (\ref{eq: iota_model}), $\Sigma_{\rm F}(x,y)$ is the face-on Freeman stellar density of each galaxy, $v_{\rm RC}(x,y)$ is the velocity assigned by the galaxy rotation curve to each pixel $(x,y)$, and $\Sigma_\star(x,y)$ is the face-on stellar density map. The parameters used to calculate $\Sigma_{\rm F}(x,y)$ are obtained by fitting the stellar radial density profile with a Freeman disc model over the range between $R_{\max}/2$ and $R_{\max}$. The rotation curve and stellar density profile are constructed by computing $\langle v_{\star\theta,k}\rangle_{m_\star}$ and $\langle m_{\star,k}\rangle$ per unit area, respectively, within concentric radial rings spanning the adopted radial interval, using a constant bin width of $\Delta R = 500$ pc, equal to the pixel size.

\begin{figure*}[ht!]
\centering
\includegraphics[width=1\textwidth]{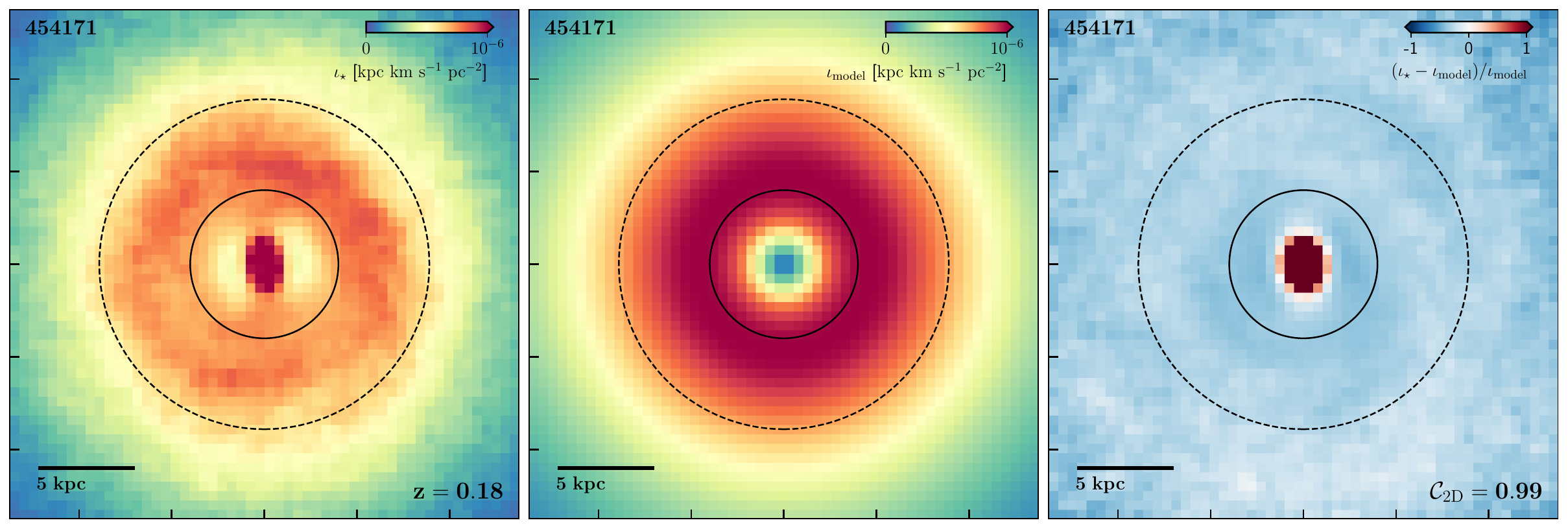}
\caption{$\iota_\star$ (left), $\iota_{\rm model}$ (middle), and $\left(\iota_\star - \iota_{\rm model}\right)/\iota_{\rm model}$ (right) for the galaxy with the highest $\mathcal{C}_{\rm 2D}$ in our sample. The solid black line on all panels shows the disc scale length, while the dashed black line shows the radius at which the maximum speed is reached on the rotation curve. The \textsc{SubHaloID} and the 5 kpc length scale are located at the top and bottom left of all face-on maps, respectively. The redshift is located at the bottom right of $\iota_\star$, and the $\mathcal{C}_{\rm 2D}$ value is located at the bottom right of the residue map.}
\label{fig: C_2D}
\end{figure*}

The Freeman discs are seen as ringed structures in $\iota_\star$ space (see \citetalias{pacheco2026} for a detailed explanation), so $\mathcal{C}_{\rm 2D}$ as defined in Eq. (\ref{eq: C2D}) is a measure of the predominance of $j_\star$-rings in our sample. This morpho-kinematic description can be clearly seen in Fig. \ref{fig: C_2D}, which shows, from left to right, $\iota_\star$, $\iota_{\rm model}$ and $\left(\iota_\star - \iota_{\rm model}\right)/\iota_{\rm model}$ for the galaxy of our sample with the highest $\mathcal{C}_{\rm 2D}$ (0.99). By construction, $\mathcal{C}_{\rm 2D}$ is bounded between 0 and 1, and is the only morpho-kinematic metric that can be affected by the resolution or smoothing of our sAMSD maps.

\subsubsection{Fourier modes decomposition}\label{Sec: Fourier_metrics}

In the same way that traditional morphological studies make use of all the information available per particle when applying the Fourier decomposition into $m_{\star,k}$ modes to identify the substructures present in $\Sigma_\star$, we define the remaining three morpho-kinematic metrics to quantify the $j_\star$-substructures present in $\iota_\star$, by computing the $j_{\star,k}$ Fourier modes of our galaxies. Specifically, we computed their Fourier coefficient profiles $\left(c_m\right)$, defined as

\begin{equation}
    c_m(R_n) = \frac{1}{\mathcal{N}_n} \displaystyle\sum_{k\in\mathcal{B}_n} j_{\star,k}\,e^{-im\theta_{\star,k}},
    \label{eq: cm_discrete}
\end{equation}

\noindent where $\mathcal{B}_n$ is the set of particles in the $n$-th radial bin centred at $R_n$, $\mathcal{N}_n$ is the total number of particles belonging to $\mathcal{B}_n$, $\theta_{\star,k}$ is the azimuthal angle of particle $k$, and $m$ is the Fourier mode. The range and width of the radial binning are the same as those used for the stellar density profile and rotation curve construction described in Sect. \ref{Sec: C_2D}. Based on Eq. (\ref{eq: cm_discrete}), and using $m \in \{0,1,2\}$, we proposed the following morpho-kinematic metrics:

\begin{equation}
    a_{1}^{\max} = \max_{R_n \in [3\epsilon_\star,R_{\max}]} \frac{\left|c_1(R_n)\right|}{\left|c_0(R_n)\right|},
    \label{eq: a1}
\end{equation}

\begin{equation}
    a_{2}^{\max} = \max_{R_n \in [3\epsilon_\star,R_{\max}]} \frac{\left|c_2(R_n)\right|}{\left|c_0(R_n)\right|},
    \label{eq: a2}
\end{equation}

\noindent and 

\begin{equation}
    R_{2}^{\max} = \frac{1}{R_{\max}} \left(\operatorname*{argmax}_{R_n \in [3\epsilon_\star,\,R_{\max}]} \frac{\left|c_2(R_n)\right|}{\left|c_0(R_n)\right|}\right).
    \label{eq: R_a2}
\end{equation}

\noindent The normalization with respect to $\left|c_0(R_n)\right|$ for $a_{1}^{\max}$ and $a_{2}^{\max}$, and with respect to $R_{\max}$ for $R_{2}^{\max}$, ensures that these metrics fall within the range 0 to 1. Equation (\ref{eq: a1}) shows how $a_{1}^{\max}$ quantifies the relative strength of the asymmetric components in the sAMSD space, providing a quantitative assessment of the significance of $j_\star$-irregularities across our sample. Equations (\ref{eq: a2}) and (\ref{eq: R_a2}) show how $a_{2}^{\max}$ and $R_{2}^{\max}$ characterize the relative strength and radial extent fraction of the bi-symmetric substructures in the sAMSD space, respectively. In particular, high $a_{2}^{\max}$ and low $R_{2}^{\max}$ values are indicative of 
$j_\star$-bars, whereas high $a_{2}^{\max}$ combined with high $R_{2}^{\max}$ values trace the predominance
$j_\star$-spirals. 

\begin{figure}
\centering
\includegraphics[width=\hsize]{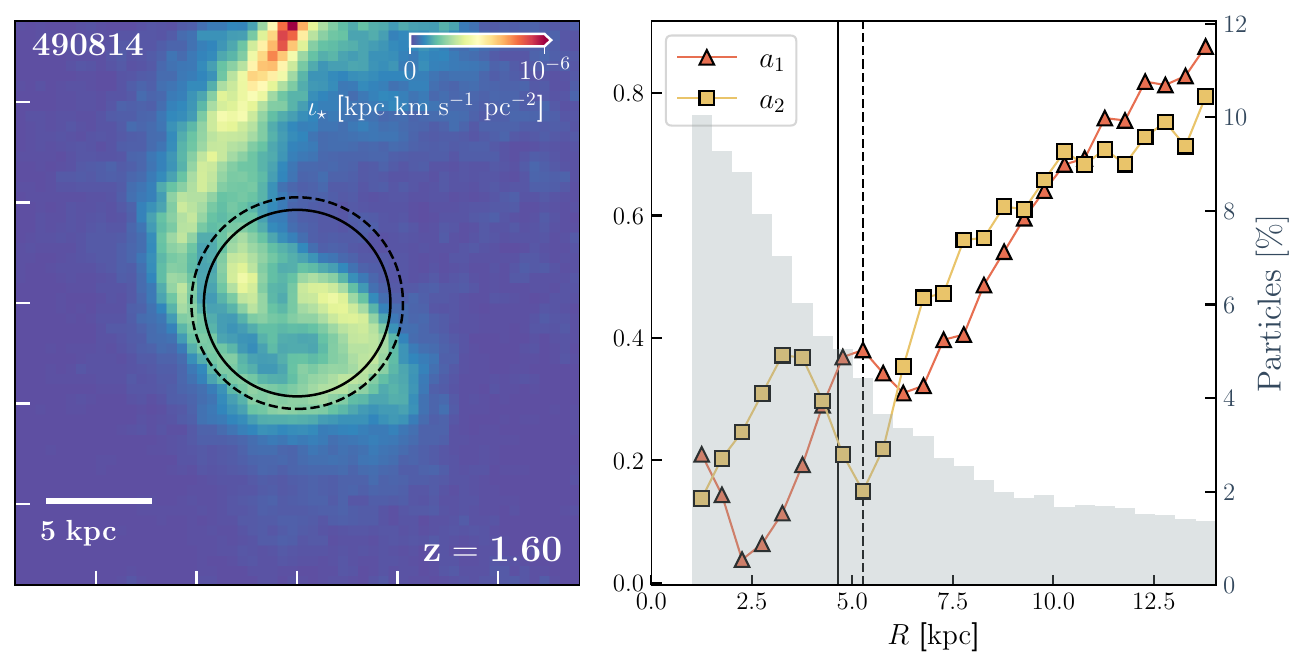}
\includegraphics[width=\hsize]{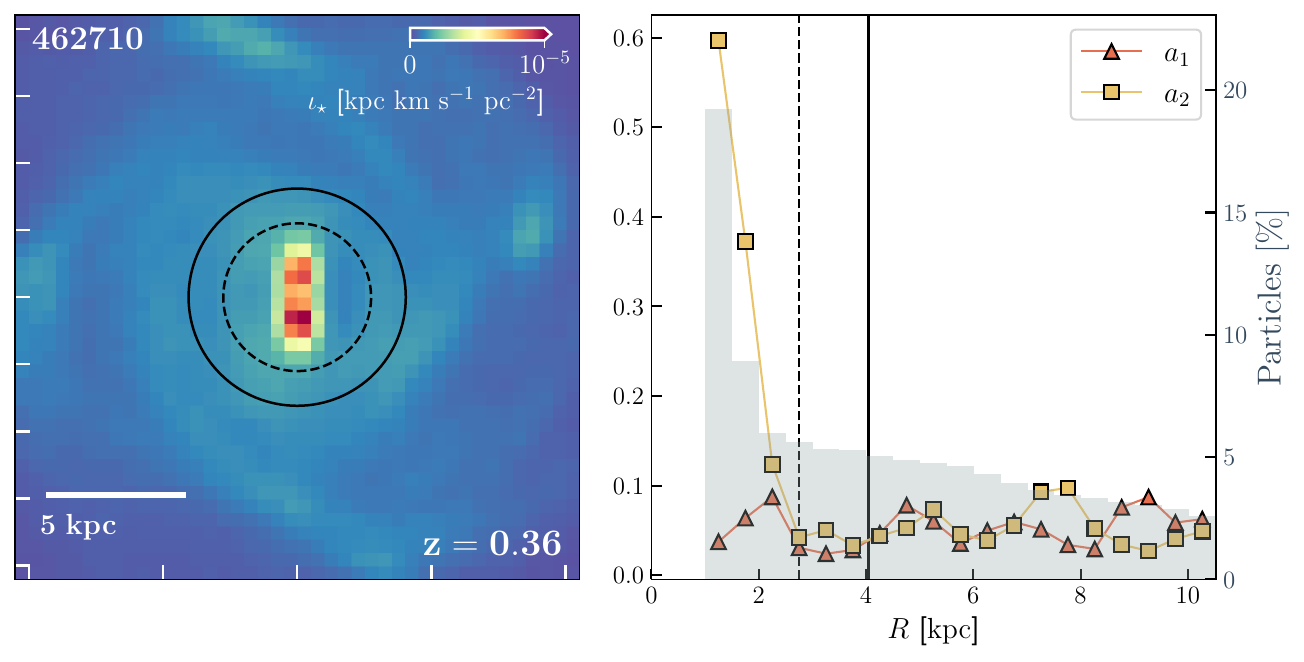}
\includegraphics[width=\hsize]{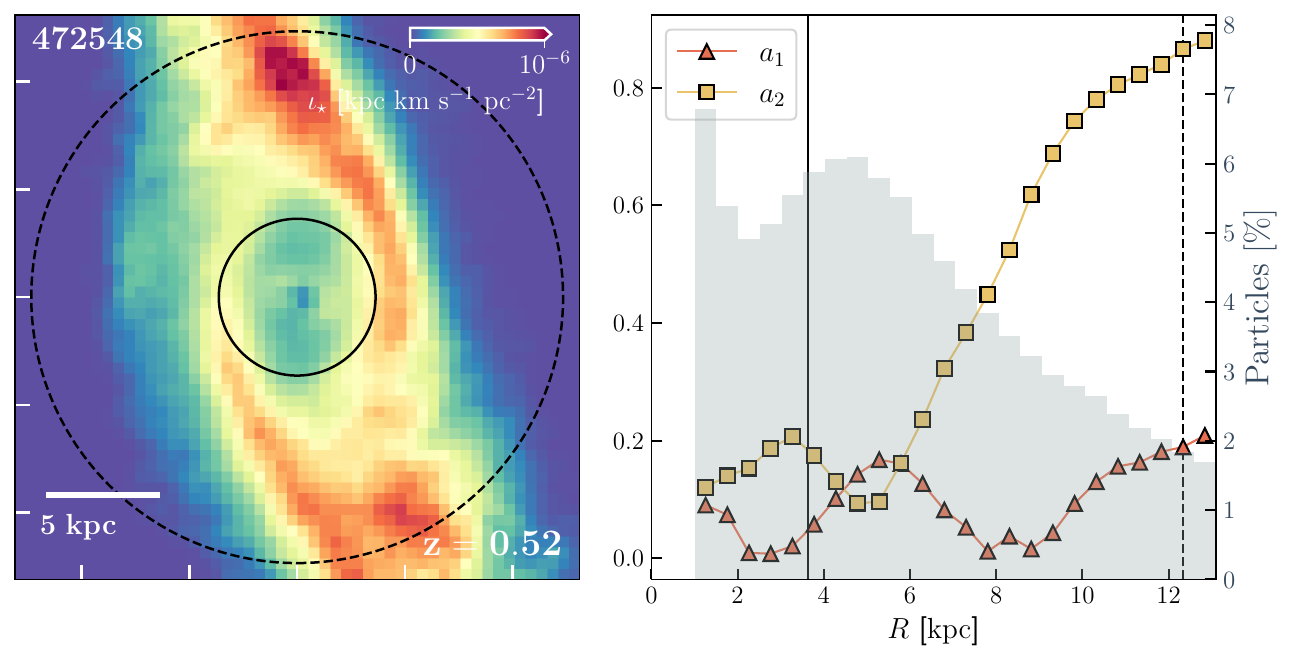}
\caption{$\iota_\star$ (left column), and their Fourier decomposition (right column), for the three galaxies in our sample that best represent the $j_\star$-irregular, $j_\star$-bar and $j_\star$-spiral morpho-kinematics, from top to bottom. The $a_{1}$ and $a_{2}$ profiles are displayed along with the radial bin stellar particle histogram in the right column. The solid black line on all panels shows the disc scale length, while the dashed black line shows the radius at which the maximum speed is reached on the rotation curve for each galaxy. The \textsc{SubHaloID} identifying each galaxy is located at the top left of all face-on maps, the 5 kpc length scale at the bottom left, and the redshift at the bottom right.}
\label{fig: morpho-kinematic_metrics}
\end{figure}

Figure \ref{fig: morpho-kinematic_metrics} presents the sAMSD maps and the normalized Fourier amplitude profiles for three representative galaxies selected to illustrate the various $j_\star$-types traced by $a_{1}^{\max}$, $a_{2}^{\max}$, and $R_{2}^{\max}$. The first panel, from top to bottom, corresponds to a system with high $a_{1}^{\max}$ (0.88), providing a clear example of a $j_\star$-irregular morpho-kinematics. The second shows a galaxy with high $a_{2}^{\max}$ (0.60) and low $R_{2}^{\max}$ (0.12), representative of a centrally concentrated $j_\star$-bar. The third panel displays high $a_{2}^{\max}$ (0.88) and high $R_{2}^{\max}$ (0.98), tracing an extended $j_\star$-spiral configuration. Together, these maps highlight the broad morpho-kinematic diversity of the sAMSD space and demonstrate how $a_{1}^{\max}$, $a_{2}^{\max}$, and $R_{2}^{\max}$ quantify the distribution of stellar angular momentum across regions of different symmetry within the galactic plane. In Appendix~\ref{app: Combined_Fourier}, we present a complementary definition of these metrics that combines the odd- and even-mode coefficients, making use of the harmonic nature of Fourier space to identify $j_\star$-substructures. Our results remain unchanged when using combined-mode Fourier metrics; therefore, we have chosen to retain the simpler definitions.

\subsection{Gaussian mixture model}\label{Sec: GMM}

\begin{figure*}[ht!]
\centering
\includegraphics[width=1\textwidth]{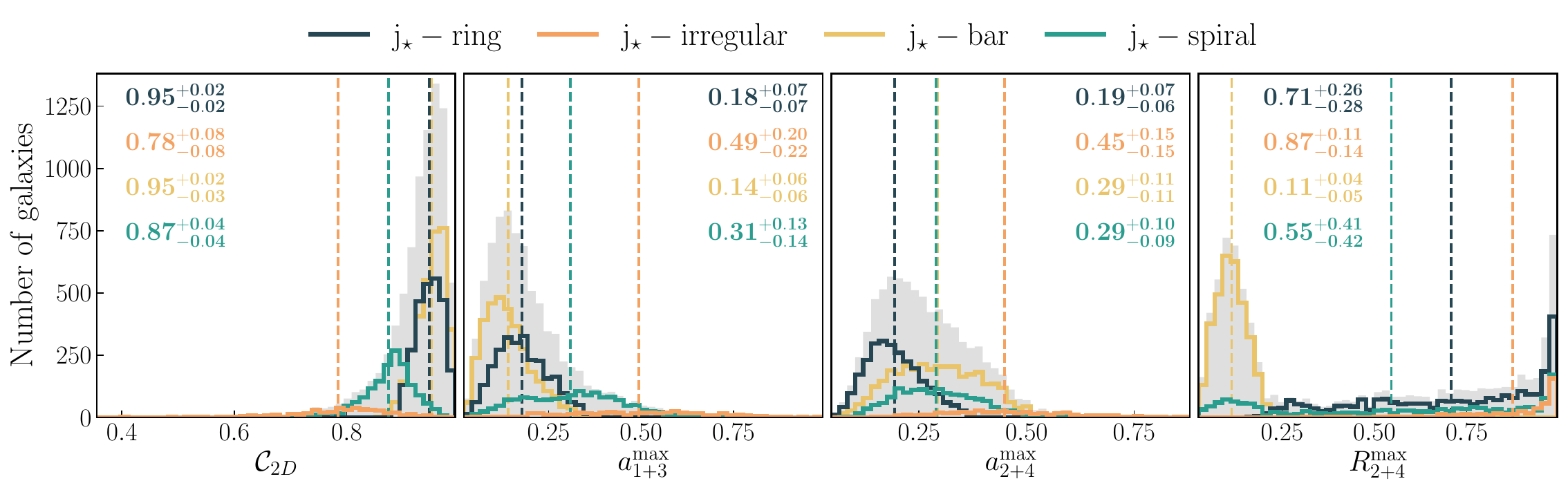}
\caption{Histograms of the morpho-kinematic metrics for each of the four components identified by the GMM. From left to right, the panels show the number of galaxies per bin for $\mathcal{C}_{\rm 2D}$, $a_1^{\max}$, $a_2^{\max}$, and $R_2^{\max}$, respectively. We assign a $j_\star$-type to each of the four components, as indicated at the top of the figure, based on the shape of their distributions and following the logic described in Sect. \ref{Sec: GMM}. The dotted lines indicate the mean values of the corresponding $j_\star$-types for each distribution, while the coloured text gives their mean values together with the ranges defined by the 16th and 84th percentiles. The grey histogram at the background of each panel represents the entire sample.}
\label{fig: GMM_distributions}
\end{figure*}

Once we have quantified the prevalence of the multiple $j_\star$-substructures in the sAMSD space, we established a probabilistic morpho-kinematic classification system, obtained by applying a Gaussian Mixture Model \citep[GMM,][]{dempster1977,pedregosa2011} to the four-dimensional parameter space defined by $\mathcal{C}_{\rm 2D}$, $a_1^{\rm max}$, $a_2^{\rm max}$, and $R_2^{\rm max}$. This model was configured with four fully covariant components, which we then associated with each of the $j_\star$-types by analysing their distributions within this parameter space, as shown in Fig.~\ref{fig: GMM_distributions}.

The interpretation of the components as distinct $j_\star$-types follows directly from the behaviour of the metrics established in Sects.~\ref{Sec: C_2D} and~\ref{Sec: Fourier_metrics}. The $j_\star$-rings correspond to the component with the highest mean value of $\mathcal{C}_{\rm 2D}$ (0.95), and display a high mean value of $R_2^{\rm max}$ (0.71), a low mean value of $a_1^{\rm max}$ (0.18), and the lowest mean value of $a_2^{\rm max}$ (0.19; dark blue histograms).The $j_\star$-irregulars correspond to the component with the highest mean $a_1^{\rm max}$ (0.49), and also have the highest mean $a_2^{\rm max}$ (0.45) and $R_2^{\rm max}$ (0.87), together with the lowest $\mathcal{C}_{\rm 2D}$ (0.78; orange histograms in Fig.~\ref{fig: GMM_distributions}). The $j_\star$-bars are associated with the component having a high mean $a_2^{\rm max}$ (0.29) together with the lowest $R_2^{\rm max}$ (0.11; yellow histograms), while exhibiting the lowest mean values of $a_1^{\rm max}$ (0.14), and sharing the same $\mathcal{C}_{\rm 2D}$ as the $j_\star$-rings (0.95; yellow histograms). Finally, the $j_\star$-spirals are associated with the component exhibiting a high mean $a_2^{\rm max}$ as the $j_\star$-bars (0.29) but with a larger $R_2^{\rm max}$ (0.55), and feature a high mean $a_1^{\rm max}$ (0.31) and a relatively low $\mathcal{C}_{\rm 2D}$ (0.87; green histograms).

Because of the probabilistic nature of the GMM, each galaxy is described, after this interpretation, by a set of posterior probabilities that quantify its association with the different $j_\star$-types. In this framework, individual galaxies contribute simultaneously to multiple $j_\star$-types, naturally accounting for transitional systems and mixed morpho-kinematic states, expanding on the rigid, single-category classification system presented in  \citetalias{pacheco2026}.

\section{Results}\label{sec: Results}

We have divided the description of our results into three sections. In Sect.~\ref{sec: sAMSD_evo}, we analyse the temporal morpho-kinematic evolution of our sample using the posterior probabilities associated with each $j_\star$-type. In Sect.~\ref{sec: sAMSD_corre}, we report the strongest correlations between our morpho-kinematic metrics and other galaxy properties using their Spearman correlation coefficients ($\rho_s$). Finally, in Sect.~\ref{sec: Fall_relation}, we describe how our galaxies are grouped into different regions of the $j_\star-M_\star$ diagram depending on their morpho-kinematic characteristics, and we report the best-fit values for the Fall relation in our sample.

\subsection{Morpho-kinematic evolution}\label{sec: sAMSD_evo}

\begin{figure}
\centering
\includegraphics[width=\hsize]{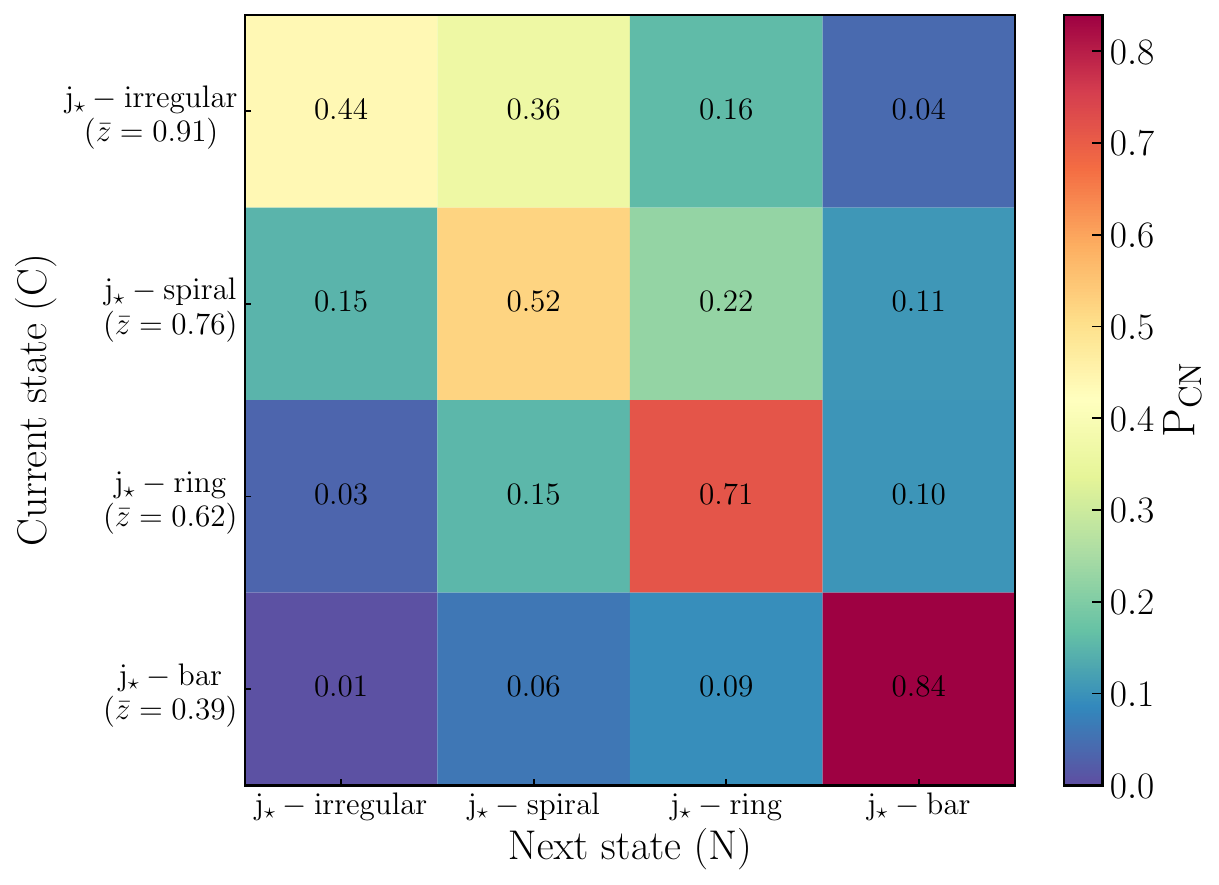}
\caption{Markov transition matrix for the different $j_\star$-types obtained from the posterior probabilities of the GMM classification. Each element $\rm P_{CN}$ represents the row-normalised probability that a galaxy snapshot associated with the current state ($\rm C$) evolves into the next state ($\rm N$) in the immediately following snapshot. The mean redshift for each $j_\star$-type is displayed below their corresponding label along the y-axis.}
\label{fig: Markov_matrix}
\end{figure}

The main aim of this section is to present and describe the temporal morpho-kinematic evolution of our sample. To do so, we used the $j_\star$-types posterior probabilities associated to each galaxy to compute the Markov transition matrix shown in Fig.~\ref{fig: Markov_matrix}. This matrix displays the $j_\star$-types transition probabilities ($\rm P_{CN}$) between consecutive snapshots of our sample. Each row of Fig.~\ref{fig: Markov_matrix} encodes the conditional probability of a galaxy evolving into the morpho-kinematic state $\rm N$ given the current state $\rm C$. By listing the largest $\rm P_{CN}$ off-diagonal values in descending order, we recover the same morpho-kinematic evolutionary sequence shown by the mean $j_\star$-types redshifts, namely $j_\star$-irregular ($\bar{z}=0.91$) $\rightarrow$ $j_\star$-spiral ($\bar{z}=0.76$) $\rightarrow$ $j_\star$-ring ($\bar{z}=0.62$) $\rightarrow$ $j_\star$-bar ($\bar{z}=0.39$). This is a clear statistical indication that this is the canonical morpho-kinematic evolutionary sequence of our sample. Furthermore, upon examining the $\rm P_{CN}$ diagonal elements, we found that $j_\star$-irregular and $j_\star$-spiral are transitional states, whereas $j_\star$-ring and $j_\star$-bar are attractors in the sAMSD space. Once a $j_\star$-ring or a $j_\star$-bar forms, it is highly likely to persist throughout the remaining evolution of the \textsc{TNG50 MW/M31} galaxies. This interpretation is further supported by the fact that $\mathcal{C}_{\rm 2D}$ and $a_1^{\max}$ exhibit the strongest anti-correlation among our morpho-kinematic metrics ($\rho_s=-0.61$), indicating that, within the framework established by the GMM, the physical processes promoting the development of $j_\star$-rings and $j_\star$-bars simultaneously suppress the formation of $j_\star$-irregularities and $j_\star$-spirals.

\begin{figure}
\centering
\includegraphics[width=\hsize]{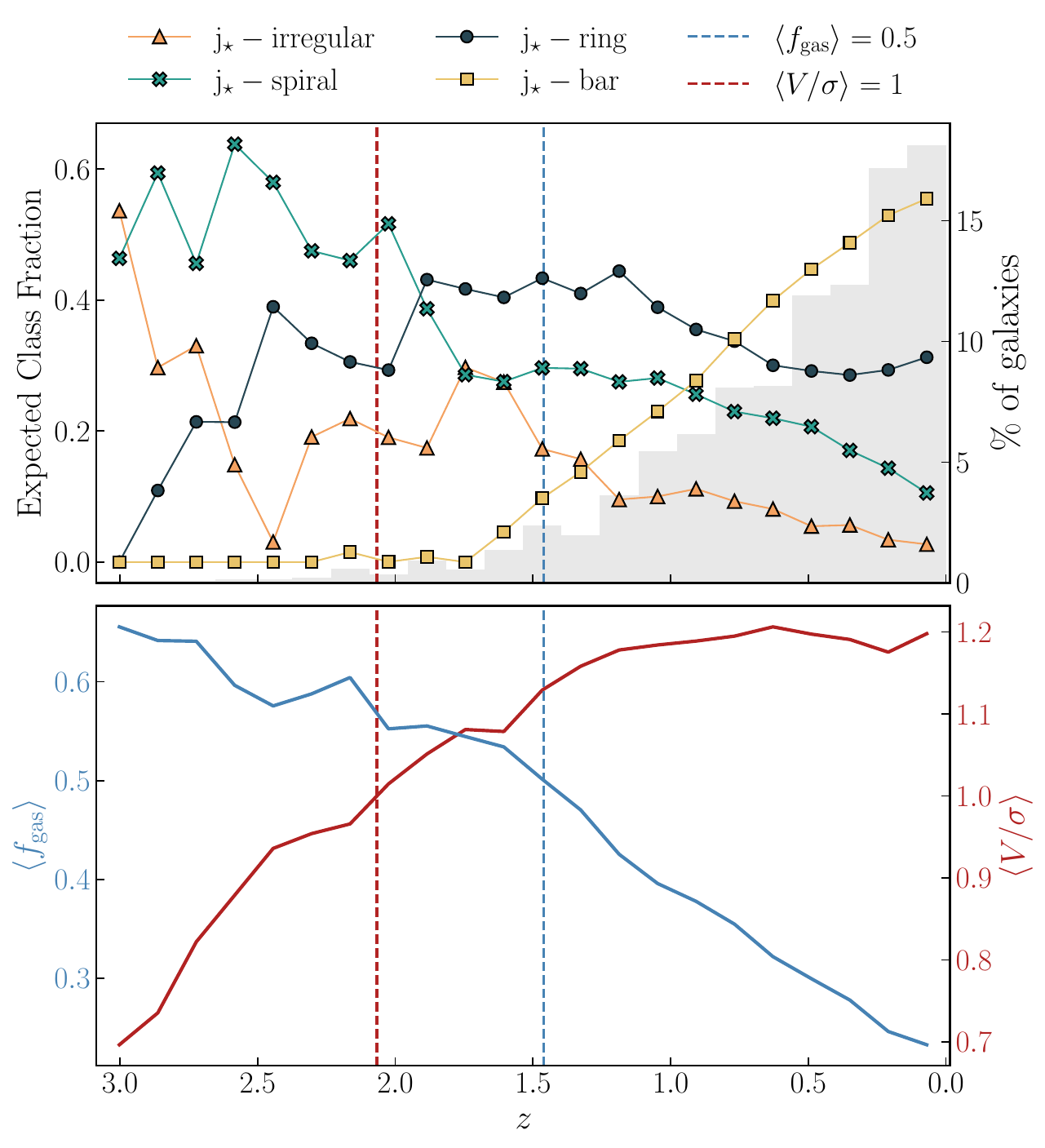}
\caption{Expected $j_\star$-types fraction (top), mean $f_{\rm gas}$ and mean $V/\sigma$ (bottom) with respect to redshift for our sample. Each $j_\star$-type is labelled at the top of the figure with its corresponding marker and colour. The blue and red dashed vertical lines in both panels mark the redshifts at which $\langle f_{\rm gas} \rangle = 0.5$ and $\langle V/\sigma \rangle = 1$, respectively. The grey histogram in the background of the top panel shows the percentage of galaxies within each redshift bin. A minimum of five galaxies per bin was ensured throughout the analysis. The x-axis is inverted such that the chronological evolution of the sample, from the past to the present, is read from left to right.}
\label{fig: Expected_class}
\end{figure}

In addition to the transitions between consecutive snapshots, we quantified the temporal evolution of the morpho-kinematic classes in our sample by computing their expected abundances as a function of redshift. The top panel of Fig.~\ref{fig: Expected_class} shows the mean posterior probability for each $j_\star$-type within a given redshift interval. This expected class fraction clearly shows that our galaxies exhibit predominantly $j_\star$-irregular/$j_\star$-spiral morpho-kinematics at the beginning of their evolution ($z\sim3$). As the redshift decreases, we observe that the abundance of $j_\star$-irregulars declines rapidly, whilst the $j_\star$-spirals remain as the dominant $j_\star$-type and the $j_\star$-rings begin to gain prominence. Around $z\sim2$, we find a significant decrease in the expected abundance of $j_\star$-spirals, leading to the emergence of $j_\star$-rings as the dominant $j_\star$-type at $z\sim1.8$. This behaviour persists until $z\sim0.8$, when $j_\star$-bars become dominant and remain the predominantly expected $j_\star$-type down to $z=0$.

\subsection{Morpho-kinematic correlations}\label{sec: sAMSD_corre}

The lower panel of Fig.~\ref{fig: Expected_class} shows the mean gas fraction (blue line) and the mean $V/\sigma$ (red line) of our sample within each redshift bin. For each galaxy, we define the gas fraction as $f_{\rm gas}=M_{\rm gas}/(M_{\rm gas}+M_{\star})$, where $M_{\rm gas}$ is the total gas mass enclosed within $R_{\max}$. $f_{\rm gas}$ is the interstellar medium property with the strongest degree of (anti-)correlation with our metrics, with $\rho_s = 0.64$ with respect to $a_1^{\max}$, $\rho_s = 0.51$ with respect to $R_2^{\max}$, and $\rho_s = -0.53$ with respect to $\mathcal{C}_{\rm 2D}$. These $\rho_s$ values suggest that gas-rich galaxies preferentially exhibit $j_\star$-irregular/$j_\star$-spiral morpho-kinematics, whereas $j_\star$-rings and $j_\star$-bars arise predominantly in systems with low gas content. This is precisely what we observe in Fig.~\ref{fig: Expected_class}, where the transition in dominance between the transient $j_\star$-types and the sAMSD attractors occurs gradually as $\langle f_{\rm gas} \rangle$ decreases, becoming fully stable once $\langle f_{\rm gas} \rangle < 0.5$ (dashed blue vertical line).

This result agrees with the angular momentum dissipation scenario, in which gas transfers its angular momentum to stars through gravitational torques \citep{lynden1972,bournaud2002}, predominantly within non-axisymmetric structures such as spiral arms and disc irregularities. This process remains active in gas-rich discs, whereas it is largely completed in gas-poor systems, where stellar angular momentum redistribution is mainly driven by bars and accumulated in rings \citep{athanassoula2002,kormendy2004}. The close link between the gas content and the stellar morpho-kinematics of our sample is one of the strongest results of this analysis.

Once the major role of $f_{\rm gas}$ had been understood, we decided to use it as a control variable for calculating partial correlations. This process revealed that only $V/\sigma$ exhibited $f_{\rm gas}$-independent correlations with our morpho-kinematic metrics, with $\rho_s = 0.43$ with respect to $\mathcal{C}_{\rm 2D}$ and $\rho_s = -0.41$ with respect to $a_2^{\max}$. This implies that, for a given amount of gas, it is the dynamical support of the stellar disc that determines its morpho-kinematics; rotationally supported discs develop $j_\star$-rings, whilst $j_\star$-bars and $j_\star$-spirals arise in the presence of non-circular motions. This can be clearly seen in Fig.~\ref{fig: Expected_class}, where the $\langle V/\sigma\rangle$ curve exhibits almost the same behaviour as that of the $j_\star$-rings, with $\langle V/\sigma \rangle = 1$ (dashed red vertical line) marking the start of the redshift range in which this $j_\star$-substructure becomes dominant.

We further investigated which properties of our sample influence the morpho-kinematics of our galaxies independently of the main drivers of galaxy evolution. To do so, we calculated the partial correlations for our metrics, this time using $M_\star$ and $z$ as control variables. Both cases yielded the same result: the mass-weighted gas metallicity ($\rm Z_{gas}$) and the three-dimensional distance to the nearest satellite ($r_{\rm sat}$) are the two properties that most strongly favour the emergence of one $j_\star$-type over another. We adopted the definition of satellite galaxies implemented in \citet{engler2021,engler2023} for which any \textsc{SubHalo} with $M_\star \geq 5 \times 10^6~\rm M_{\odot}$ (more than 100 stellar particles) within a radius of 300 kpc from the main \textsc{SubHalo} was classified as a satellite. The scenario with $M_\star$ ($z$) as the control variable shows that $\rm Z_{gas}$ has a $\rho_s = 0.42~(0.49)$ with respect to $\mathcal{C}_{\rm 2D}$, whilst $r_{\rm sat}$ has $\rho_s = -0.46~(-0.47)$ and $\rho_s = 0.58~(0.61)$ with respect to $\mathcal{C}_{\rm 2D}$ and $a_1^{\max}$, respectively. In summary, for galaxies with the same stellar mass or at the same redshift, systems with low metallicity and nearby satellites predominantly exhibit $j_\star$-irregular or $j_\star$-spiral morpho-kinematics which are transitional because more dynamically disturbed, whilst those with high metallicity and distant satellites tend to develop $j_\star$-rings and $j_\star$-bars, which are long-lived dynamical configurations. This suggests that the balance between circumgalactic gas accretion—providing fresh gas, merger-driven growth, and angular momentum—and stellar feedback—regulating the redistribution and recycling of gas—is the key physical mechanism driving the dichotomy between transitional states and attractors in the sAMSD space in the \textsc{TNG50} simulations. The other properties of our galaxies for which we did not detect any significant correlation with respect to our morpho-kinematic metrics are shown in Appendix~\ref{app: correlation_properties}.

\subsection{Metrics in the $j_\star-M_\star$ diagram}\label{sec: Fall_relation}

\begin{figure}
\centering
\includegraphics[width=.86\hsize]{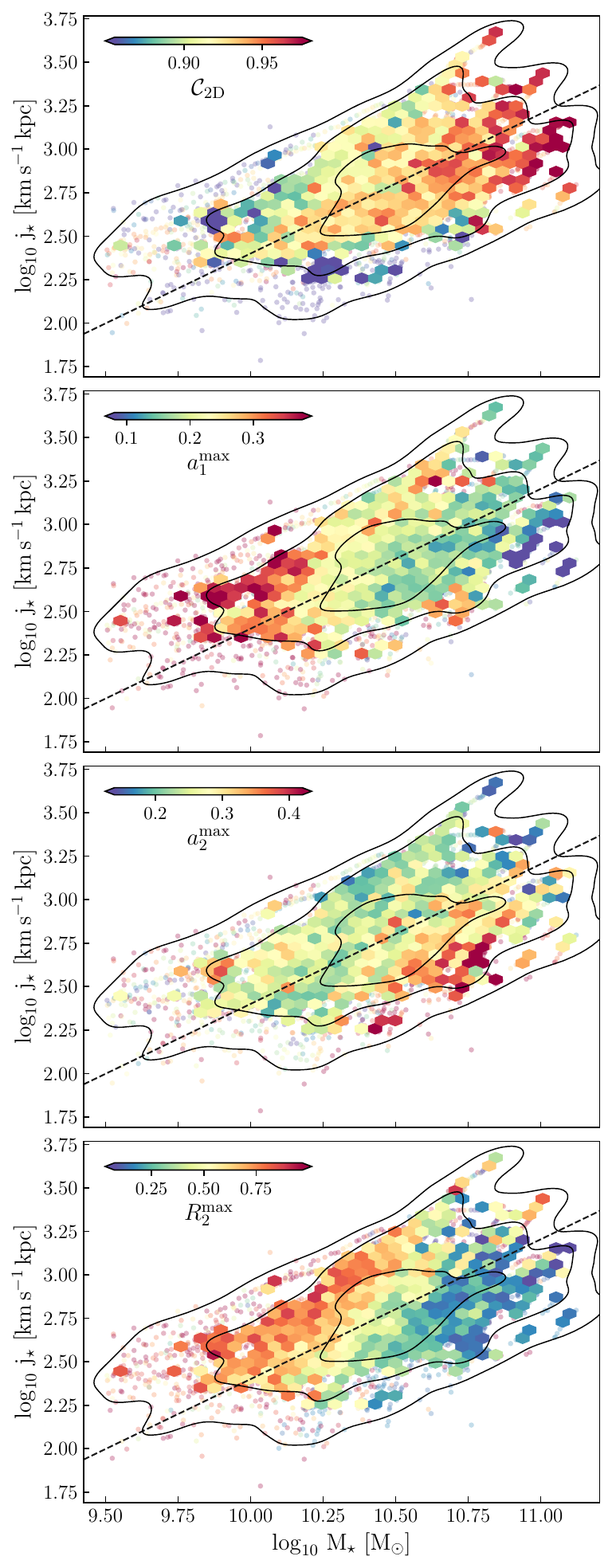}
\caption{Morpho-kinematic metrics in the $j_\star$--$M_\star$ diagram for our sample. The x-axis and y-axis show the total stellar mass and total stellar sAM, respectively. Hexagonal bins containing at least five galaxies are coloured according to the mean value of the corresponding metric, whilst individual galaxies outside the bins are shown with smaller and semi-transparent symbols. The minimum and maximum values of the colour bars correspond to the 10th and 90th percentiles of each metric distribution. The dashed black line represents the best-fitting Fall relation for our sample. The contours show the smoothed 2D density distribution of the galaxies, enclosing approximately the 1$\sigma$, 2$\sigma$, and 3$\sigma$ Gaussian-equivalent credible regions.}
\label{fig: Fall_relation_TNG50}
\end{figure}

Figure \ref{fig: Fall_relation_TNG50} shows, combining all $z$, the distribution of our galaxies in the $j_\star-M_\star$ diagram, colour-coded according to each of the four morpho-kinematic metrics. The black dashed line in all panels corresponds to our best-fitting Fall relation, written as

\begin{equation}
    \log_{10}{j_{\star}} = \alpha \left(\log_{10} {M_{\star}} - \log_{10} {\widetilde{M_{\star}}}\right) + \beta,
    \label{eq: Fall_relation}
\end{equation}

\noindent where $\widetilde{M_{\star}} = 3.19\times10^{10}~\rm M_\odot$ is the median stellar mass of our sample. The best-fit parameters, $\alpha = 0.80 \pm 0.01$ and $\beta = 2.81 \pm 0.01$, together with an intrinsic scatter of $\sigma_{\rm int} = 0.18 \pm 0.01$ dex, were obtained through a maximum-likelihood regression that explicitly includes a scatter term, assuming a Gaussian distribution of residuals about the relation. The fit was performed using stellar mass as a pivot variable to reduce covariance between slope and intercept. Parameter uncertainties were estimated from the 16th--84th percentile range of 1 000 bootstrap resamplings of the sample.

Compared with other reported values of $\alpha$, $\beta$, and $\sigma_{\rm int}$ for different subsamples of \textsc{IllustrisTNG50} galaxies, the slope of our stellar Fall relation is steeper than the range $0.30 \leq \alpha \leq 0.50$ reported by \cite{bouche2021} and the value $\alpha = 0.55$ reported by \cite{du2022}. This difference can be understood in terms of our sample selection, which excludes dispersion-dominated galaxies through a kinematic cut of $V/\sigma \geq 0.5$. This criterion preferentially retains rotation-supported systems and removes the population of low-angular-momentum galaxies that is known to contribute to a flattening of the $j_\star-M_\star$ relation \citep{bouche2021}. In contrast, our intercept at $M_\star = \widetilde{M_{\star}}$ shows a very good agreement ($0.20$ dex difference) with the $\beta = 3.01$ one of the relation inferred by \cite{du2022} from disc-dominated \textsc{TNG50} galaxies in the mass range $10^{9} \leq M_\star \leq 10^{11.5}~\rm M_\odot$. Likewise, our inferred intrinsic scatter is consistent with the regime-dependent dispersion reported by \cite{bouche2021}, which shows that the scatter of the $j_\star-M_\star$ sequence is closely linked to $V/\sigma$ for \textsc{TNG50}.

The four panels in Fig. \ref{fig: Fall_relation_TNG50} illustrate how galaxies with similar structures in their sAMSD maps cluster in different regions of the $j_\star-M_\star$ diagram.The top panel, coloured by $\mathcal{C}_{\rm 2D}$, shows that galaxies with higher $M_\star$ exhibit, on average, the largest values of this metric, whereas systems with lower $M_\star$ display the smallest. This trend shows that $j_\star$-rings and $j_\star$-bars are the predominant substructures in massive \textsc{TNG50} disc-type galaxies (see Sect.~\ref{Sec: GMM}). The second panel, from top to bottom, shows the distribution of $a_{1}^{\max}$. This metric is the strongest above the best fitting line in the low-$M_\star$ region, 
and gradually decreases towards the right-hand side of the diagram, reaching its lowest values in the high-$M_\star$ regime below the best fitting line. This behaviour indicates that $j_\star$-irregularities are most prevalent in low-mass systems and become progressively less significant as the simulated galaxies grow more massive, with the exception of the intermediate-mass $j_\star$-spirals galaxies with high $j_\star$, where $a_1^{\max}$ rises again. The third panel displays the values of $a_{2}^{\max}$. In contrast to the behaviour of $\mathcal{C}_{\rm 2D}$ and $a_{1}^{\max}$, no clear trend is observed with $M_\star$. Instead, the highest values of this metric are found below the best fitting line in the intermediate-to-high $M_\star$ region. The location of these systems in the $j_\star-M_\star$ plane is also consistent with the region occupied by observed $j_\star$-bars and $j_\star$-spirals. Finally, the bottom panel presents the $j_\star-M_\star$ diagram coloured by $R_{2}^{\max}$. Among our four morpho-kinematic metrics, $R_{2}^{\max}$ most clearly exhibits a gradual diagonal decline in its mean values. The $j_\star$-spirals are concentrated above the best fitting line in the lower-intermediate $M_\star$ region, whereas the $j_\star$-bars are located below the best fitting line in the upper-intermediate $M_\star$ region. Overall, the distribution of these substructures across the $j_\star-M_\star$ plane broadly agrees with the observational trends reported by \citetalias{pacheco2026}. We assessed the robustness of these mean trends against variations in the number of galaxies per hexagonal bin using bootstrap resampling. The results, presented in Appendix \ref{app: bootstraping} and illustrated in Fig. \ref{fig: bootstraping_std_TNG50}, provide statistical support for the trends described above.

\section{Discussion}\label{sec: Discussion}

We begin our discussion in Sect. \ref{sec: secular_model} by introducing the angular momentum redistribution scenario for secularly evolving discs, where we examine the impact of our sample selection criteria on the obtained results (see Sect.~\ref{sec: sample_effects}) together with the role of the galactic environment (see Sect.~\ref{sec: environment}). We conclude the discussion in Sect. \ref{sec: morphology_ comparison} by highlighting the advantages of our morpho-kinematic approach over traditional morphological analyses.

\subsection{Angular momentum redistribution within galactic discs}\label{sec: secular_model}

\begin{figure}
\centering
\includegraphics[width=1\hsize]{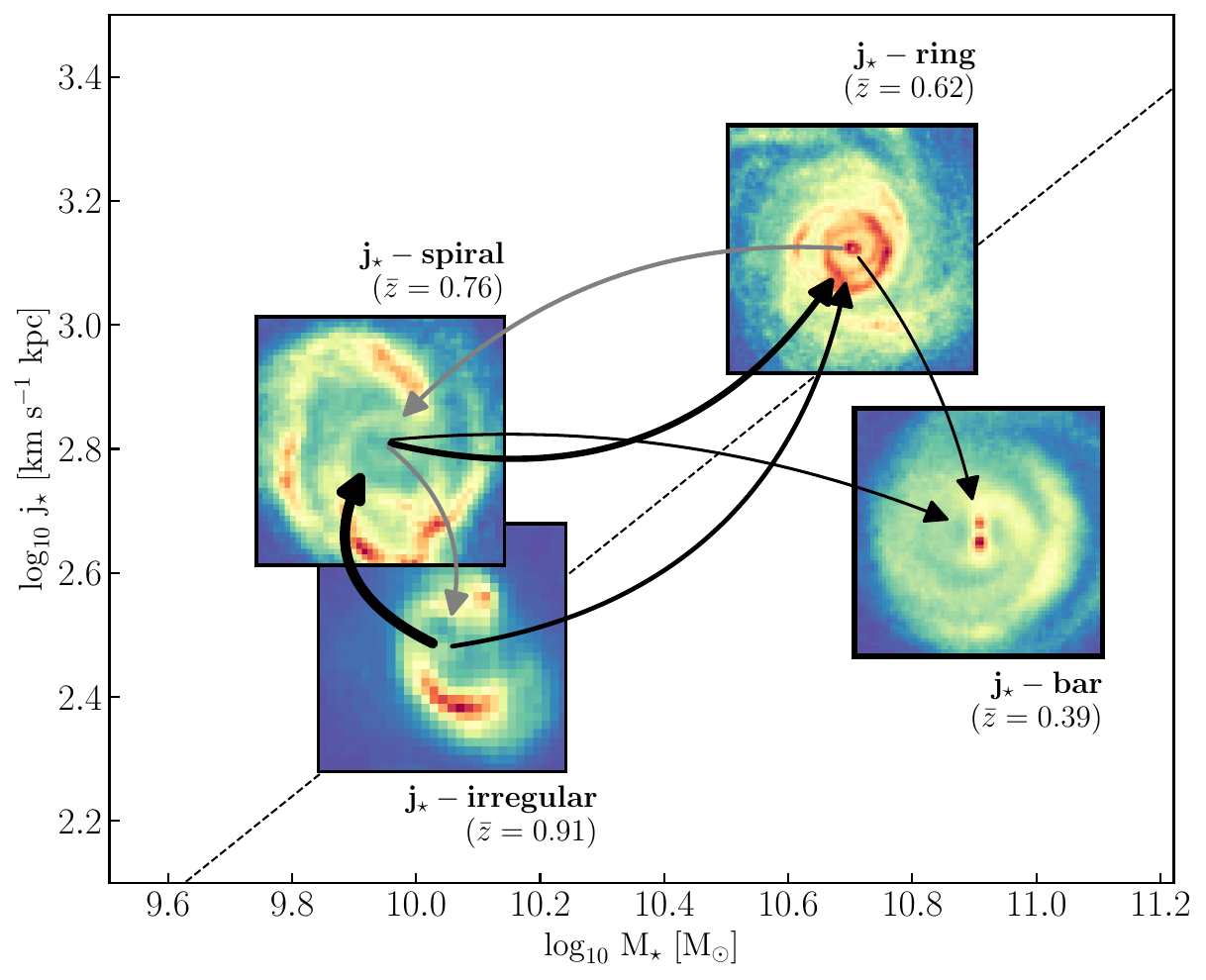}
\caption{The $j_\star$-types transition diagram in the $j_\star-M_\star$ plane. Each sAMSD map representing a morpho-kinematic class corresponds to the galaxy with the minimum Euclidean distance to the associated GMM centroid in the scaled four-dimensional parameter space, and is centred on its corresponding $M_\star$ and $j_\star$ values. The value of redshift, corresponding to the mean redshift from each class, is shown below the label indicating its $j_\star$-type. The arrows between maps represent transitions with $\rm P_{CN} \geq 0.10$ between $j_\star$-types. Black arrows indicate transitions from a $j_\star$-type with higher mean redshift to one with lower mean redshift, whilst grey arrows represent the opposite behaviour. The arrow width scales with $\rm P_{CN}$, whereas the frame width is proportional to the self-transition probability $\rm P_{CC}$. The dashed black line is the same best-fitting Fall relation shown in Fig.~\ref{fig: Fall_relation_TNG50}.}
\label{fig: Transition_flow}
\end{figure}

Using all the results presented so far, we have reconstructed the history of stellar angular momentum redistribution within the secularly evolving discs simulated in \textsc{TNG50}. To illustrate this clearly, we have shown in Fig.~\ref{fig: Transition_flow} 
the sAMSD maps of the best representatives of each $j_\star$-type are in their $j_\star-M_\star$ parameter space, connected by the lines defined by their $\rm P_{CN} \geq 0.10$ transitions in Fig. \ref{fig: Markov_matrix}. Figure~\ref{fig: Transition_flow} provides a visual representation of the preferred evolutionary pathways linking the different morpho-kinematic states identified throughout our analysis. It is important to remember that, as mentioned at the end of Sect.~\ref{Sec: GMM}, the probabilistic nature of our labels allows individual galaxies to simultaneously host multiple $j_\star$-substructures. Consequently, the intrinsically mixed nature of galaxy morpho-kinematics is naturally incorporated into our framework, avoiding the need to force galaxies into a single discrete class at any stage of their evolution.

In this scenario, discs typically begin their morpho-kinematic evolution as a $j_\star$-irregular ($\bar{z}=0.91$), owing to the large amount of primordial gas characteristic of its formation environment. From this point onwards, the two possible evolutionary paths are, in descending order of probability: (i) if the disc retains a large proportion of its gas content through gentle dynamical interactions or gas inflow from its surroundings, the stellar angular momentum will come to be stored mainly in its spiral arms, transforming the disc into a $j_\star$-spiral; (ii) if, on the other hand, there are no mechanisms to feed its gas reservoir, and there is a moderate decrease in its $f_{\rm gas}$ as a result of star formation, the galaxy will redistribute its angular momentum mainly to a ringed region characteristic of an unperturbed Freeman disc, and will become a $j_\star$-ring.

Following the chronological sequence of the most likely path, once our disc is classified as a $j_\star$-spiral ($\bar{z}=0.76$), it has three options: to form a $j_\star$-ring, to come back as a $j_\star$-irregular, or to form a $j_\star$-bar, listed in order of decreasing probability. Once again, the factor that tips the balance in favour of any of the scenarios mentioned is the galaxy's gas reservoir. If our disc maintains an average star formation rate without accreting gas from its intergalactic medium, it will form a $j_\star$-ring. If, on the other hand, our disc accretes pristine gas due to interactions with its nearby satellites, it will revert to being a $j_\star$-irregular. Finally, if the disc experiences an abrupt decrease in $f_{\rm gas}$, due to highly efficient star formation or intense supermassive black hole feedback, it will be classified as a $j_\star$-bar. Continuing the sequence, once our disc meets the conditions to transform into a $j_\star$-ring ($\bar{z}=0.62$), it has only two further evolutionary steps: either to revert to a $j_\star$-spiral or to develop a $j_\star$-bar ($\bar{z}=0.39$), the former being slightly more likely. Once again, gas accretion and stellar feedback will tip the balance one way or the other.

The persistence of $j_\star$-rings throughout the evolution of our discs, together with their natural prevalence as the final stage in the morpho-kinematic evolution of most of our galaxies, is fully consistent with what has been reported observationally \citepalias{pacheco2026}, despite we did not expect as many $j_\star$-bars, nor the absence of the $j_\star$-clump category in the simulations. Regarding the former, we have identified that, once a $j_\star$-bar appears in our simulations, it facilitates the consumption of pristine gas by channelling it towards the centre of the disc. This process efficiently depletes the galaxy gas reservoir, further enhancing the prominence of the $j_\star$-bar throughout its subsequent evolution. The absence of effective regulatory mechanisms for stellar bars is also consistent with the results reported by \cite{frosst2025} for \textsc{TNG50}. Regarding the second point, it is clear that the formation of stellar clumps in \textsc{TNG50} is a process highly limited by resolution, meaning that their appearance, scales and lifetimes do not reflect physical processes but are numerical artefacts, making them incomparable with observations \citep{celiz2025}. These two discrepancies may be a clear indicator of the need to refine the feedback processes within galactic discs in the \textsc{TNG50} simulation.

\subsubsection{Sample selection effects}\label{sec: sample_effects}

The transient and attractor nature of our $j_\star$-types may be partially linked to the construction of our sample. The grey histogram in the first panel of Fig.~\ref{fig: GMM_distributions} shows that the majority of galaxies in our sample (84\%) have $\mathcal{C}_{\rm 2D} \geq 0.88$. The distribution of  $\mathcal{C}_{\rm 2D}$ is strongly skewed toward high values. This concentration is a direct consequence of the sample selection. $\mathcal{C}_{\rm 2D}$ is, by definition, maximised for Freeman-type disc morphologies (see Eq.~\ref{eq: C2D}) while all systems in our parent \textsc{TNG50 MW/M31} sample are disc galaxies at $z=0$ \citep{pillepich2024milky}.

We find that the inclusion of dynamically stable galaxies at $z>0$ only modestly broadens the $\mathcal{C}_{\rm 2D}$ distribution. The low median value of $a_1^{\max}$ and its relatively narrow distribution (see the grey histogram in the second panel of Fig.~\ref{fig: GMM_distributions}) can also be understood as a consequence of both the morphological cutoff at $z=0$ and the imposed dynamical stability of our sample. These selection criteria reduce the prevalence of irregular substructures within the discs by preferentially excluding low-mass and gas-rich galaxies, thereby suppressing high $a_1^{\max}$ values. Maximising $\mathcal{C}_{\rm 2D}$ in conjunction with reducing $a_1^{\max}$ directly favour the prevalence of $j_\star$-rings and $j_\star$-bars over $j_\star$-irregulars and $j_\star$-spirals in our sample.

\subsubsection{Impact of the environment on morpho-kinematics}\label{sec: environment}

The position of galaxies within the cosmic web has been shown to play an important role in regulating their angular momentum acquisition and alignment with the surrounding large-scale structure \citep[see e.g.][]{codis2012,tempel2013,codis2015,lee2018,kraljic2020}. Motivated by this, we investigated whether the environmental conditions of our sample influence the formation of $j_\star$-substructures within galactic discs. To do this, for each galaxy in our sample at $z=0$ (128 galaxies in total), we computed its associated local density, in Mpc$^{-3}$, using the Delaunay tessellation field estimator ($\rho_{\rm dtfe}$), together with its distances, in Mpc, to the nearest filament ($D_{\rm fil}$), node ($D_{\rm node}$), and wall ($D_{\rm wall}$). We also calculated the cosine of the alignment angle between the total angular momentum vector of each galaxy and that of its nearest filament ($\cos\theta_{j_\star}$). All these quantities were extracted from the \textsc{TNG50} cosmic web reconstruction performed following the procedure described in Appendix~\ref{app: cosmic_web}.

When comparing these parameters with our morpho-kinematic metrics, we found that only $\mathcal{C}_{\rm 2D}$ and $a_1^{\max}$ show emerging correlations ($|\rho_s| \geq 0.20$) with the location of galaxies within the cosmic web. In general, $\rho_{\rm dtfe}$ correlates only with $a_1^{\max}$ ($\rho_s = 0.31$), suggesting that locally dense environments favour the emergence of $j_\star$-irregularities. Regarding the distance measurements, $a_1^{\max}$ exhibits anticorrelations with $D_{\rm node}$ ($\rho_s = -0.25$), $D_{\rm fil}$ ($\rho_s = -0.25$), and $D_{\rm wall}$ ($\rho_s = -0.22$), whilst $\mathcal{C}_{\rm 2D}$ correlates solely with $D_{\rm fil}$ ($\rho_s = 0.22$). These preliminary results suggest that $j_\star$-irregular morpho-kinematics preferentially arise in galaxies located farther from nodes, filaments, and walls, whereas $j_\star$-rings tend to inhabit environments closer to filaments. The absence of any correlation with $a_2^{\max}$ and $R_2^{\max}$ seems to indicate that the environment has little influence on the formation of $j_\star$-bars.

\begin{figure}
\centering
\includegraphics[width=0.9\hsize]{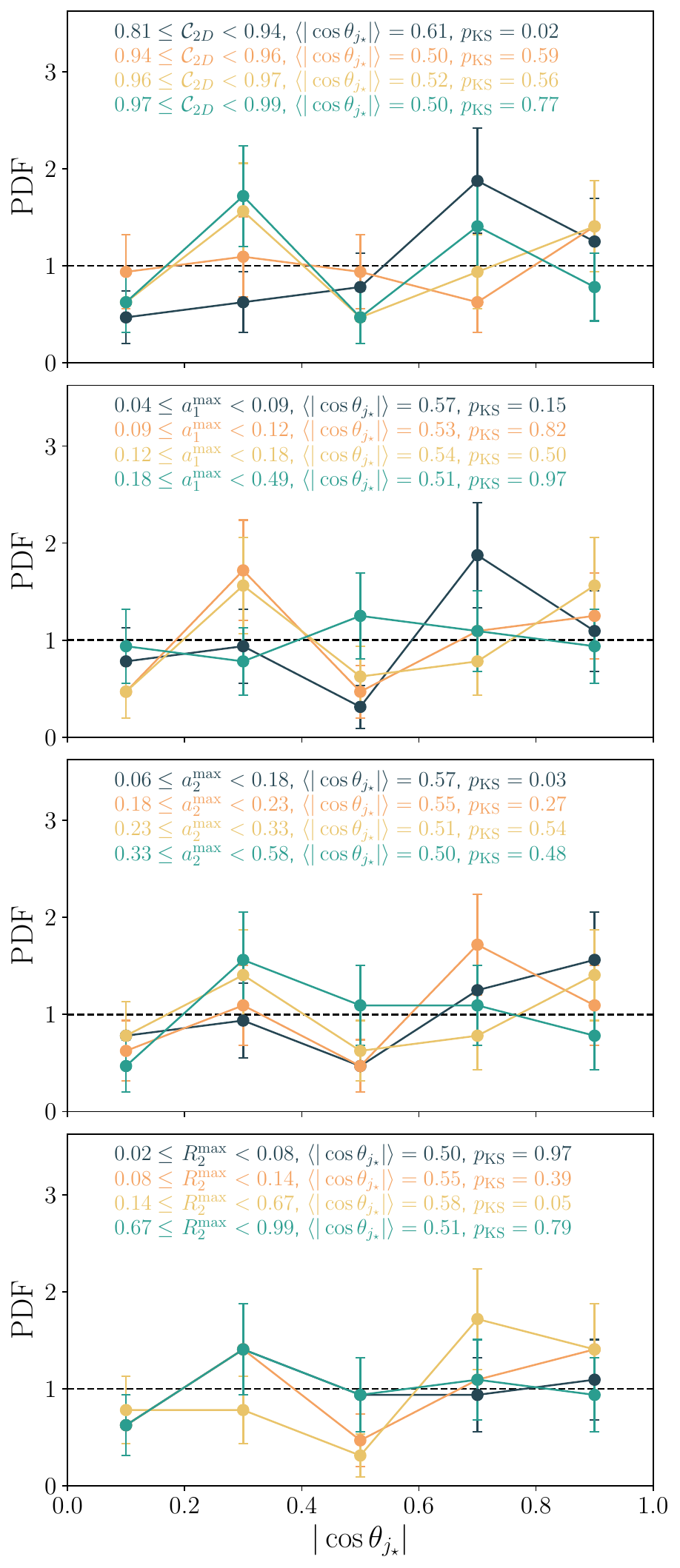}
\caption{Angular momentum orientation of our galaxies with respect to their nearest filament for different morpho-kinematic bins. $\theta_{j_\star}$ is the angle between the total angular momentum vector of each galaxy and that of its nearest filament. From top to bottom, the PDF of $\cos\theta_{j_\star}$ is computed in intervals of $\mathcal{C}_{\rm 2D}$, $a_1^{\max}$, $a_2^{\max}$, and $R_2^{\max}$. The intervals were defined by the quartiles of each metric, corresponding to the 0th, 25th, 50th, 75th, and 100th percentiles of their distributions. The horizontal black dashed line represents a random distribution. Error bars correspond to the Poisson noise in each bin, providing an estimate of the number of galaxies. The values of $\langle\cos\theta_{j_\star}\rangle$ and $p_{\rm KS}$-test are shown in each panel using the same colour as their corresponding metric interval.}
\label{fig: cos_theta}
\end{figure}

In order to determine whether there is any dependence between $\cos\theta_{j_\star}$ and our $j_\star$-types, we computed the probability distribution function (PDF) of $\cos\theta_{j_\star}$ in different bins of our morpho-kinematic metrics, together with $\langle\cos\theta_{j_\star}\rangle$ and the corresponding Kolmogorov–Smirnov (KS) test p-value ($p_{\rm KS}$) relative to a random distribution, as shown in the four panels of Fig.~\ref{fig: cos_theta}. Examining the values of $\langle\cos\theta_{j_\star}\rangle$, $p_{\rm KS}$, and the overall behaviour of the PDFs, we find that the distributions of $\cos\theta_{j_\star}$ are consistent with a random orientation for most of our $j_\star$-types. This result indicates that the substructures in which angular momentum is redistributed within our discs are largely independent of the alignment between the galaxy angular momentum vector and their nearest filaments. The absence of a significant correlation between our morpho-kinematic metrics and $\cos\theta_{j_\star}$, could be explained by the  stellar mass range covered by our sample. As shown in  hydrodynamic cosmological simulations \citep[see e.g.][]{codis2018,kraljic2020} this mass range corresponds roughly to a transition mass at $z=0$ from the parallel to perpendicular alignment for which a nearly random spin-filament orientation is expected.

\subsection{Morphokinematics vs morphology}\label{sec: morphology_ comparison}

Unlike in \citetalias{pacheco2026}, we cannot directly compare the morphological types of all our dis cs with their morpho-kinematic counterparts, since a traditional morphological classification is not available for galaxies at $z > 0$ in our sample. Nevertheless, we can still perform such a comparison at $z=0$, where we contrast our $j_\star$-types with the morphological classification of the parent sample. We recall that between 65\% and 75\% of galaxies in \textsc{TNG50 MW/M31} are classified as barred, depending on whether the classification of \citet{zana2022} or \citet{rosas2022} are adopted, while all systems are classified as spiral galaxies \citep{pillepich2024milky}. These fractions differ from the morpho-kinematic class distribution at $z=0$, where we find 55\% $j_\star$-bars and 11\% $j_\star$-spirals (see Fig.~\ref{fig: Expected_class}). More specifically, among galaxies classified as morphological bars by \citet{zana2022} and included in our sample, 1\% are identified as $j_\star$-irregular, 6\% are $j_\star$-spiral, 20\% are $j_\star$-ring and 73\% are identified as $j_\star$-bars. When adopting the criteria of \citet{rosas2022}, these fractions become 1\%, 7\%, 23\%, and 69\%, respectively. This comparison shows that the presence of morphological bars or spiral arms in simulations does not necessarily imply the existence of their morpho-kinematic counterparts. This result, already established observationally by \citetalias{pacheco2026}, applies to all our $j_\star$-types. 

Although the fraction of stellar mass accumulated within morphological substructures contributes directly to the redistribution of angular momentum throughout galactic discs, the kinematic information of galaxies plays a dominant role in shaping their structure within the sAMSD space. Key phenomena in galaxy evolution—such as fluctuations in the gravitational potential \citep{lynden1967,binney2013}, the emergence of resonances \citep{lynden1972,antoja2018}, and the propagation of density waves \citep{lin1964,lin1966,binney2008}—are directly encoded in stellar dynamics. A purely morphological description of galaxies therefore fails to capture them, as morphology is merely the surface expression of large- and small-scale density fluctuations in the disc, induced by gravity, hydrodynamical forces, and turbulence at all scales (including gas components). On the other hand, kinematics capture these phenomena as the direct manifestation of the galaxy's underlying dynamics. Angular momentum, as a fundamental and essential physical quantity, complements both morphological and kinematic information \citep[see e.g.][]{binney2008}, and as a consequence, these phenomena emerge naturally in our morpho-kinematic description.

\section{Conclusions}\label{sec: Conclusions}

Using a methodology that enables us to visualise and quantify the distribution of stellar angular momentum within galactic discs, we have developed a probabilistic morpho-kinematic classification system for a subsample of galaxies from the \textsc{TNG50} simulations based on four metrics. The $\sim$ 8 000 galaxies we analysed cover a stellar mass range of $10^{9.5}$ to $10^{11.1}$ $\rm M_\odot$, with a maximum redshift of $z = 3.50$. They all resemble the morphological, kinematic, and environmental characteristics of the Milky Way and M31 at $\rm z = 0$ \citep{pillepich2024milky}. We investigate how stellar angular momentum is stored and redistributed within simulated discs, providing the first direct confirmation in simulations of the rich diversity of $j_\star$-substructures previously identified in observations. The morpho-kinematic structures found in our sample are fully consistent with the classification framework proposed by \citetalias{pacheco2026} for late-type galaxies. In particular, the sAMSD maps of our galaxies predominantly contain four of the five observationally identified $j_\star$-substructures: $j_\star$-irregularities, $j_\star$-spirals, $j_\star$-rings, and $j_\star$-bars. The absence of $j_\star$-clumps, when compared with the observations, is most likely a consequence of the limited spatial resolution and numerical constraints of \textsc{TNG50} \citep{celiz2025}.

By tracking the temporal evolution of simulated galaxies within the sAMSD space, we determined the canonical morpho-kinematic evolutionary pathway followed by our sample. This result, together with the correlations between our morpho-kinematic metrics and other galaxy properties, enabled us to reconstruct a scenario for the redistribution of angular momentum within secularly evolving galactic discs. Within this framework, $f_{\rm gas}$ emerges as the property most strongly correlated with our metrics and therefore as the primary driver of the stellar morpho-kinematics of our discs. In this picture, galaxies initially store a significant fraction of their angular momentum in $j_\star$-irregularities as a consequence of their high gas content. As their gas reservoirs become depleted, angular momentum is progressively redistributed into $j_\star$-spirals, and, once ordered stellar rotational support is maximised, the discs develop their characteristic $j_\star$-rings. During the final stages, when only small amounts of gas remain, stellar angular momentum becomes increasingly concentrated within $j_\star$-bars. Nevertheless, this evolutionary pathway is neither strictly sequential nor linear. Depending on the physical properties and evolutionary history of each galaxy, systems may return to previous morpho-kinematic categories or skip intermediate stages altogether. Moreover, different $j_\star$-substructures can coexist within the same galaxy over extended periods, highlighting the continuous and complex nature of angular momentum redistribution in disc galaxies.

Within this scenario, the transition probabilities between $j_\star$-types reveal a clear dichotomy in the sAMSD space. While $j_\star$-irregularities and $j_\star$-spirals behave as transitional states, $j_\star$-rings and $j_\star$-bars act as attractors, with $f_{\rm gas}$ driving this separation. In a similar way, partial correlations with respect to our metrics indicate that gas accretion from the circumgalactic medium and the efficiency of stellar feedback in the \textsc{TNG50} simulations are the most significant physical processes associated with each group of $j_\star$-types, respectively. This dichotomy naturally explains the high expected class fractions of $j_\star$-rings and $j_\star$-bars at $z=0$ for our discs. However, the prominence of these classes is likely influenced, at least in part, by the selection criteria adopted to construct our sample, together with the absence of efficient regulatory mechanisms for bar destruction in \textsc{TNG50} \citep{frosst2025}. Despite these limitations, the secular morpho-kinematic evolution scenario described here is consistent with the angular momentum dissipation framework, in which stellar angular momentum redistribution is primarily driven by the gravitational torques exerted by non-axisymmetric structures in gas-rich discs \citep[see e.g.][]{lynden1972,bournaud2002}, and by bars and rings in gas-poor systems \citep[see e.g.][]{athanassoula2002,kormendy2004}. The analysis of correlations between our metrics and the environment of our sample at $z=0$ indicates that environmental conditions around our discs do not significantly influence their shape in the sAMSD space.

The morpho-kinematic evolution of our sample is also reflected in the $j_\star$--$M_\star$ diagram, where our galaxies naturally cluster into regions defined by our metrics. Stellar mass and total stellar sAM are strongly linked to the $j_\star$-substructures and their transformations in the \textsc{TNG50} discs, as is also seen in observations \citepalias{pacheco2026}. The parameters of the best-fit Fall relation for our sample lie within the ranges reported by other authors using different \textsc{TNG50} subsamples \citep{bouche2021,du2022}. It is important to emphasise that, as already seen at $z=0$, there is no one-to-one correspondence between morphological substructures and $j_\star$-substructures in our sample. The description of galaxies in the sAMSD space therefore provides a natural framework that integrates dynamical and morphological information, making the morpho-kinematic analysis a more detailed tool than traditional morphology for understanding the processes that shape galaxies.

Finally, it is necessary to extend the scope of our sAMSD analysis in order to confirm the morpho-kinematic trends presented here. To this end, we need to expand the sample of galaxy simulations using different cosmological hydrodynamic suites that would allow us to explore other ranges of redshift and stellar mass, as well as new subgrid physical models. Increasing the numerical resolution of the analysed simulations, together with a detailed study of the correlation between the environment of galactic discs and their morpho-kinematic diversity, will be key to understanding the main drivers of angular momentum accumulation and redistribution within galaxies. Similarly, it is essential to continue characterising the observational $j_\star$-substructures. The sample of galactic disc observations analysed in the sAMSD space must be expanded to include galaxies at $z > 0$. Only through the synergy of observational and simulated data will it be possible to construct the definitive scenario of morpho-kinematic evolution for the galaxies in our Universe.

\begin{acknowledgements}
We acknowledge the use of data from the IllustrisTNG simulations, undertaken with the computing time awarded by the Gauss Centre for Supercomputing (GCS) under GCS Large-Scale Projects GCS-ILLU and GCS-DWAR on the GCS share of the supercomputer Hazel Hen at the High Performance Computing Center Stuttgart (HLRS), as well as on the machines of the Max Planck Computing and Data Facility (MPCDF) in Garching, Germany. We also acknowledge the Python community for developing and maintaining open-source tools that enabled this research. In particular, we made use of \textsc{Astropy}\footnote{\url{https://astropy.org}} \citep{astropy2013, astropy2018}, \textsc{Scipy}\footnote{\url{https://scipy.org/}} \citep{scipy2020}, \textsc{Numpy}\footnote{\url{https://numpy.org/}} \citep{numpy2020}, and \textsc{Matplotlib}\footnote{\url{https://matplotlib.org/}} \citep{matplotlib2007}. K.K. acknowledges the funding of the French Agence Nationale de la Recherche for the project GALBAR (grant ANR-25-CE31-4684).
\end{acknowledgements}

\bibliographystyle{aa} 
\bibliography{References}

\begin{appendix}

\section{Comparison with observations}\label{app: Obs_Vs_Simu}

\begin{figure}[h!]
\centering
\includegraphics[width=\hsize]{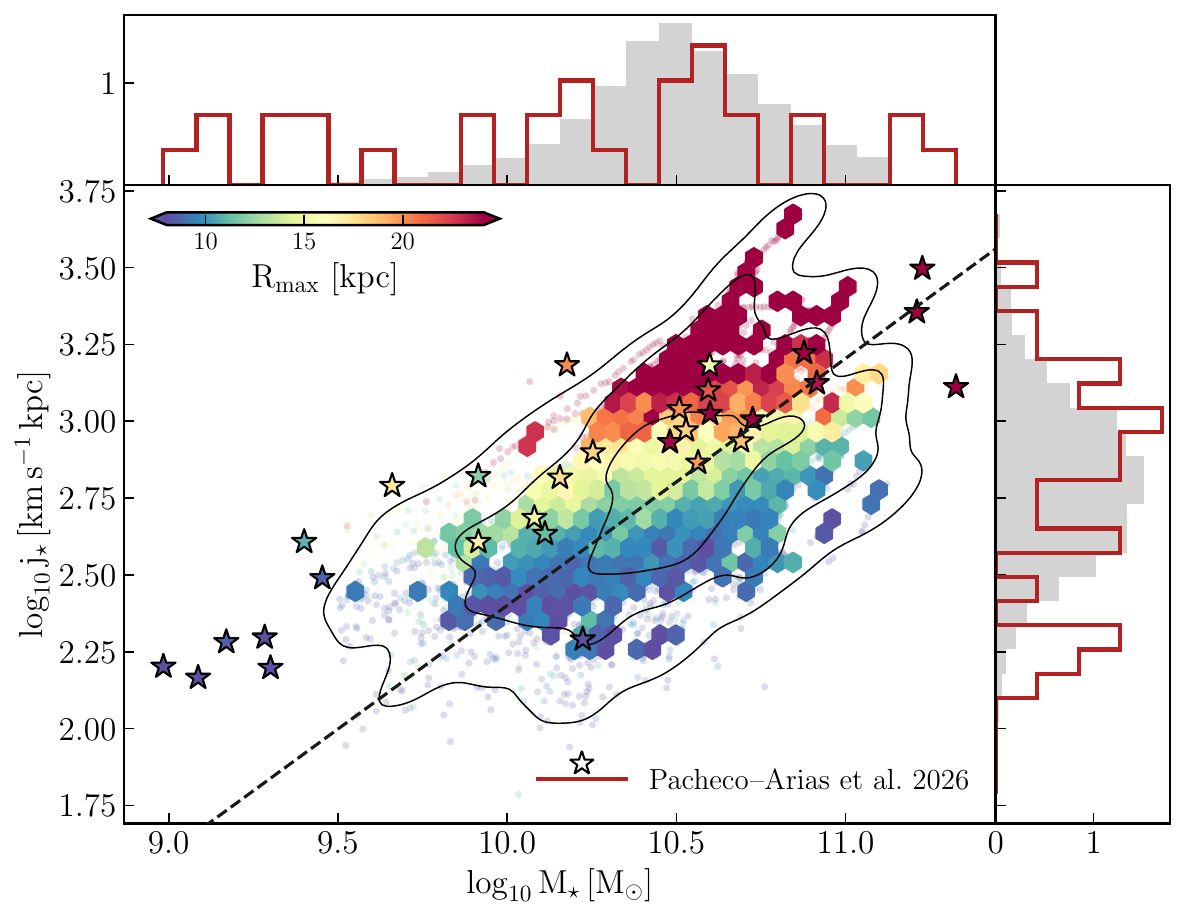}
\caption{Comparison between observations and simulations in the $j_\star-M_\star$ diagram colored by $R_{\max}$. The star markers of the central panel correspond to the galaxy observations analysed in \citetalias{pacheco2026}, whilst the hexagonal bins and scattered points represent the TNG50 simulation galaxies, following Fig.~\ref{fig: Fall_relation_TNG50}. For the central panel the axes, the minimum and maximum values of the coloured bar, the black dashed line, and the contours levels are described in Fig. \ref{fig: Fall_relation_TNG50}. The density normalised histograms for the stellar mass (top panel) and stellar sAM (right panel) show the counts for our galaxies using filled light grey bars, whilst the observations are display using the red line contours.}
\label{fig: Obs_Vs_Simu}
\end{figure}

The sample of observations to be compared with the TNG50 simulations of the present analysis consists of 30 late-type galaxies in the local universe that form part of GHASP \citep{epinat2008ghaspb,epinat2008ghaspa,korsaga2019ghasp}, and which are described in detail in \citetalias{pacheco2026}. To visualise how the simulations compare in size, mass and stellar sAM with these observations, Fig.~\ref{fig: Obs_Vs_Simu} shows both populations overlaid on the $j_\star-M_\star$ diagram, coloured by their $R_{\max}$. There we can see how the star-like markers corresponding to the \citetalias{pacheco2026} galaxies span the full range of values for the colour bar and coordinate axes, as the scattered points and hexagonal bins representing our galaxies do. Similarly, the $j_\star$ histograms for both samples, shown on the right-hand side of Fig.~\ref{fig: Obs_Vs_Simu}, are equivalent in shape and range, whilst the histograms for $M_\star$, located at the top of Fig.~\ref{fig: Obs_Vs_Simu}, show how the observations include low-mass discs that are not represented in the simulations due to their numerical resolution limitations. In morphological terms, the observations by \citetalias{pacheco2026} consist of 28 spiral galaxies and 2 irregulars. Of these 30 discs, 18 are barred, accounting for 60\% of the total sample. These proportions are entirely consistent with the morphology of our galaxies at $z=0$. 

\newpage

\section{Morpho-kinematic zoo}\label{app: Morpho_zoo}

\FloatBarrier

\begin{figure}[ht!]
    \centering
    \begin{subfigure}[b]{0.49\textwidth}
        \includegraphics[width=\textwidth]{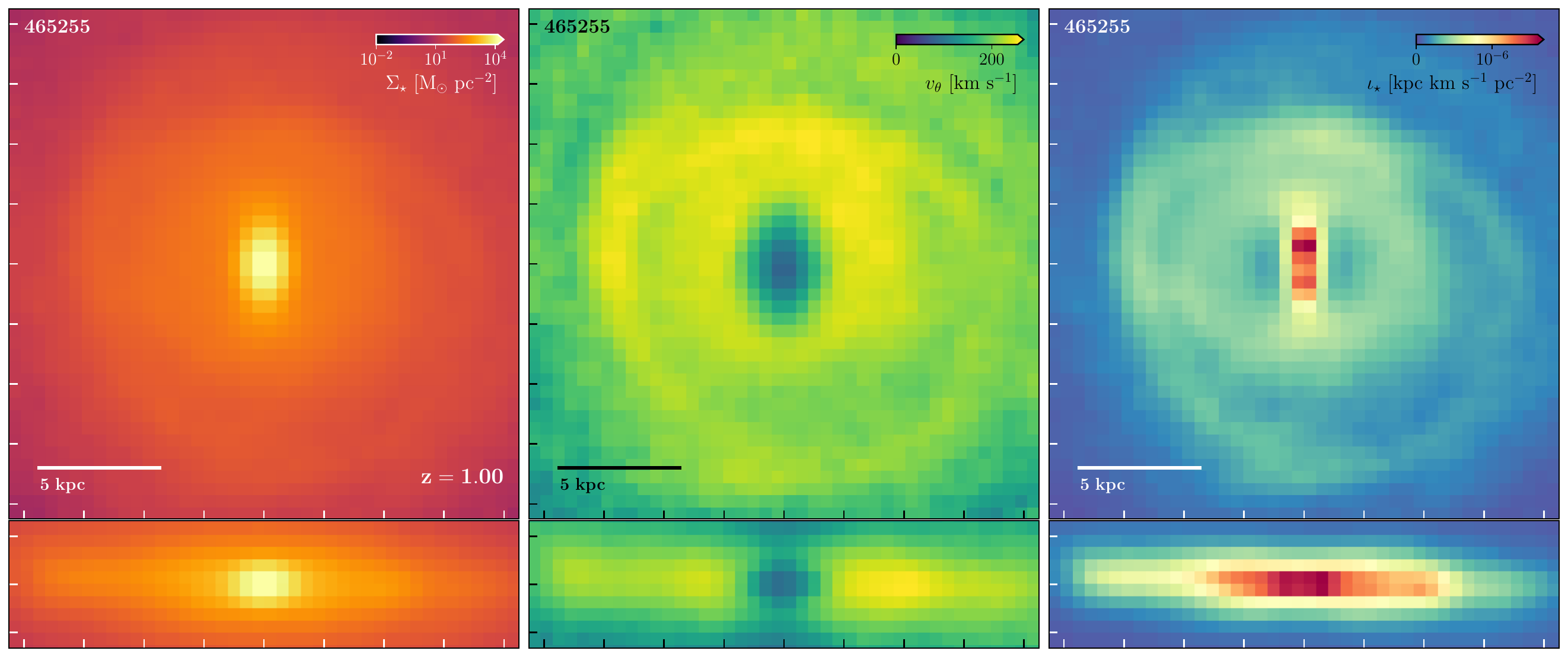}
    \end{subfigure}
    \\
    \begin{subfigure}[b]{0.49\textwidth}
        \includegraphics[width=\textwidth]{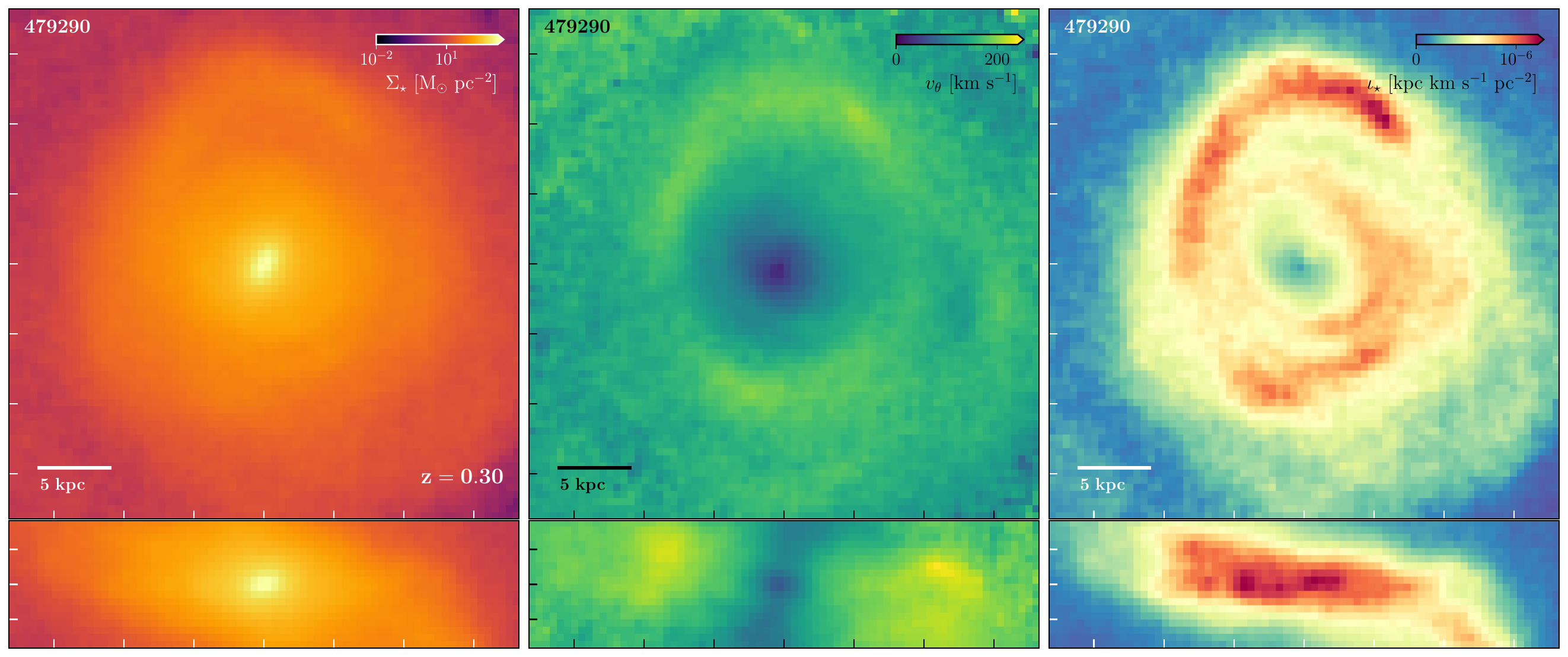}
    \end{subfigure}
    \\
    \begin{subfigure}[b]{0.49\textwidth}
        \includegraphics[width=\textwidth]{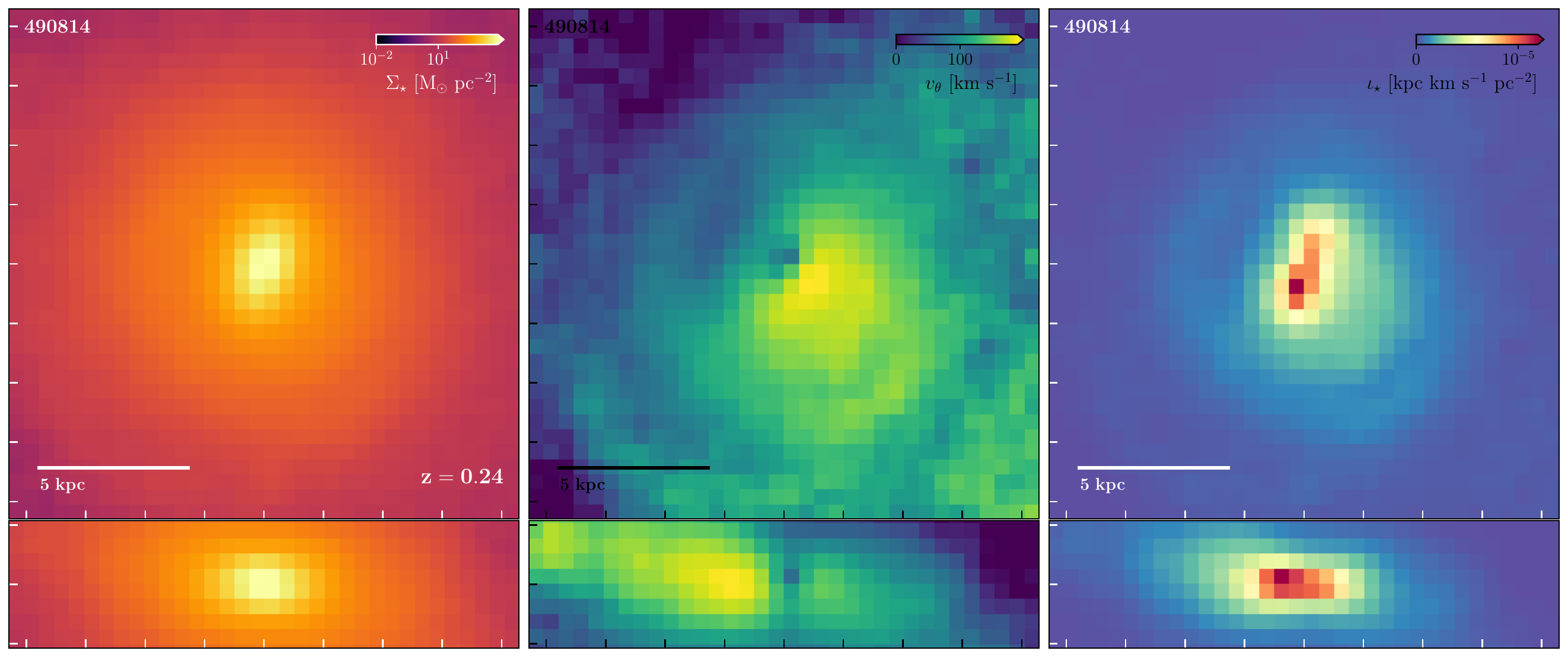}
    \end{subfigure}
    \\
    \begin{subfigure}[b]{0.49\textwidth}
        \includegraphics[width=\textwidth]{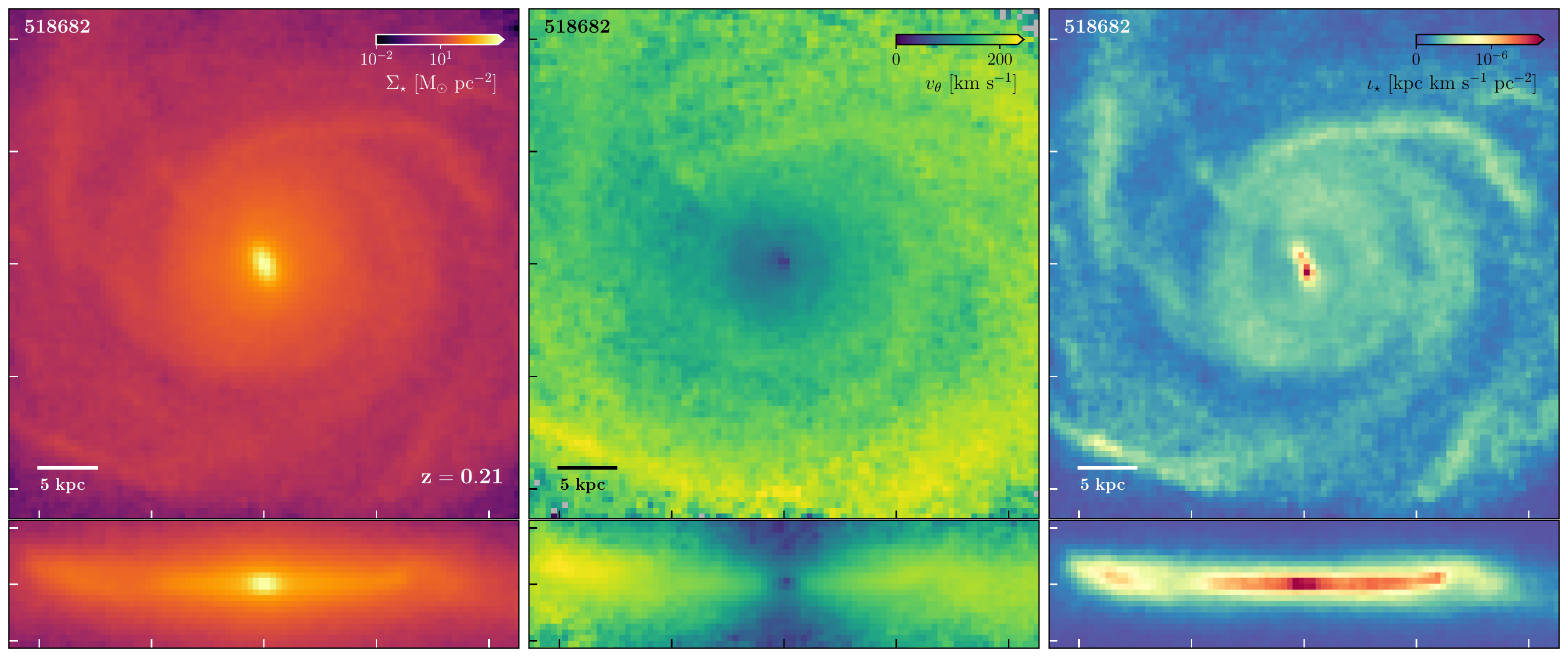}
    \end{subfigure}
    \\
    \begin{subfigure}[b]{0.49\textwidth}
        \includegraphics[width=\textwidth]{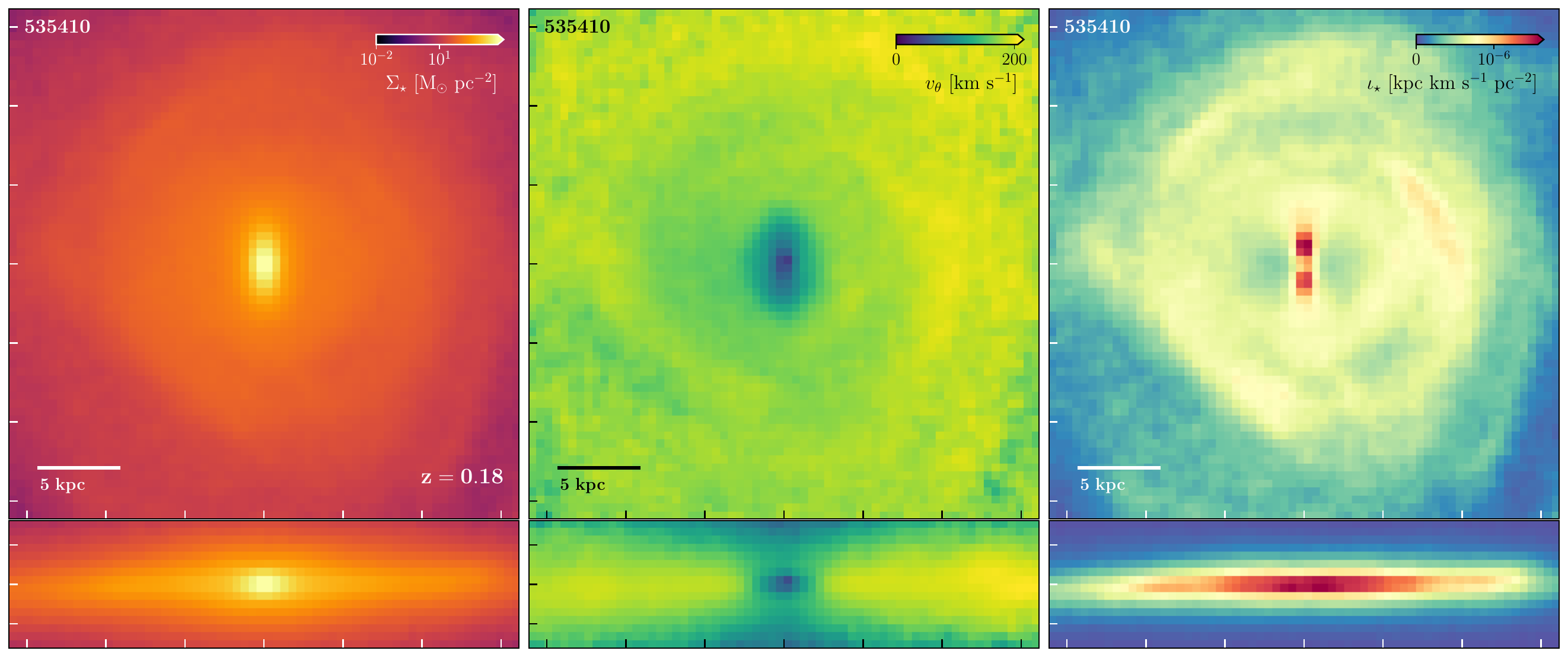}
    \end{subfigure}
    \caption{$\Sigma_\star$, $v_{\star\theta}$, and $\iota_\star$ maps for 53 randomly selected galaxies of our sample. The description for the remaining details is the same as for Fig.~\ref{fig: density_vel_j_maps}.}
    \label{fig: zoo}
\end{figure}

\clearpage
\onecolumn

\begin{figure*}[ht!]\ContinuedFloat
    \centering
    \begin{subfigure}[b]{0.49\textwidth}
        \includegraphics[width=\textwidth]{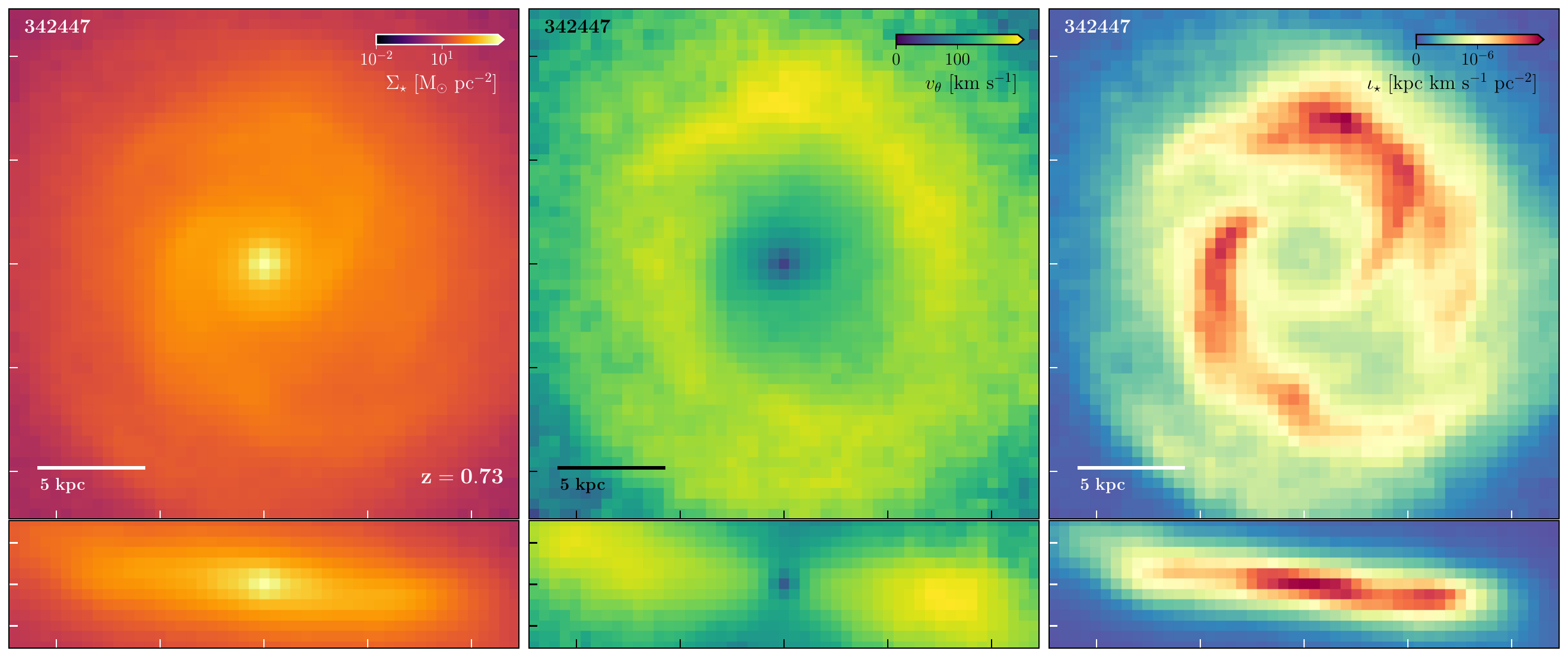}
    \end{subfigure}
    \begin{subfigure}[b]{0.49\textwidth}
        \includegraphics[width=\textwidth]{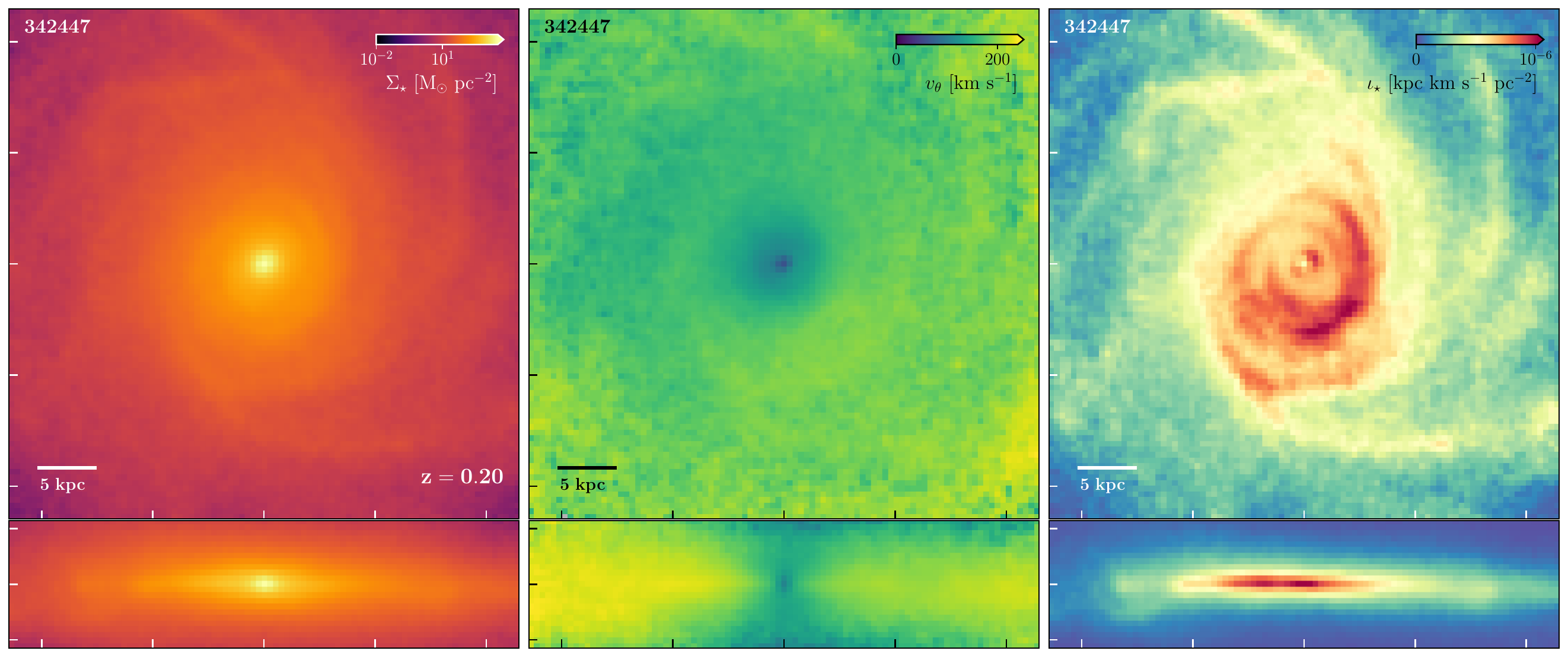}
    \end{subfigure}
    \\
    \begin{subfigure}[b]{0.49\textwidth}
        \includegraphics[width=\textwidth]{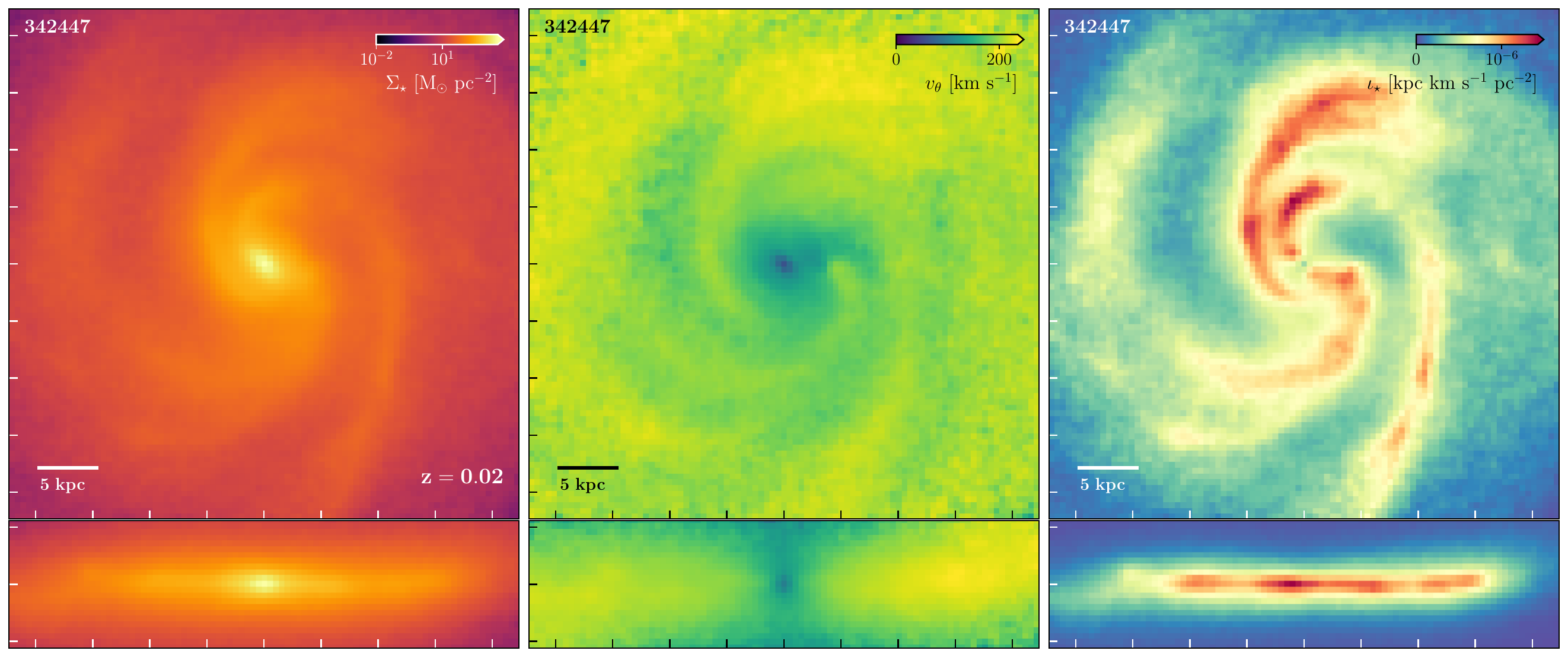}
    \end{subfigure}
    \begin{subfigure}[b]{0.49\textwidth}
        \includegraphics[width=\textwidth]{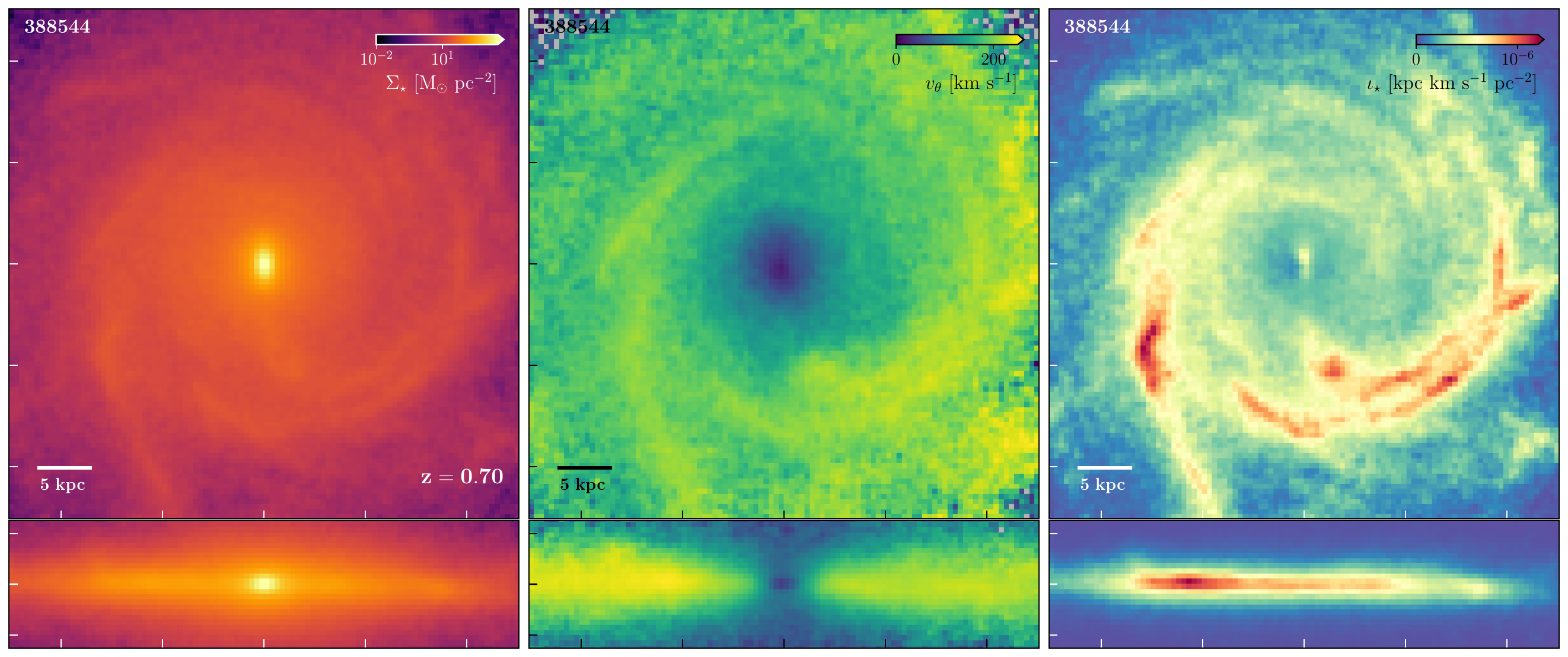}
    \end{subfigure}
    \\
    \begin{subfigure}[b]{0.49\textwidth}
        \includegraphics[width=\textwidth]{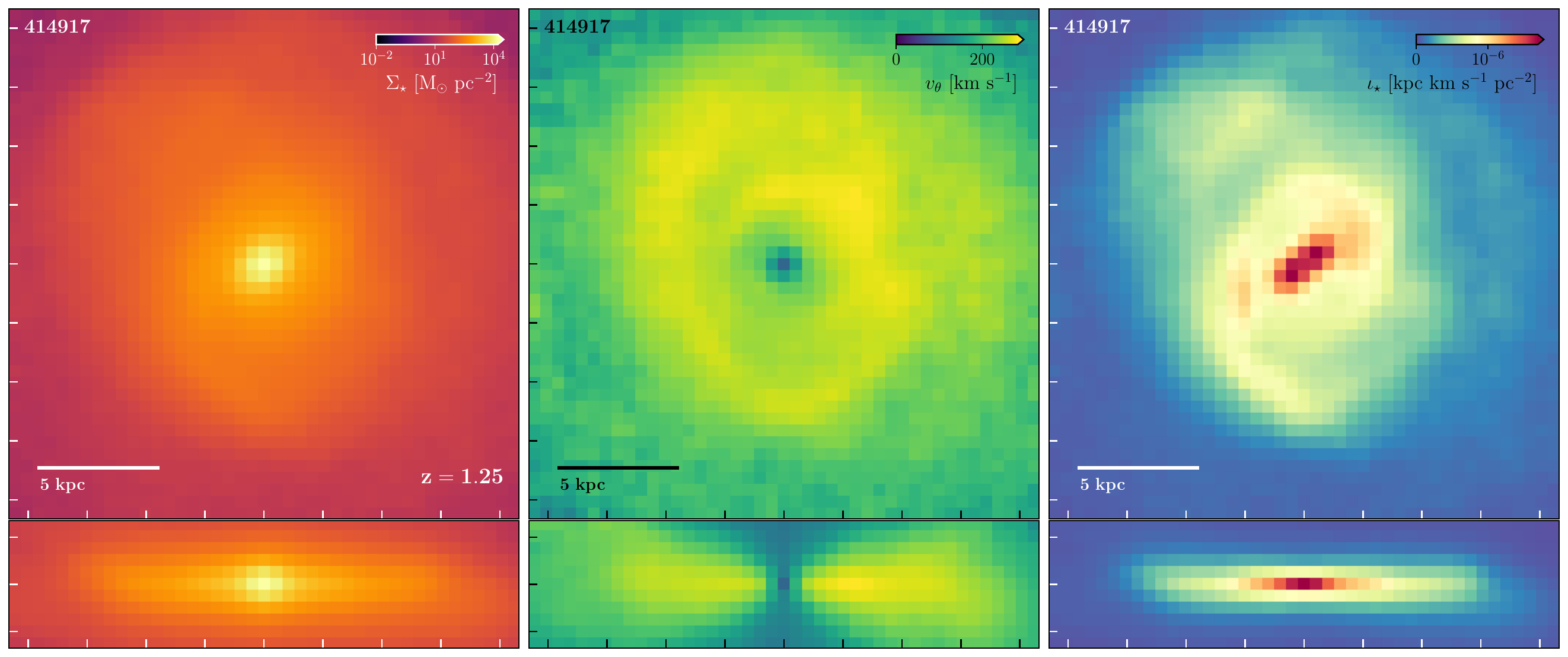}
    \end{subfigure}
    \begin{subfigure}[b]{0.49\textwidth}
        \includegraphics[width=\textwidth]{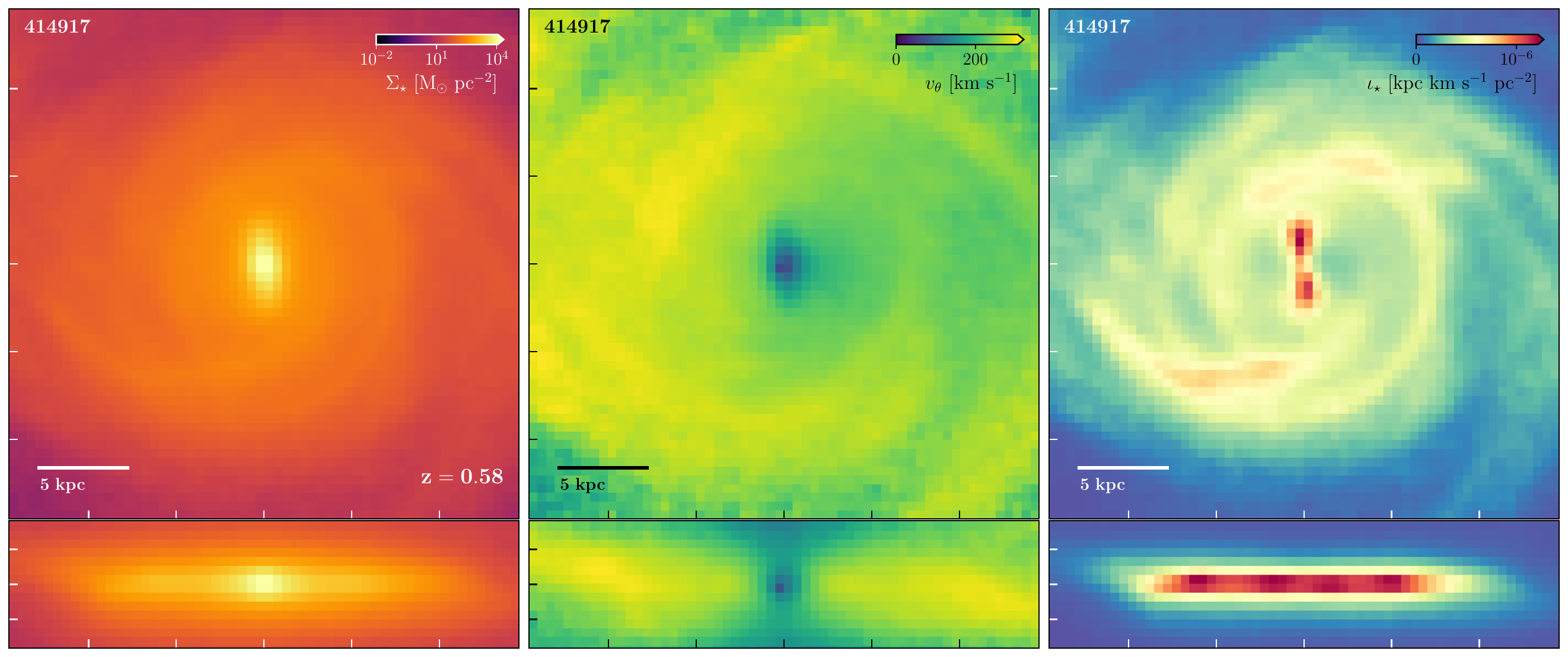}
    \end{subfigure}
    \\
    \begin{subfigure}[b]{0.49\textwidth}
        \includegraphics[width=\textwidth]{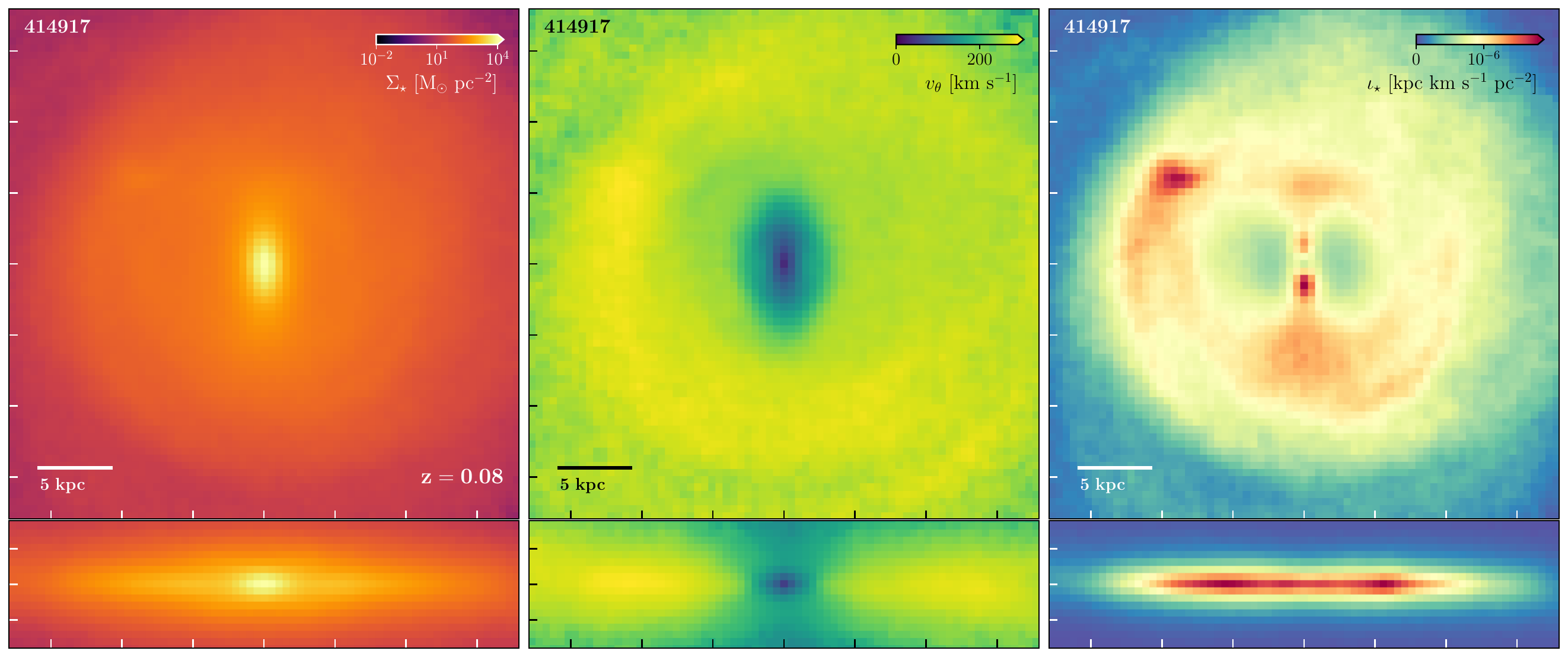}
    \end{subfigure}
    \begin{subfigure}[b]{0.49\textwidth}
        \includegraphics[width=\textwidth]{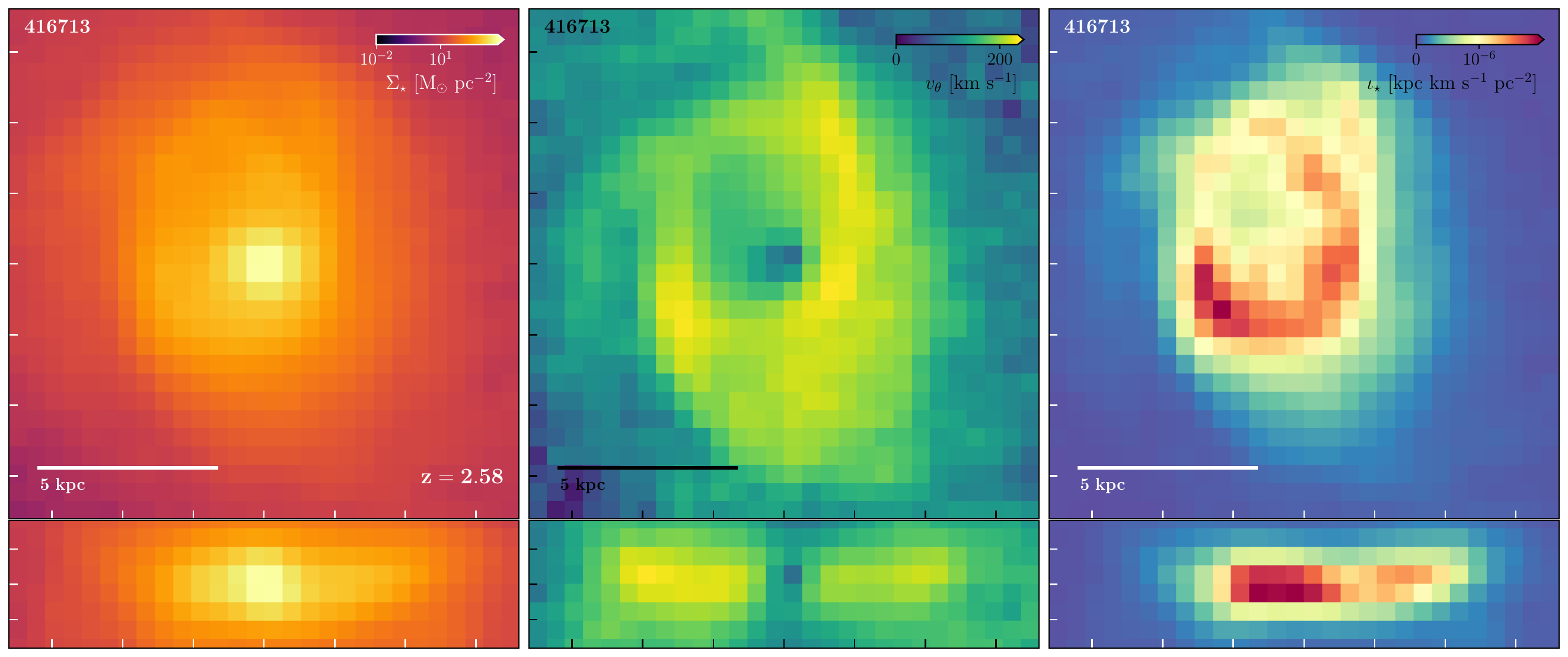}
    \end{subfigure}
    \\
    \begin{subfigure}[b]{0.49\textwidth}
        \includegraphics[width=\textwidth]{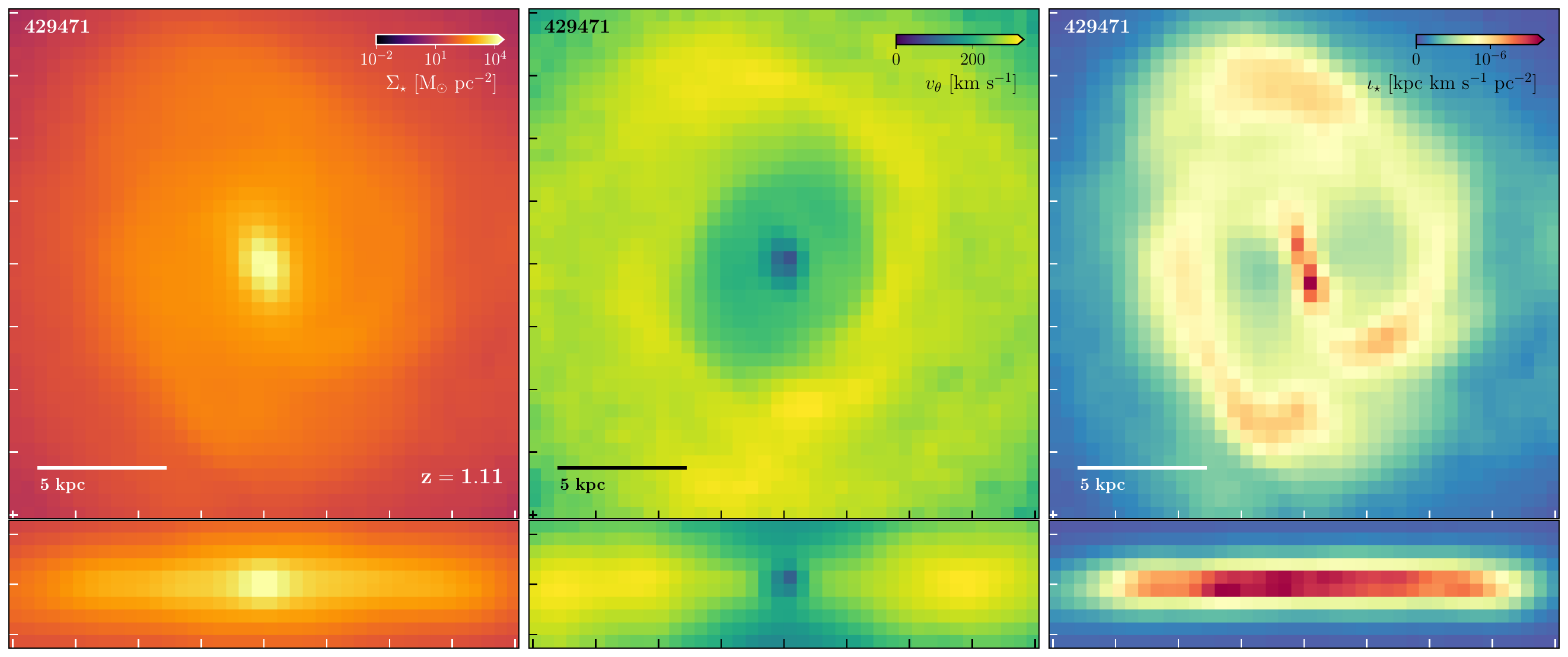}
    \end{subfigure}
    \begin{subfigure}[b]{0.49\textwidth}
        \includegraphics[width=\textwidth]{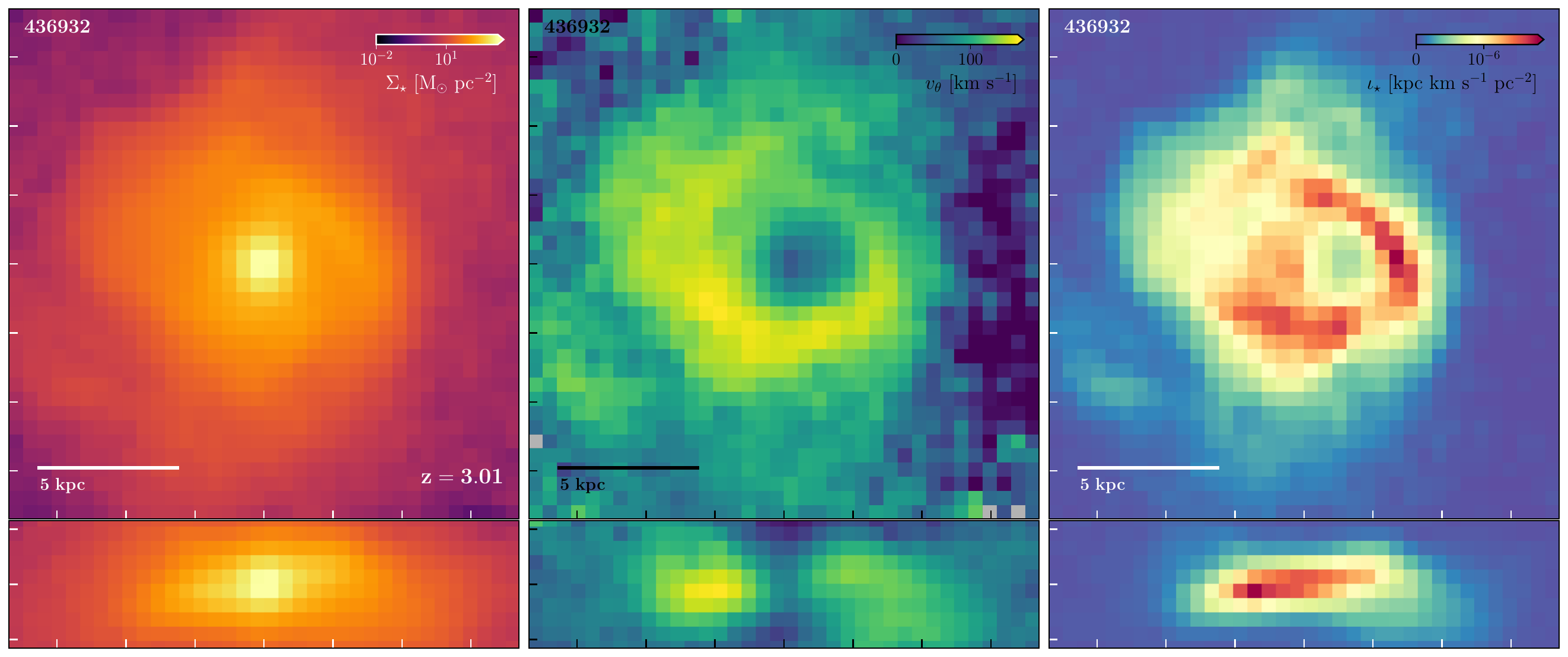}
    \end{subfigure}
    \\
    \begin{subfigure}[b]{0.49\textwidth}
        \includegraphics[width=\textwidth]{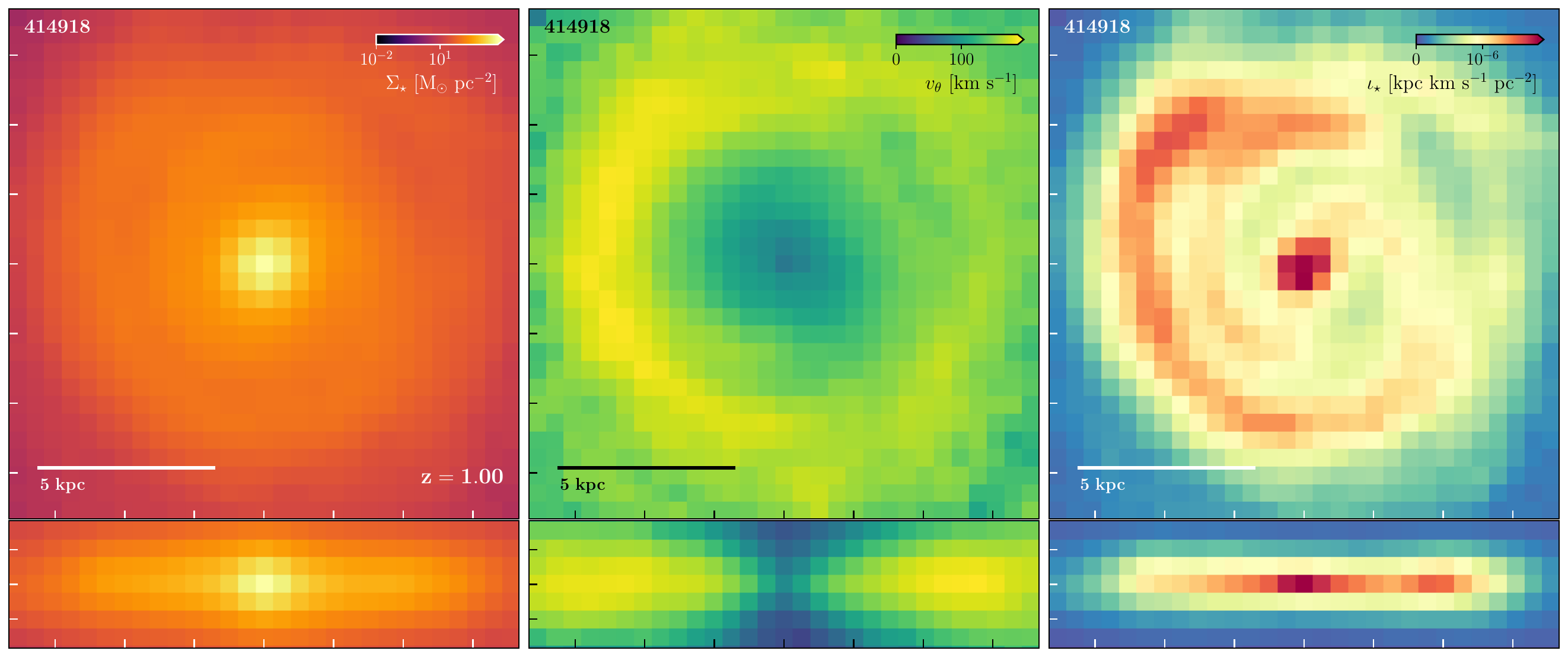}
    \end{subfigure}
    \begin{subfigure}[b]{0.49\textwidth}
        \includegraphics[width=\textwidth]{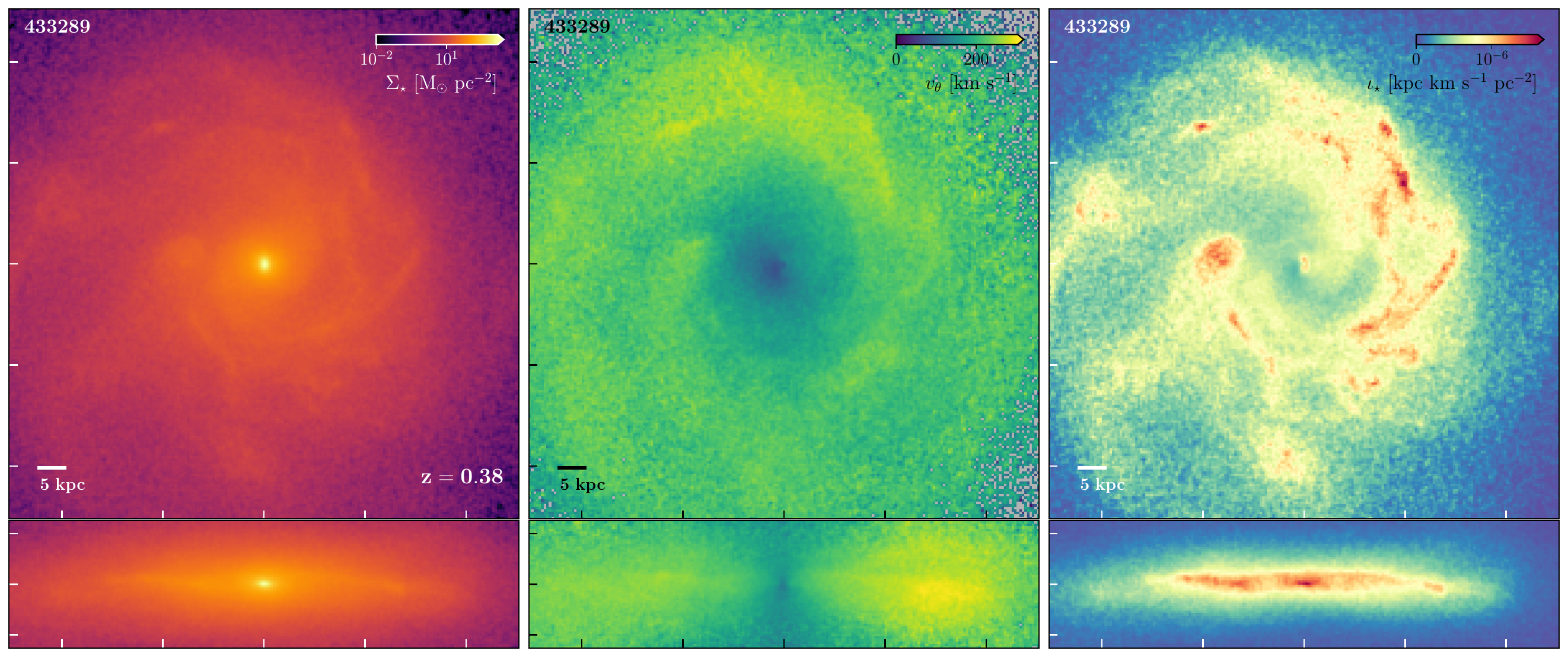}
    \end{subfigure}
    \caption{Continued.}
    \label{fig: zoo}
\end{figure*}

\begin{figure*}[ht!]\ContinuedFloat
    \centering
    \begin{subfigure}[b]{0.49\textwidth}
        \includegraphics[width=\textwidth]{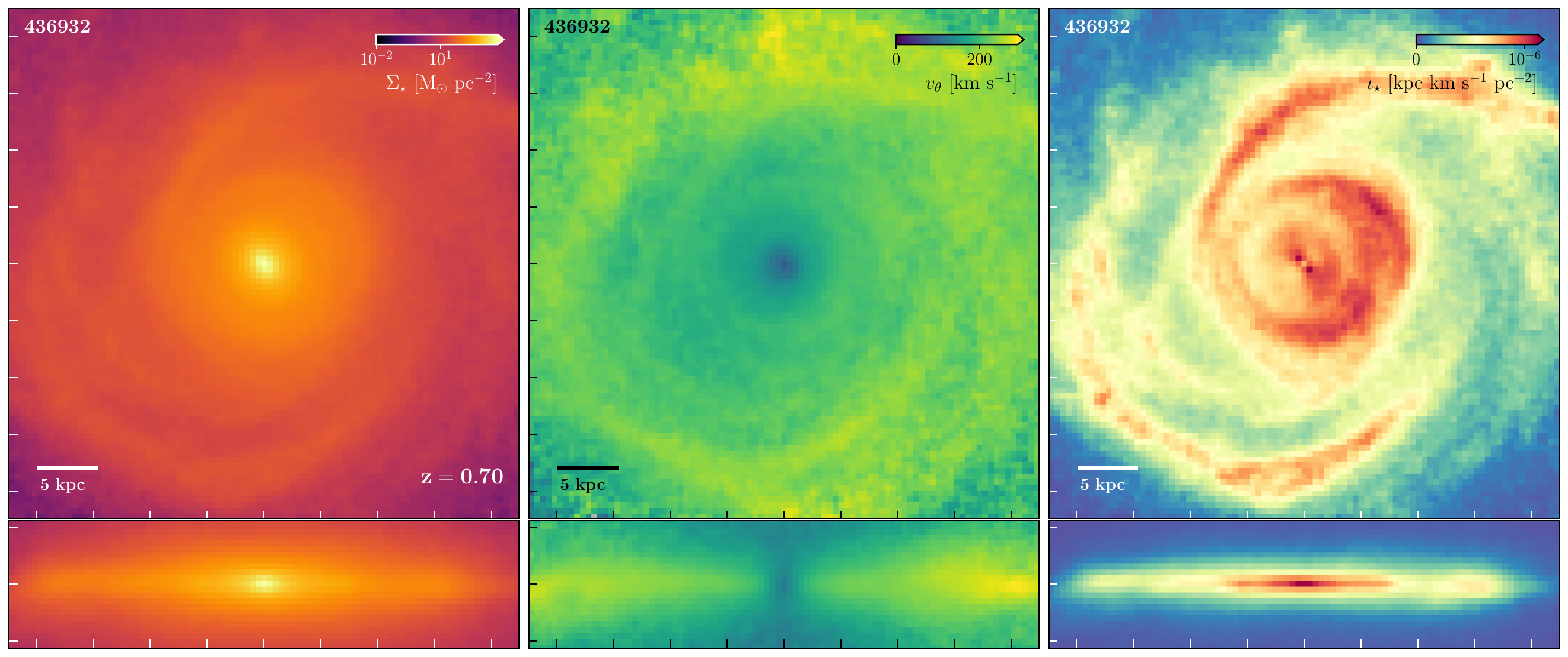}
    \end{subfigure}
    \begin{subfigure}[b]{0.49\textwidth}
        \includegraphics[width=\textwidth]{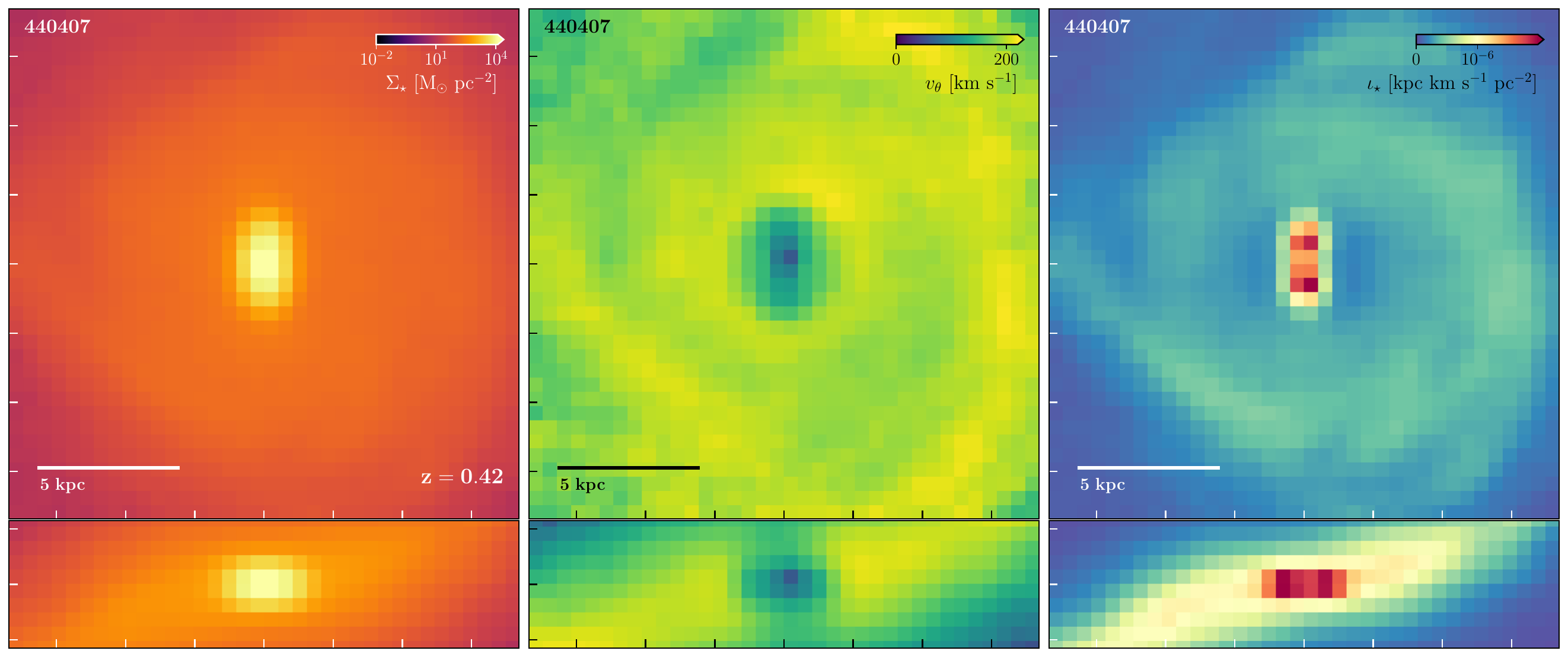}
    \end{subfigure}
    \\
    \begin{subfigure}[b]{0.49\textwidth}
        \includegraphics[width=\textwidth]{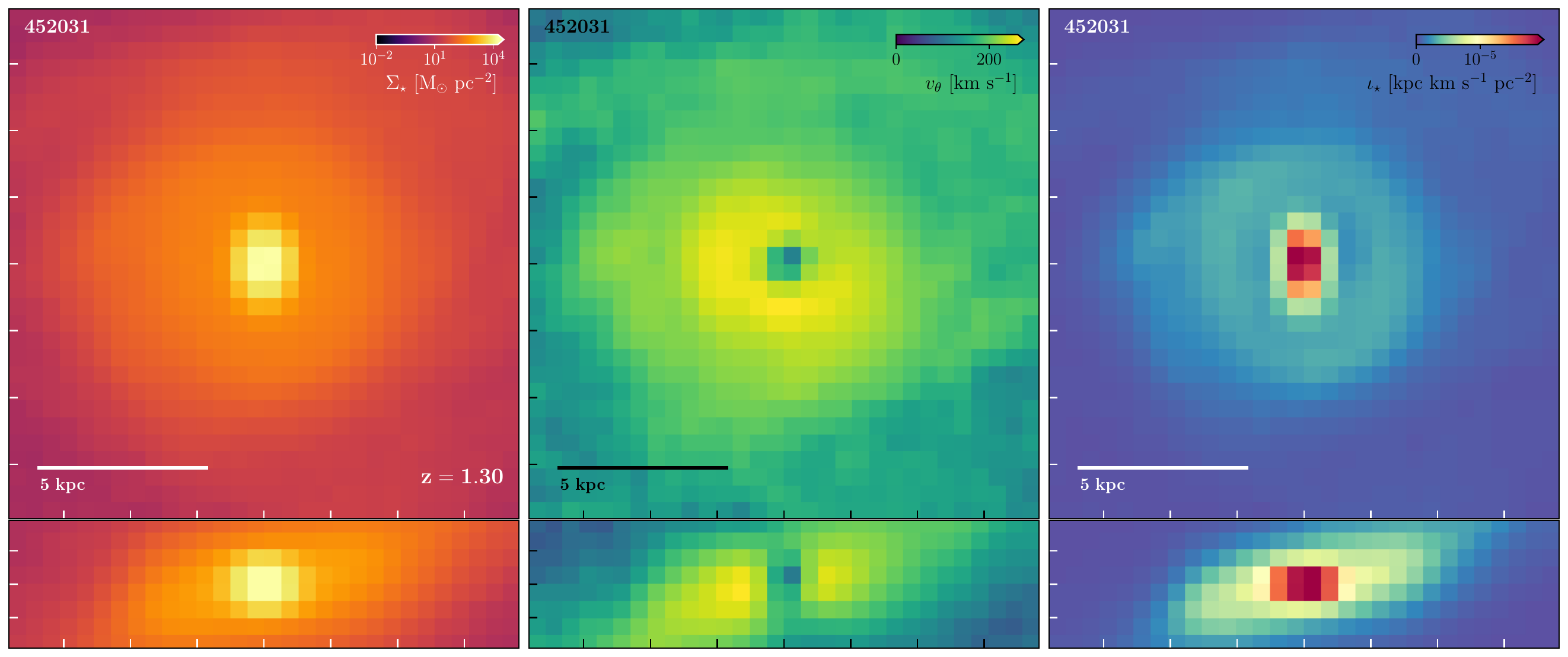}
    \end{subfigure}
    \begin{subfigure}[b]{0.49\textwidth}
        \includegraphics[width=\textwidth]{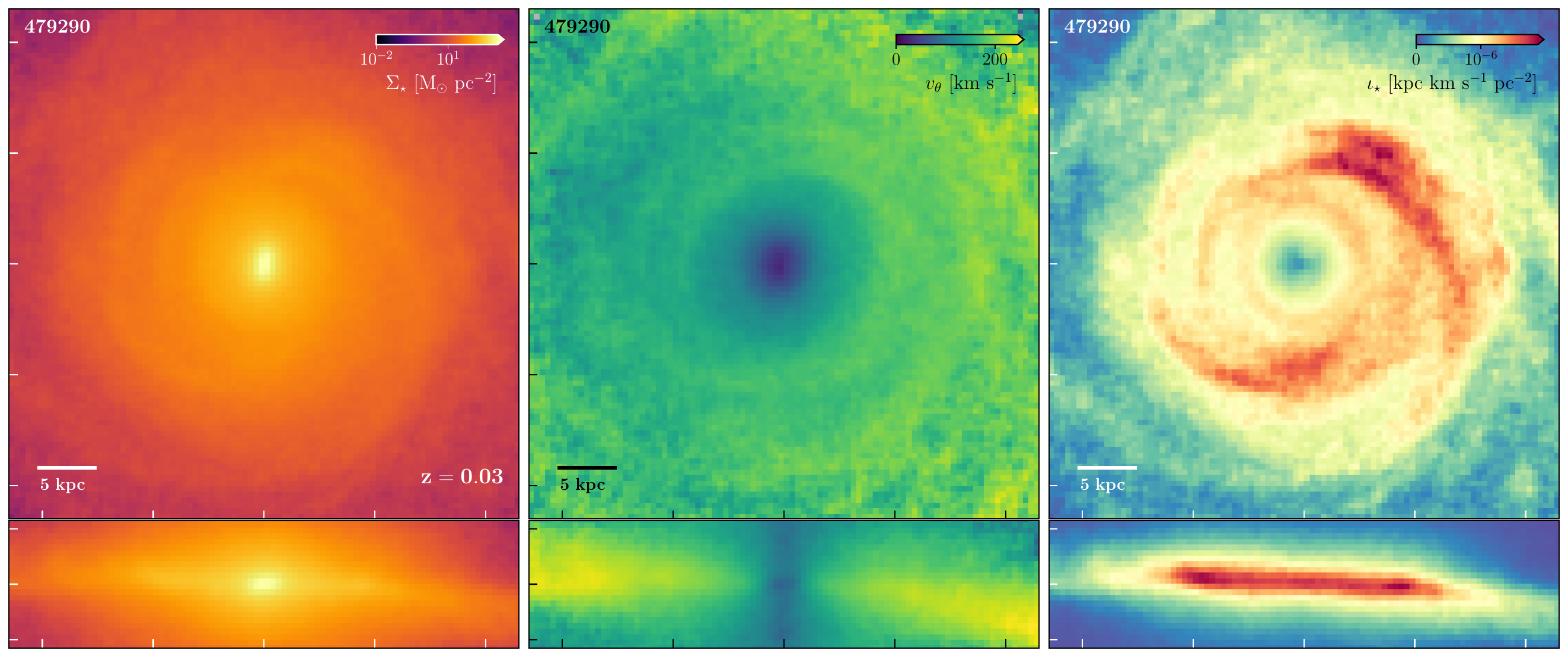}
    \end{subfigure}
    \\
    \begin{subfigure}[b]{0.49\textwidth}
        \includegraphics[width=\textwidth]{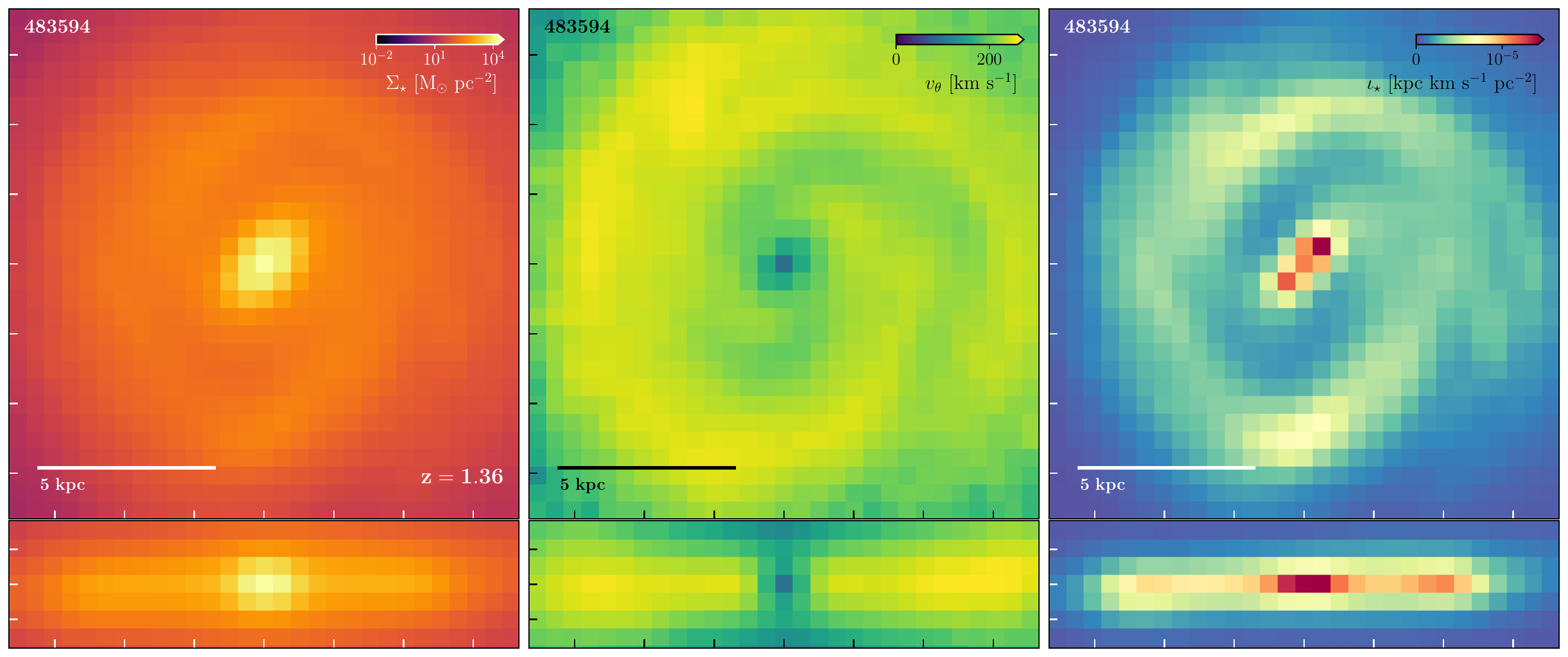}
    \end{subfigure}
    \begin{subfigure}[b]{0.49\textwidth}
        \includegraphics[width=\textwidth]{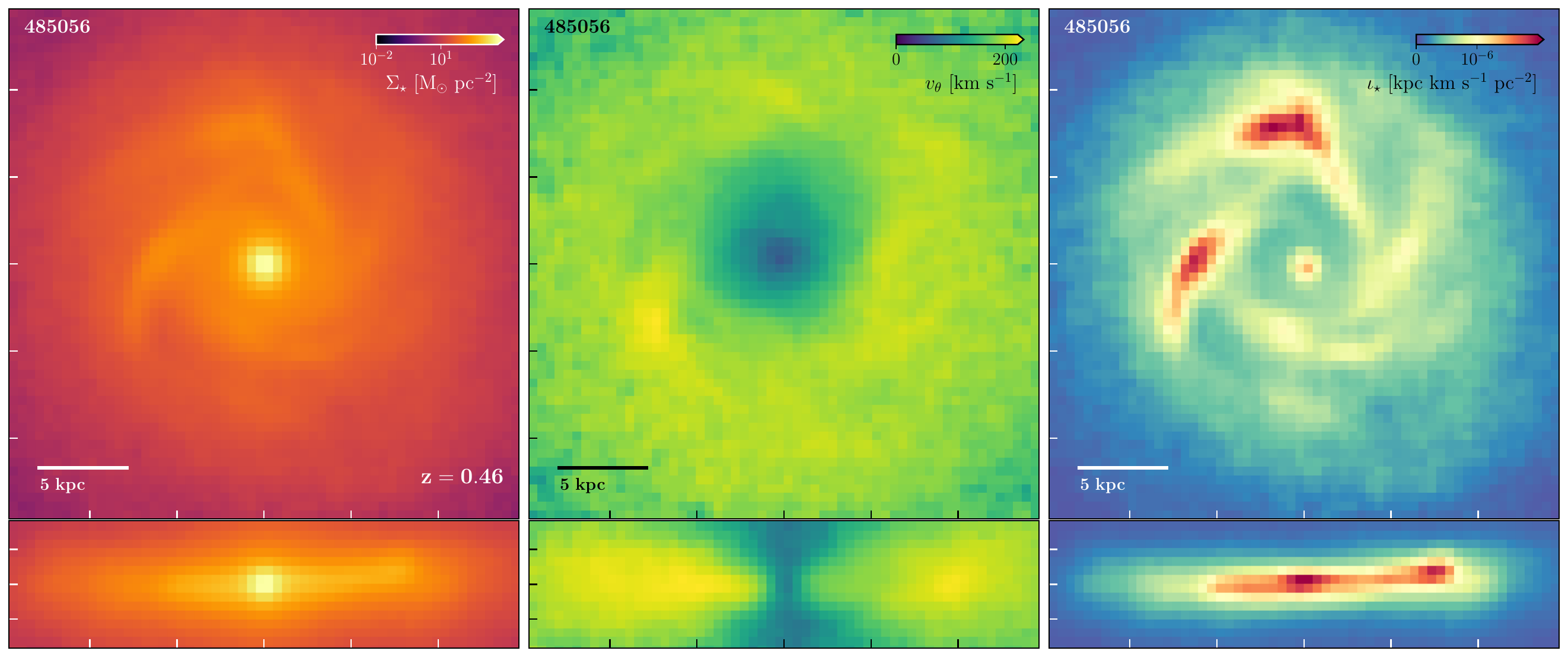}
    \end{subfigure}
    \\
    \begin{subfigure}[b]{0.49\textwidth}
        \includegraphics[width=\textwidth]{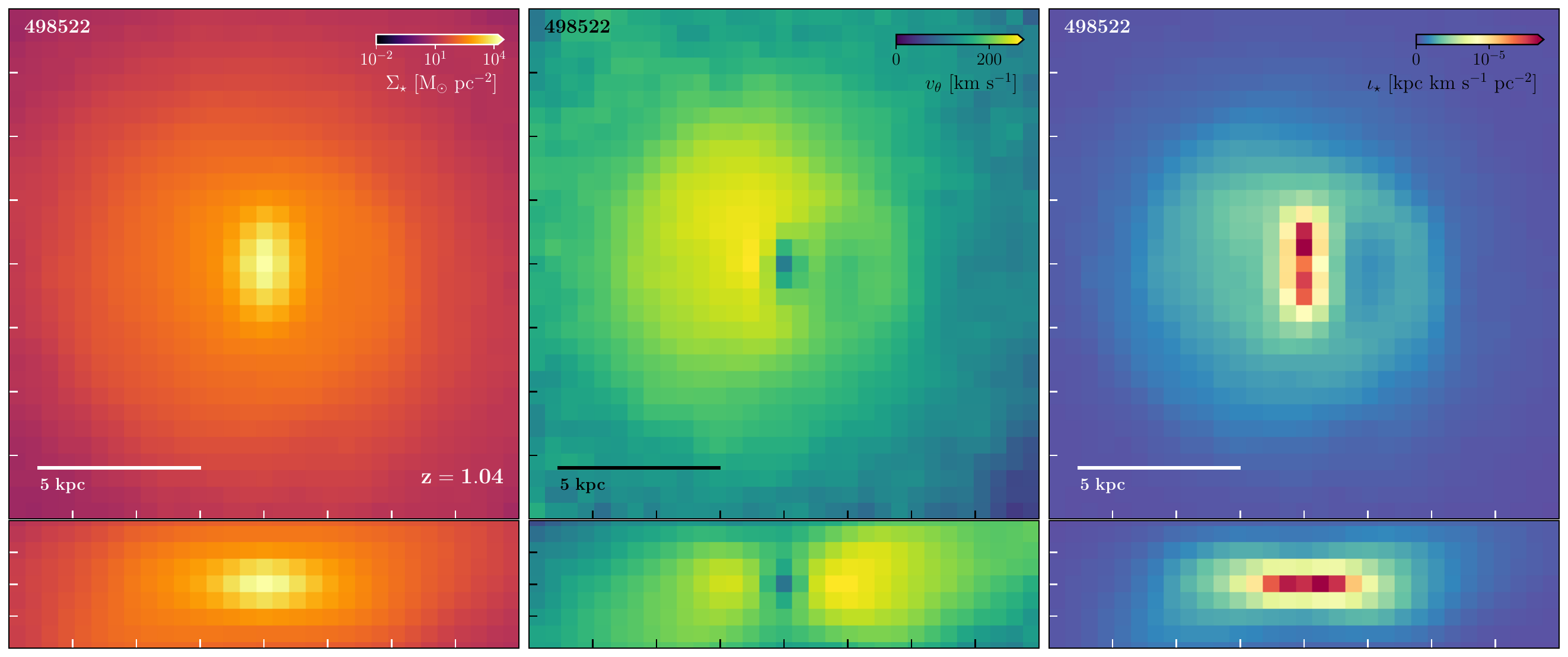}
    \end{subfigure}
    \begin{subfigure}[b]{0.49\textwidth}
        \includegraphics[width=\textwidth]{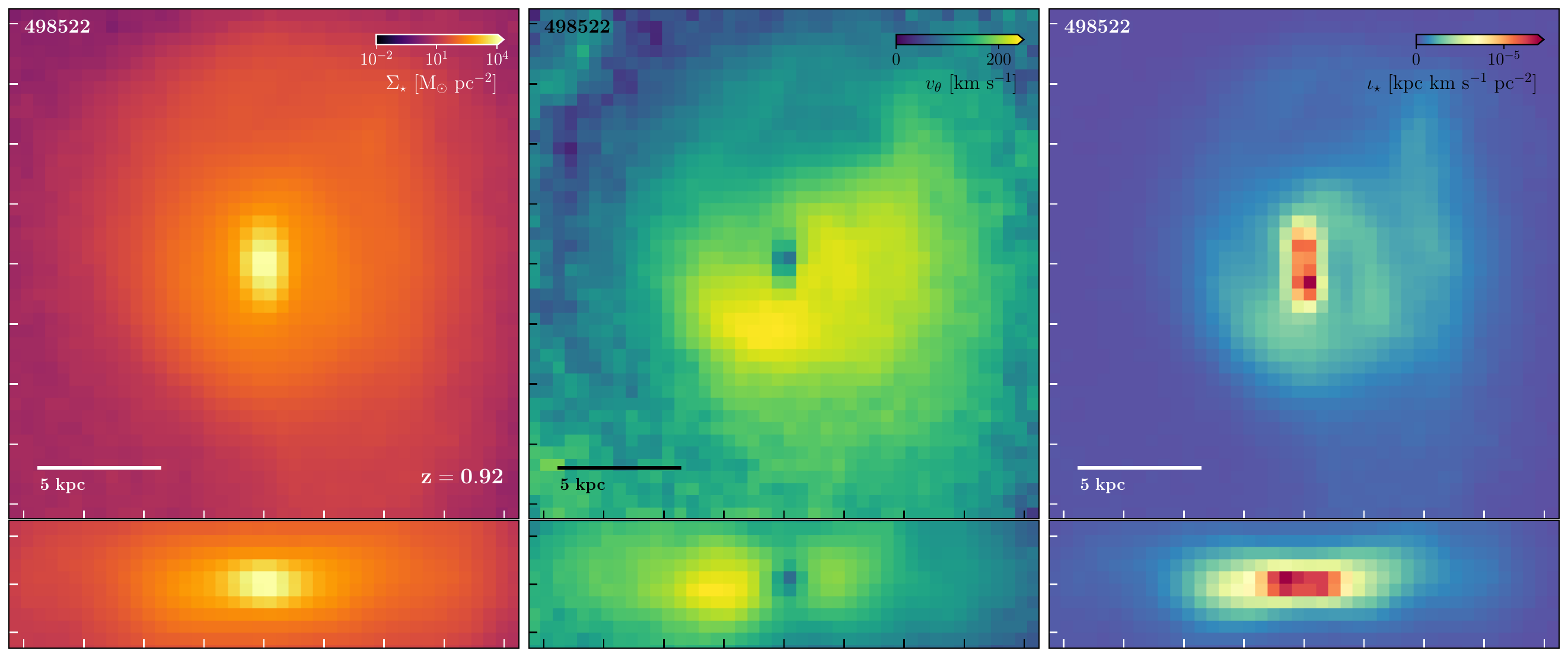}
    \end{subfigure}
    \\
    \begin{subfigure}[b]{0.49\textwidth}
        \includegraphics[width=\textwidth]{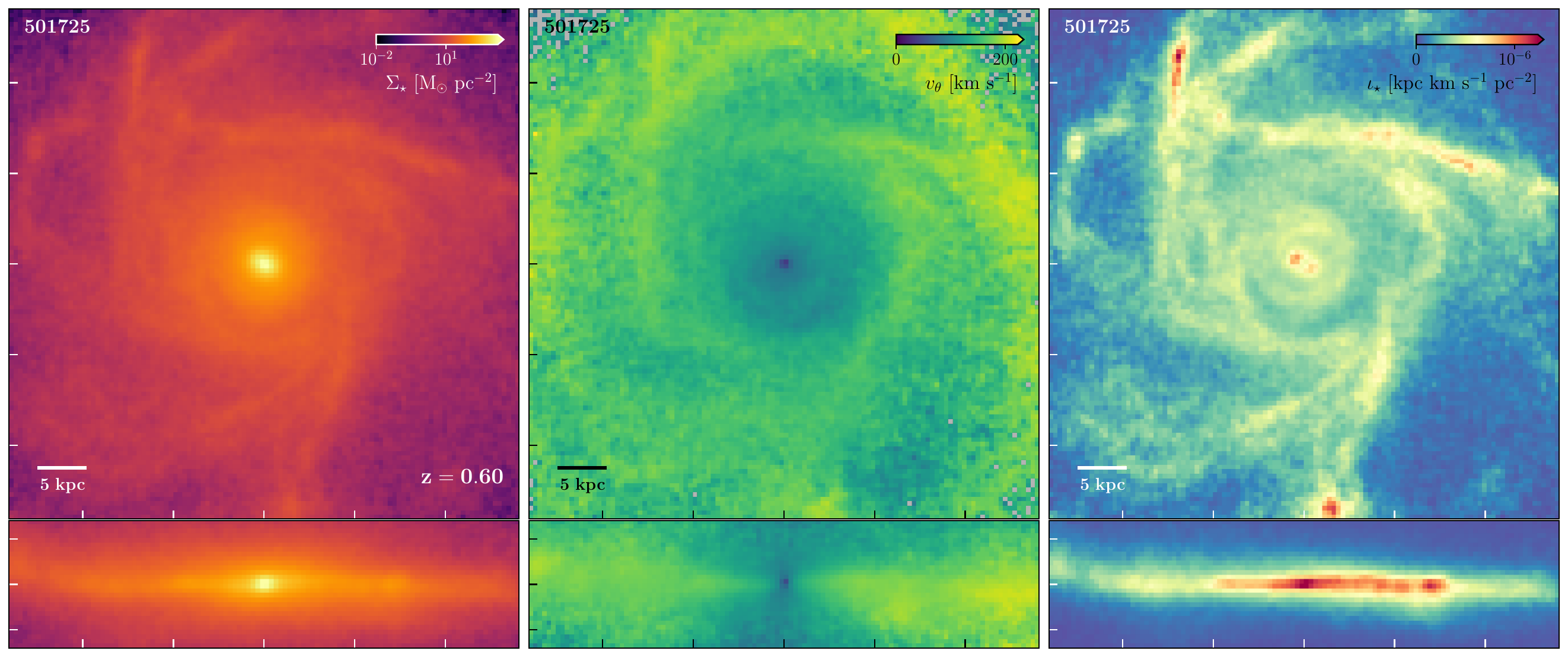}
    \end{subfigure}
    \begin{subfigure}[b]{0.49\textwidth}
        \includegraphics[width=\textwidth]{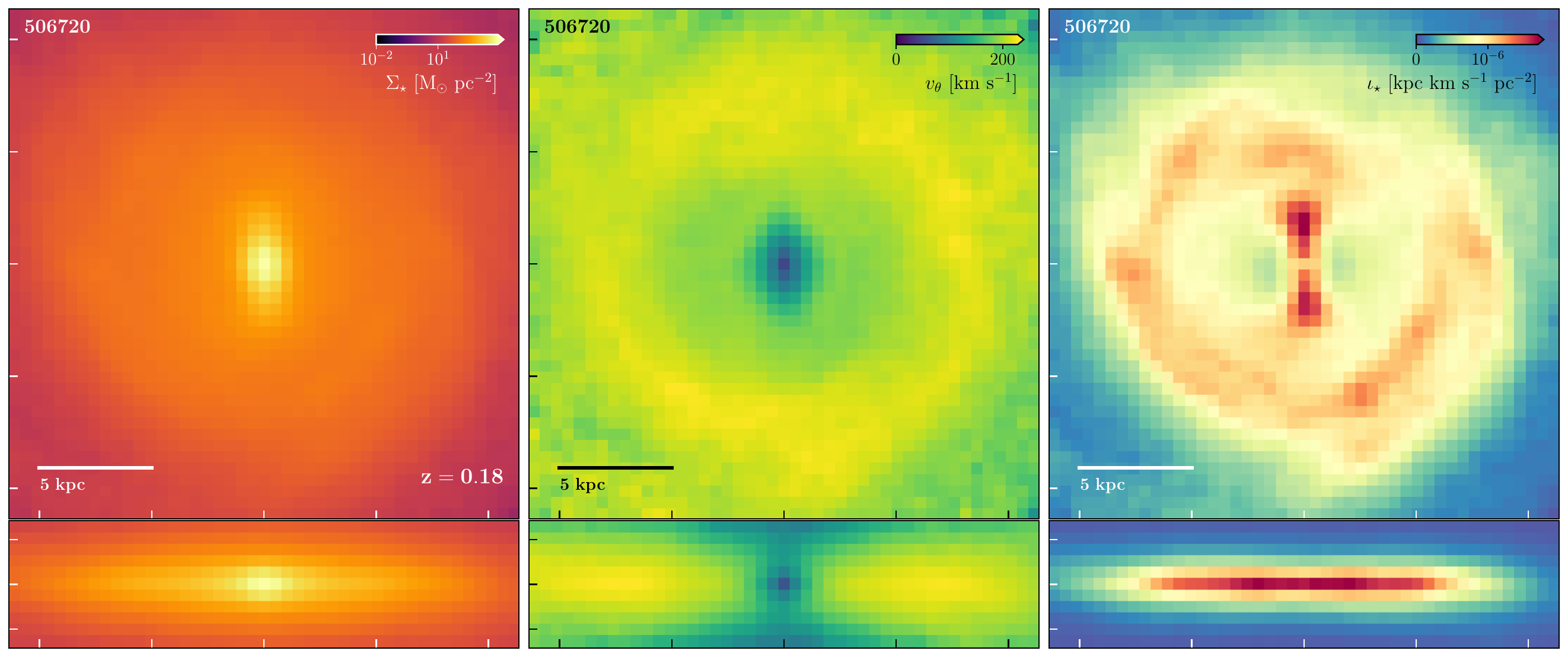}
    \end{subfigure}
    \\
    \begin{subfigure}[b]{0.49\textwidth}
        \includegraphics[width=\textwidth]{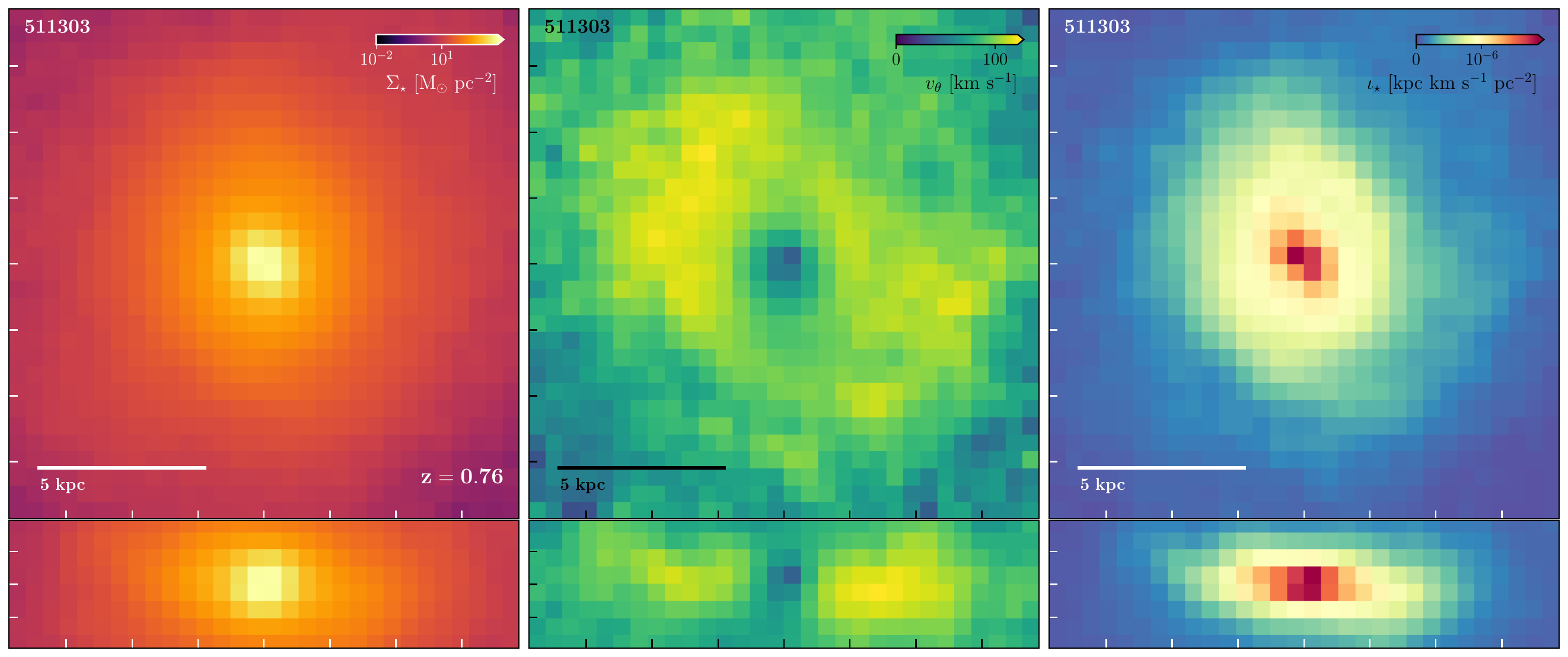}
    \end{subfigure}
    \begin{subfigure}[b]{0.49\textwidth}
        \includegraphics[width=\textwidth]{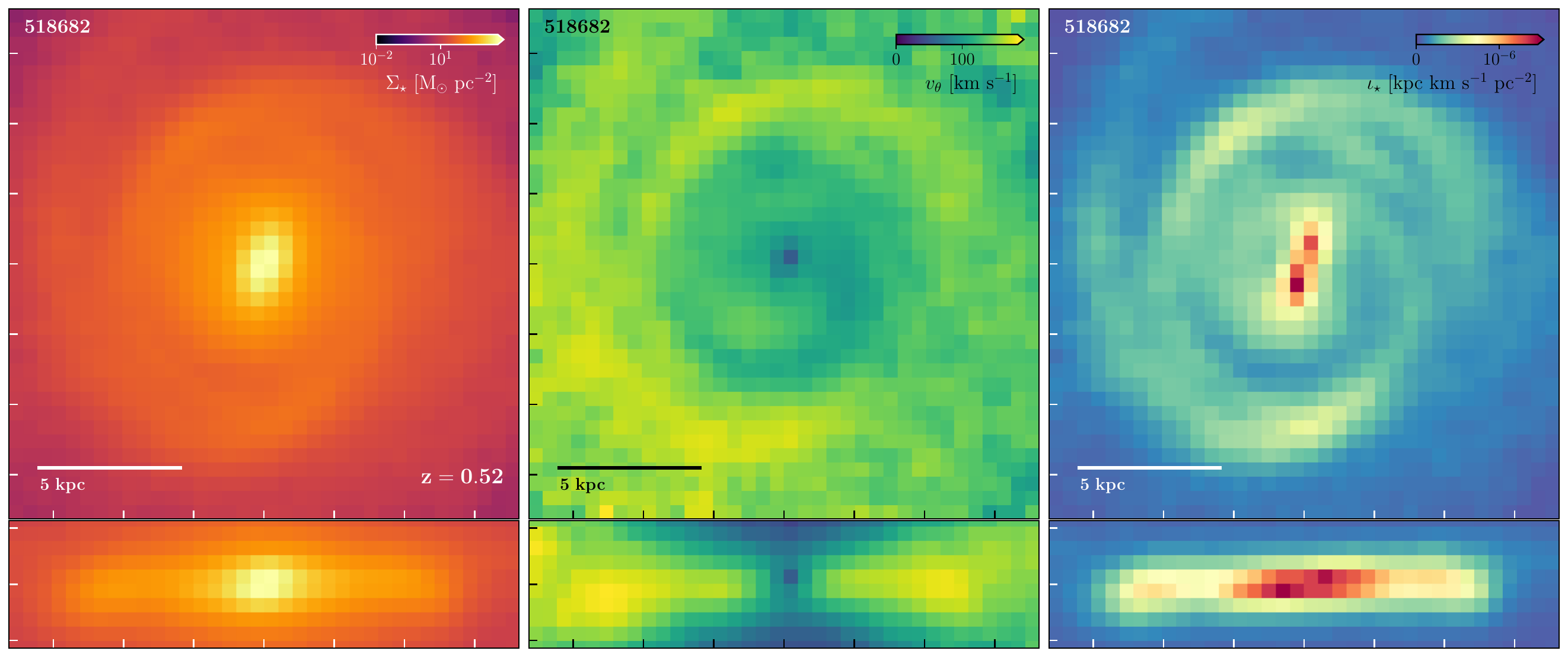}
    \end{subfigure}
    \caption{Continued.}
    \label{fig: zoo}
\end{figure*}

\begin{figure*}[ht!]\ContinuedFloat
    \centering
    \begin{subfigure}[b]{0.49\textwidth}
        \includegraphics[width=\textwidth]{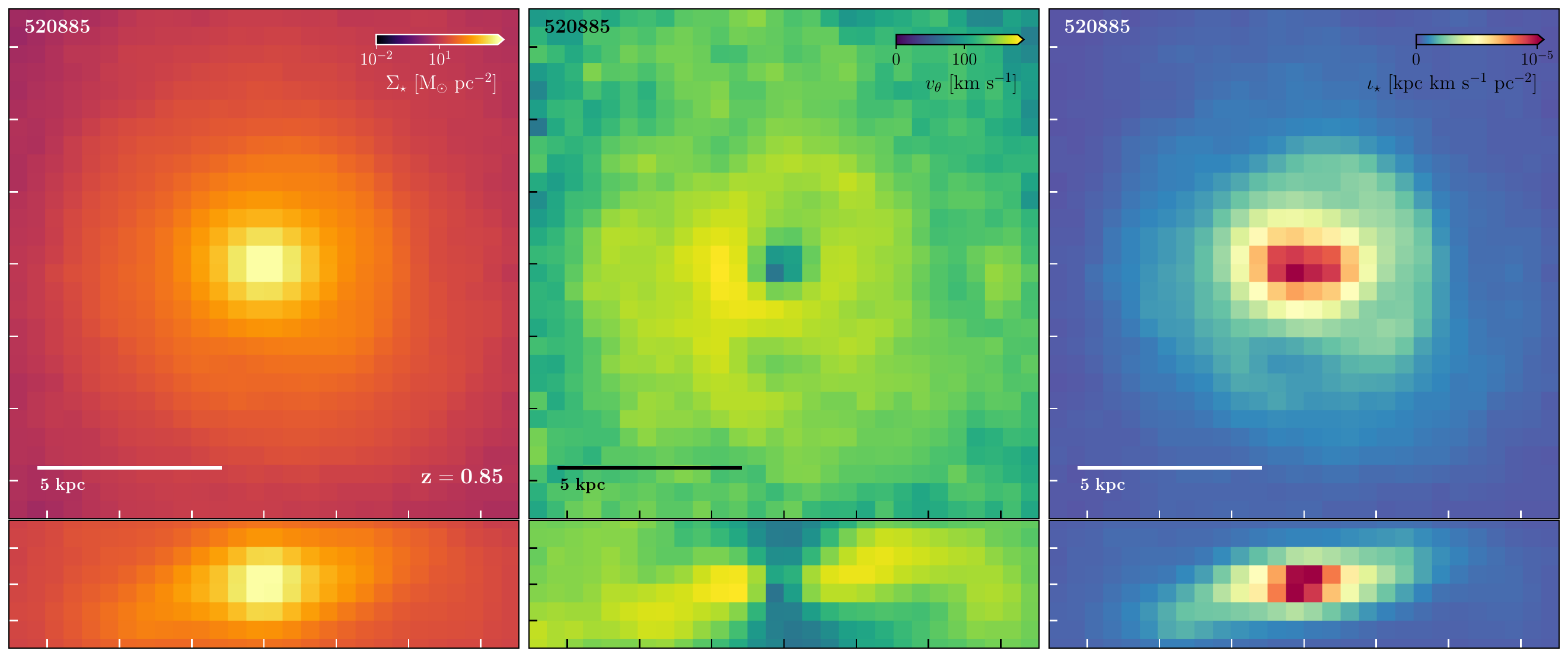}
    \end{subfigure}
    \begin{subfigure}[b]{0.49\textwidth}
        \includegraphics[width=\textwidth]{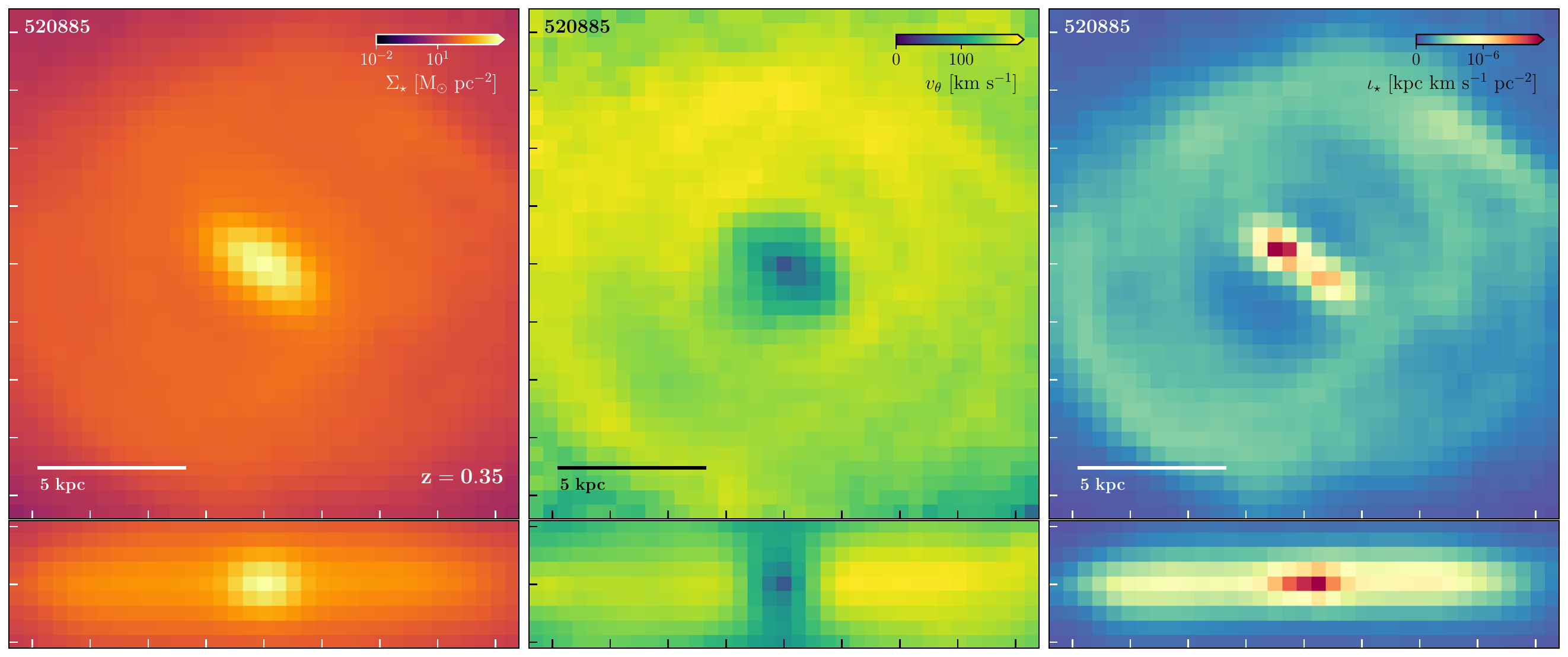}
    \end{subfigure}
    \\
    \begin{subfigure}[b]{0.49\textwidth}
        \includegraphics[width=\textwidth]{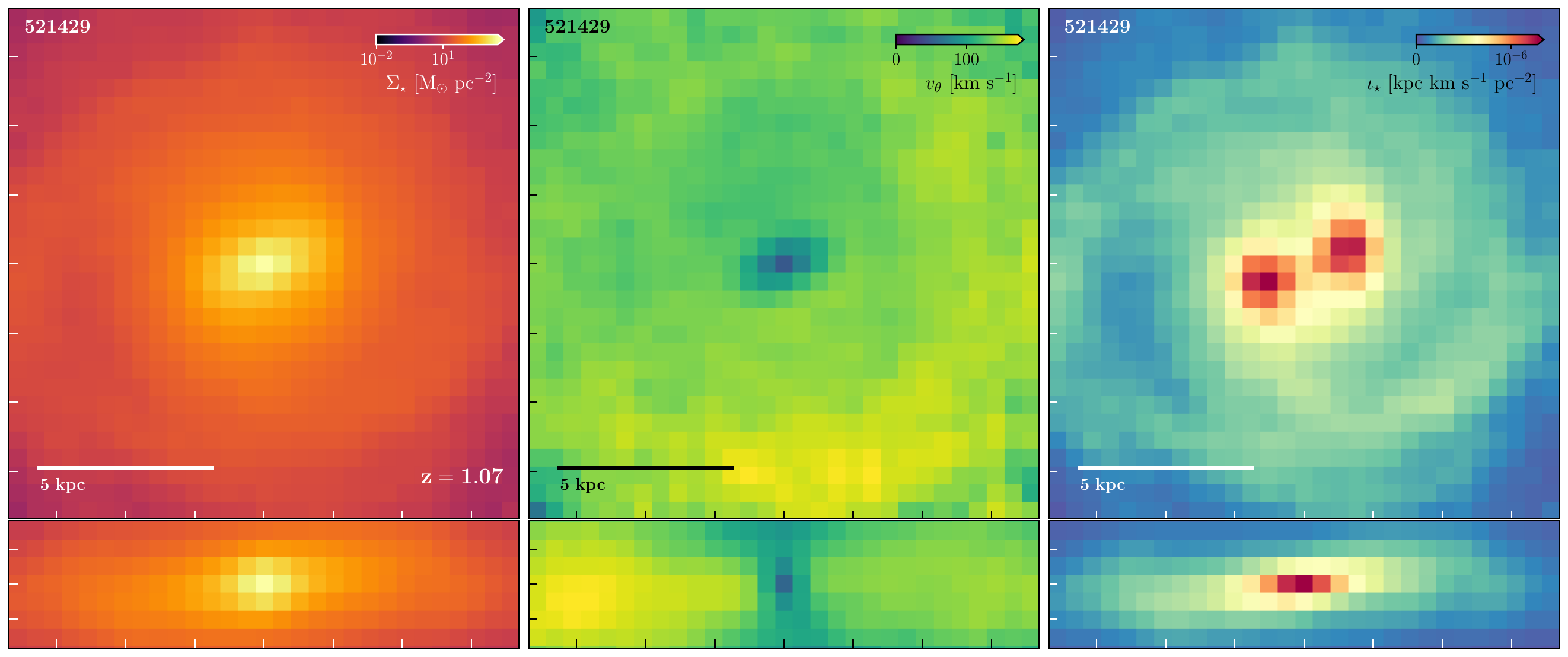}
    \end{subfigure}
    \begin{subfigure}[b]{0.49\textwidth}
        \includegraphics[width=\textwidth]{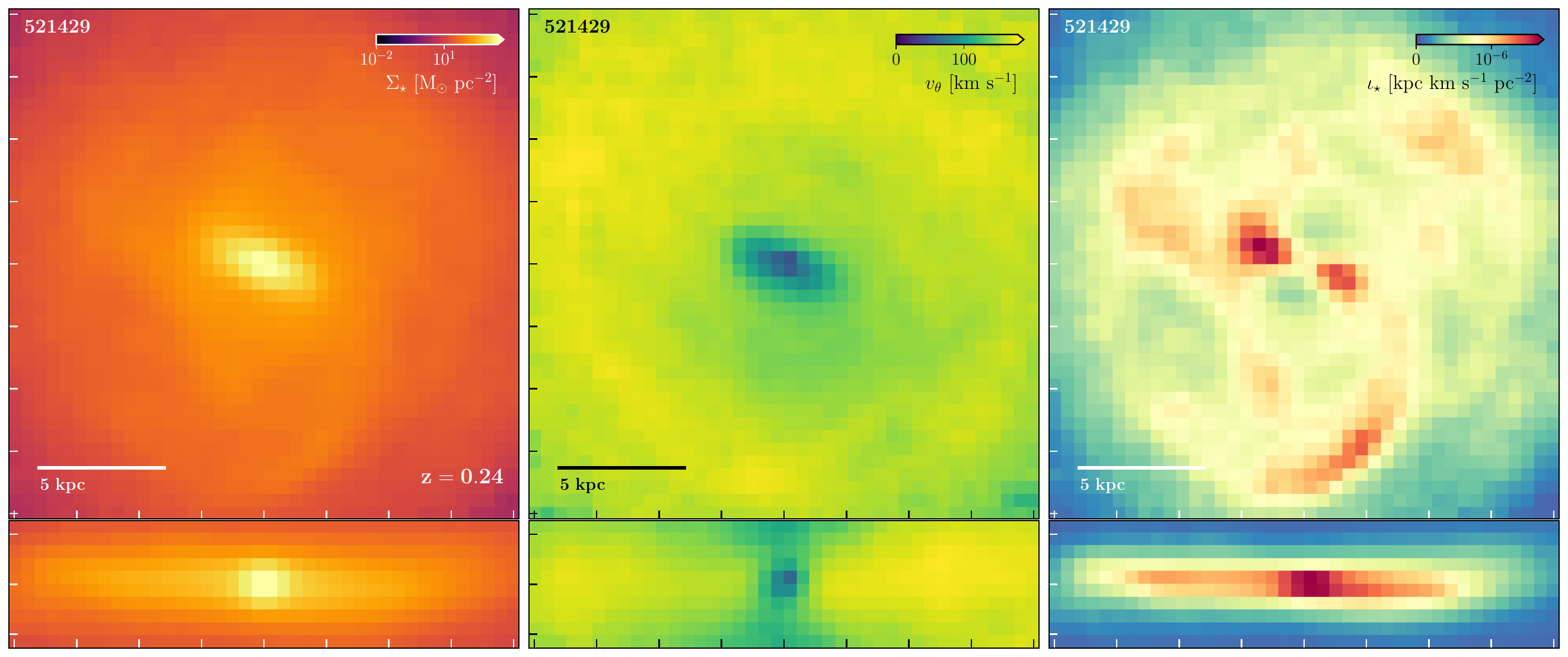}
    \end{subfigure}
    \\
    \begin{subfigure}[b]{0.49\textwidth}
        \includegraphics[width=\textwidth]{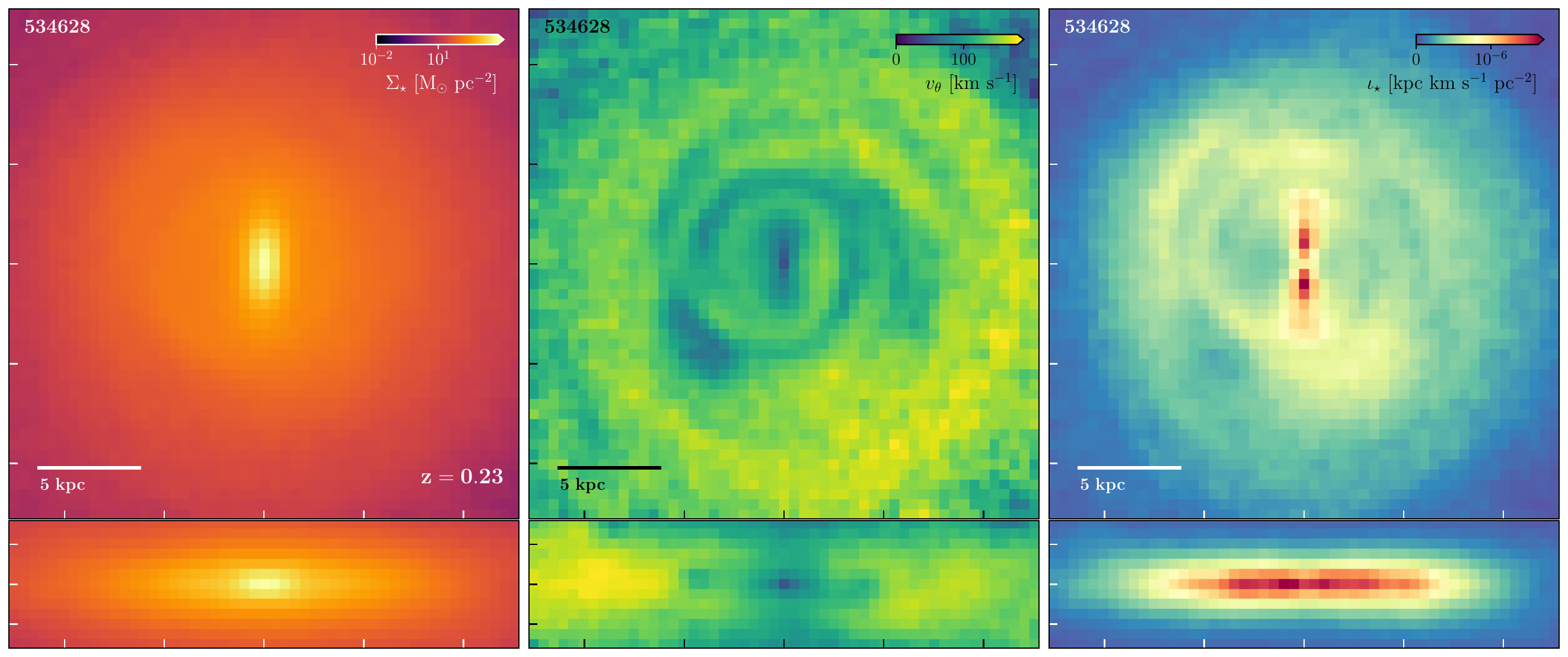}
    \end{subfigure}
    \begin{subfigure}[b]{0.49\textwidth}
        \includegraphics[width=\textwidth]{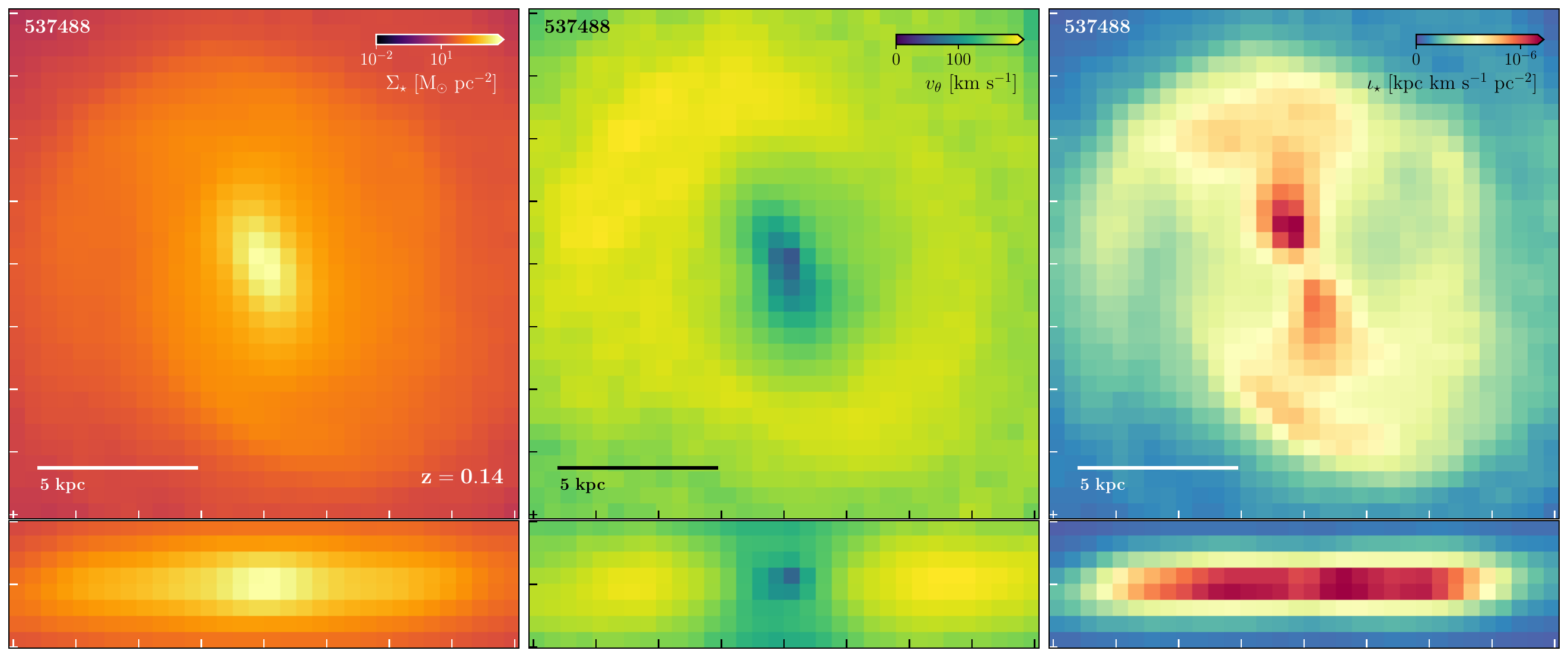}
    \end{subfigure}
    \\
    \begin{subfigure}[b]{0.49\textwidth}
        \includegraphics[width=\textwidth]{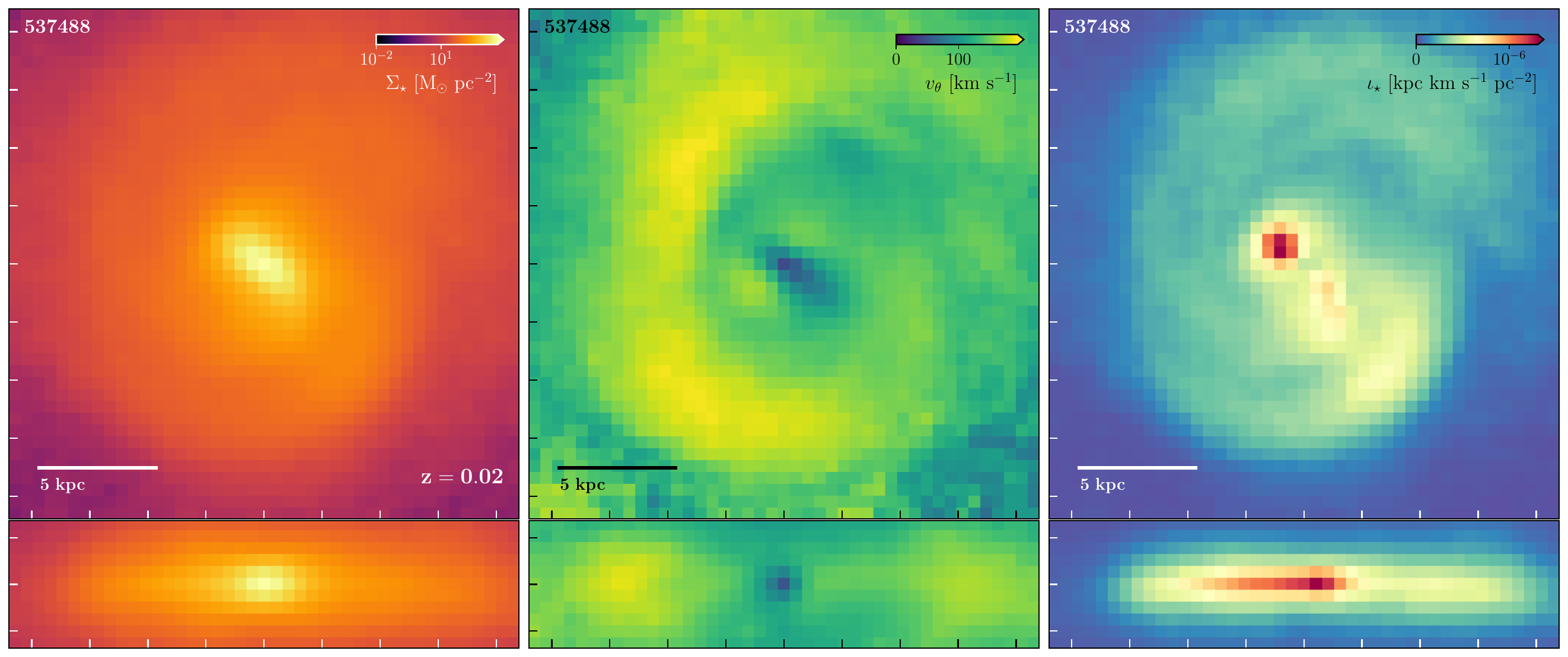}
    \end{subfigure}
    \begin{subfigure}[b]{0.49\textwidth}
        \includegraphics[width=\textwidth]{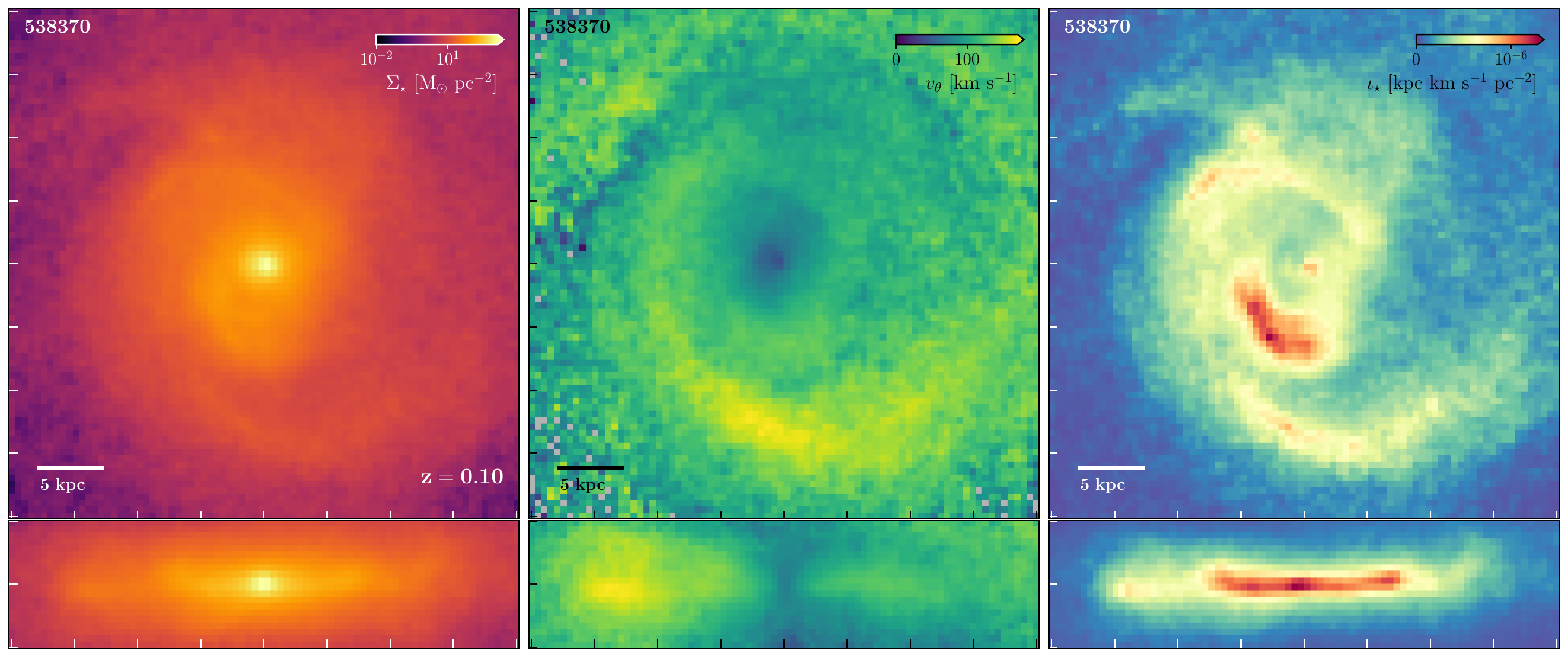}
    \end{subfigure}
    \\
    \begin{subfigure}[b]{0.49\textwidth}
        \includegraphics[width=\textwidth]{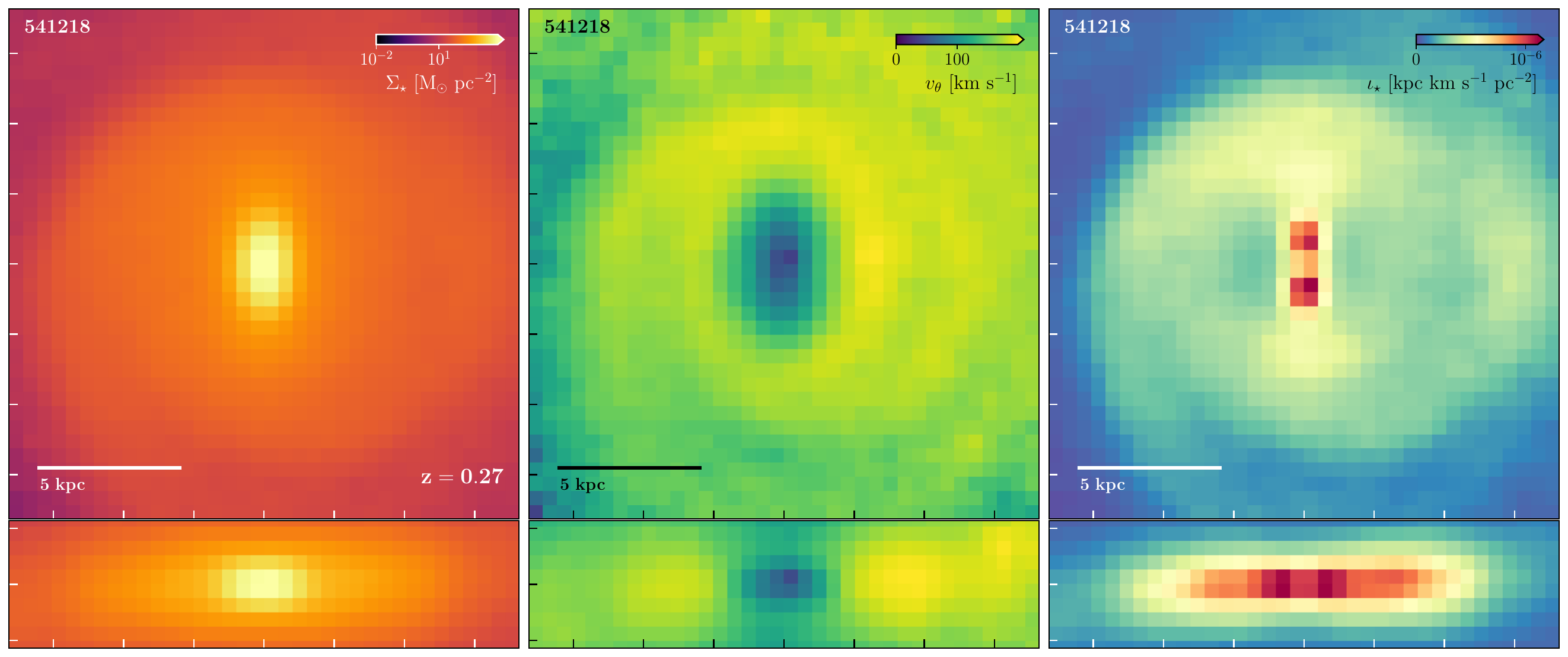}
    \end{subfigure}
    \begin{subfigure}[b]{0.49\textwidth}
        \includegraphics[width=\textwidth]{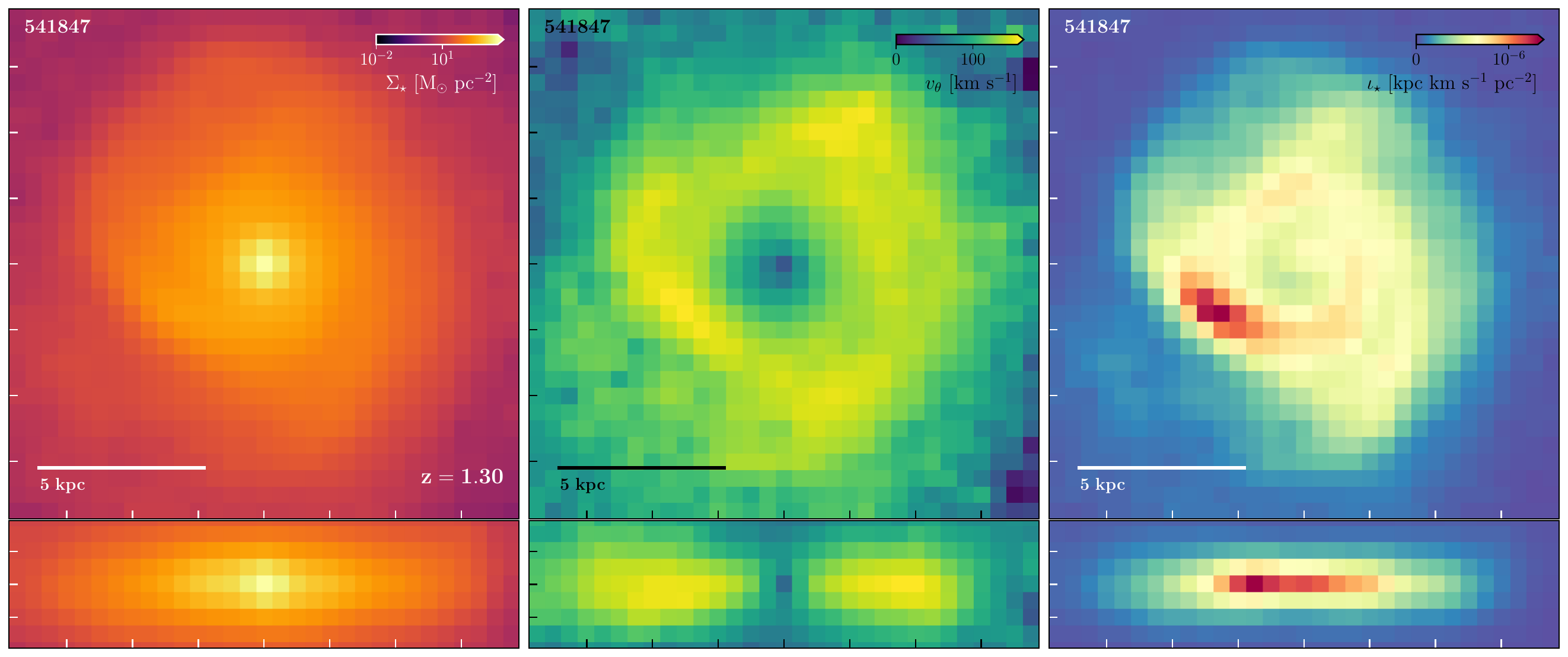}
    \end{subfigure}
    \\
    \begin{subfigure}[b]{0.49\textwidth}
        \includegraphics[width=\textwidth]{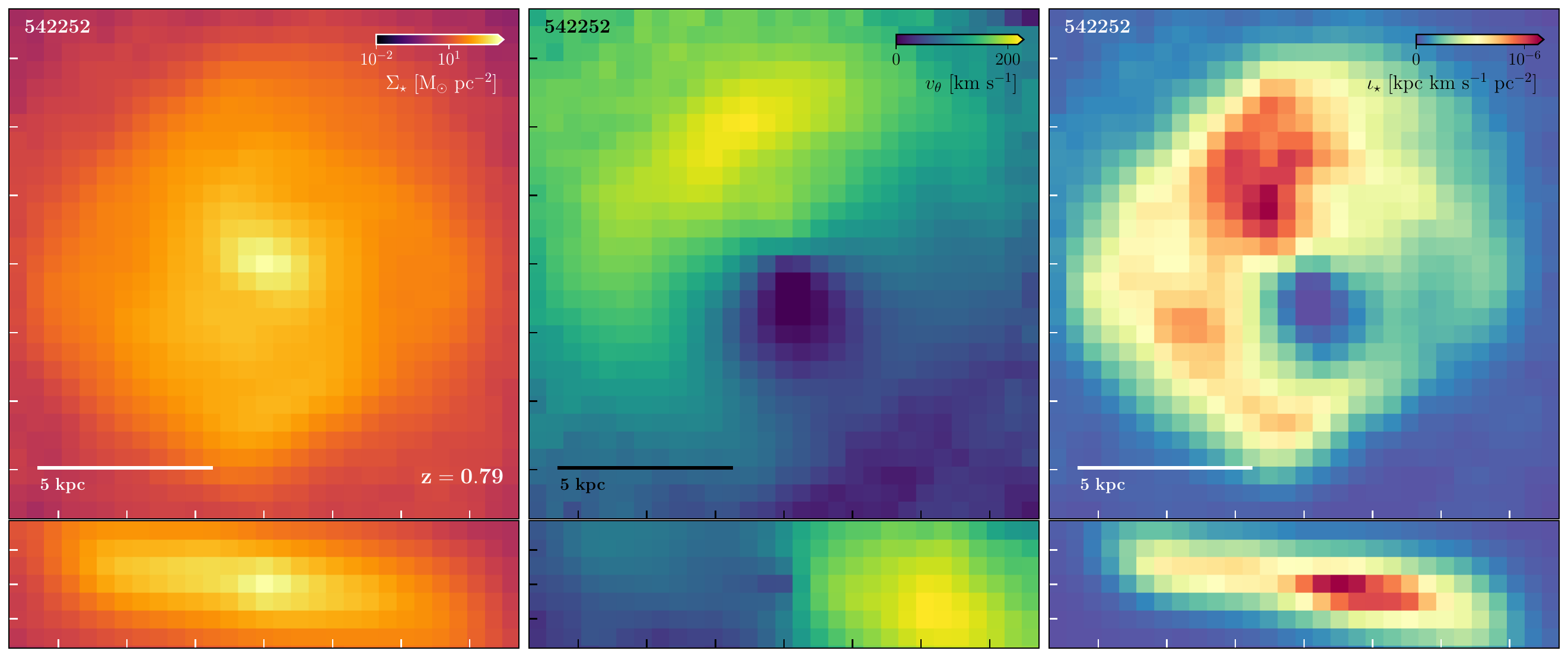}
    \end{subfigure}
    \begin{subfigure}[b]{0.49\textwidth}
        \includegraphics[width=\textwidth]{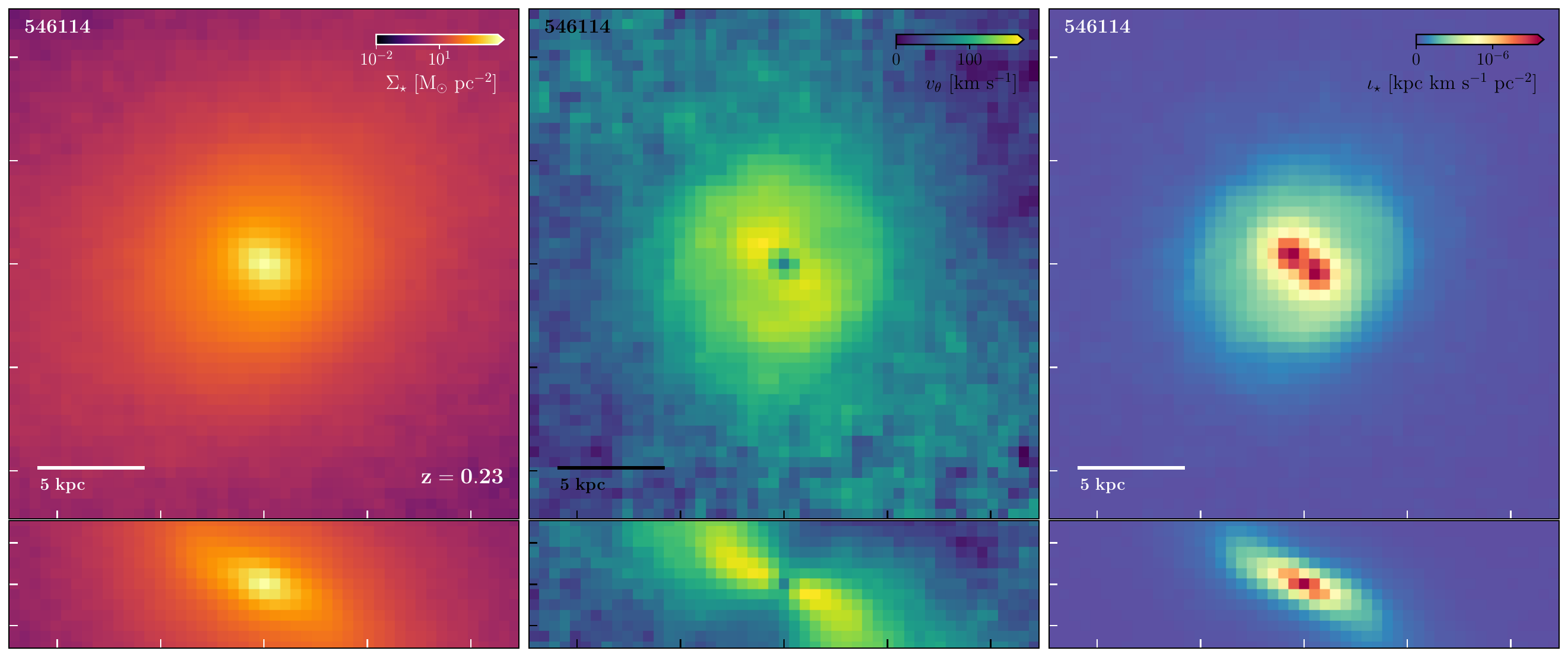}
    \end{subfigure}
    \caption{Continued.}
    \label{fig: zoo}
\end{figure*}

\begin{figure*}[ht!]\ContinuedFloat
    \centering
    \begin{subfigure}[b]{0.49\textwidth}
        \includegraphics[width=\textwidth]{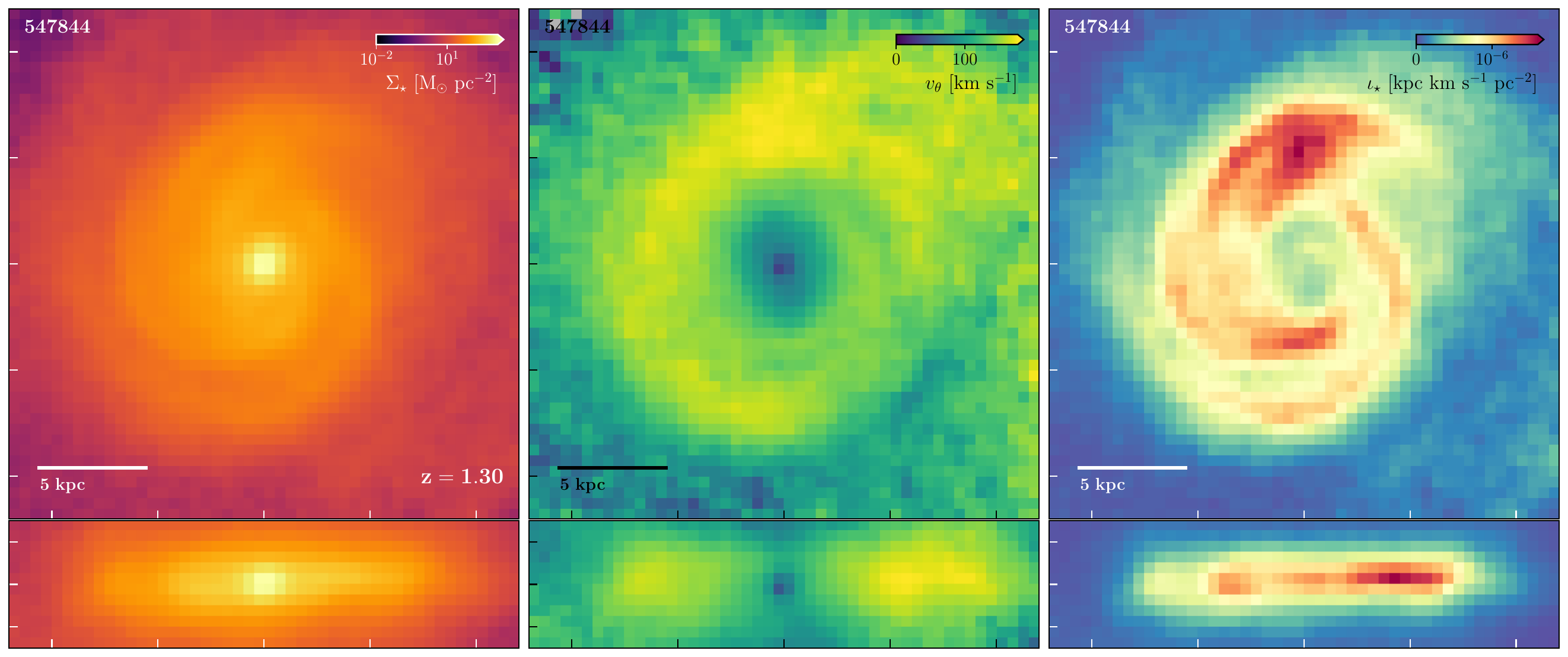}
    \end{subfigure}
    \begin{subfigure}[b]{0.49\textwidth}
        \includegraphics[width=\textwidth]{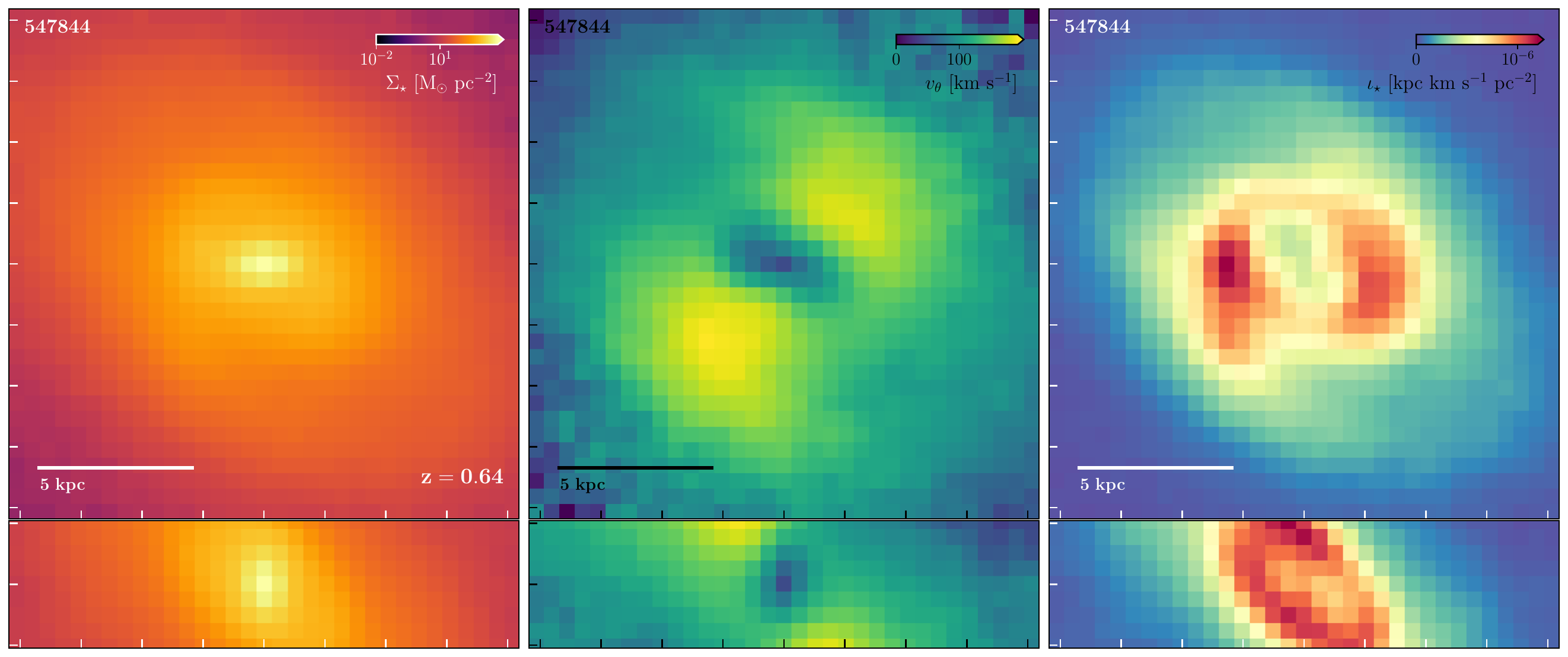}
    \end{subfigure}
    \\
    \begin{subfigure}[b]{0.49\textwidth}
        \includegraphics[width=\textwidth]{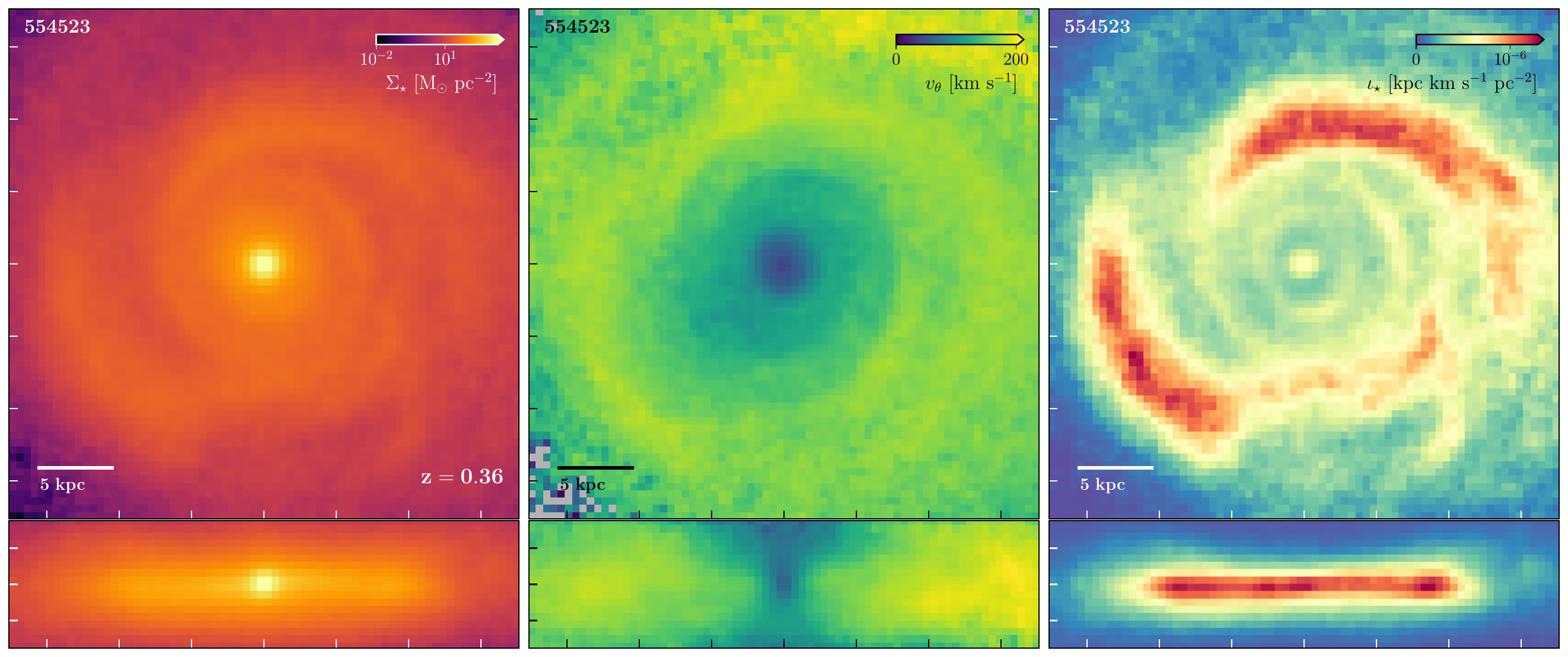}
    \end{subfigure}
    \begin{subfigure}[b]{0.49\textwidth}
        \includegraphics[width=\textwidth]{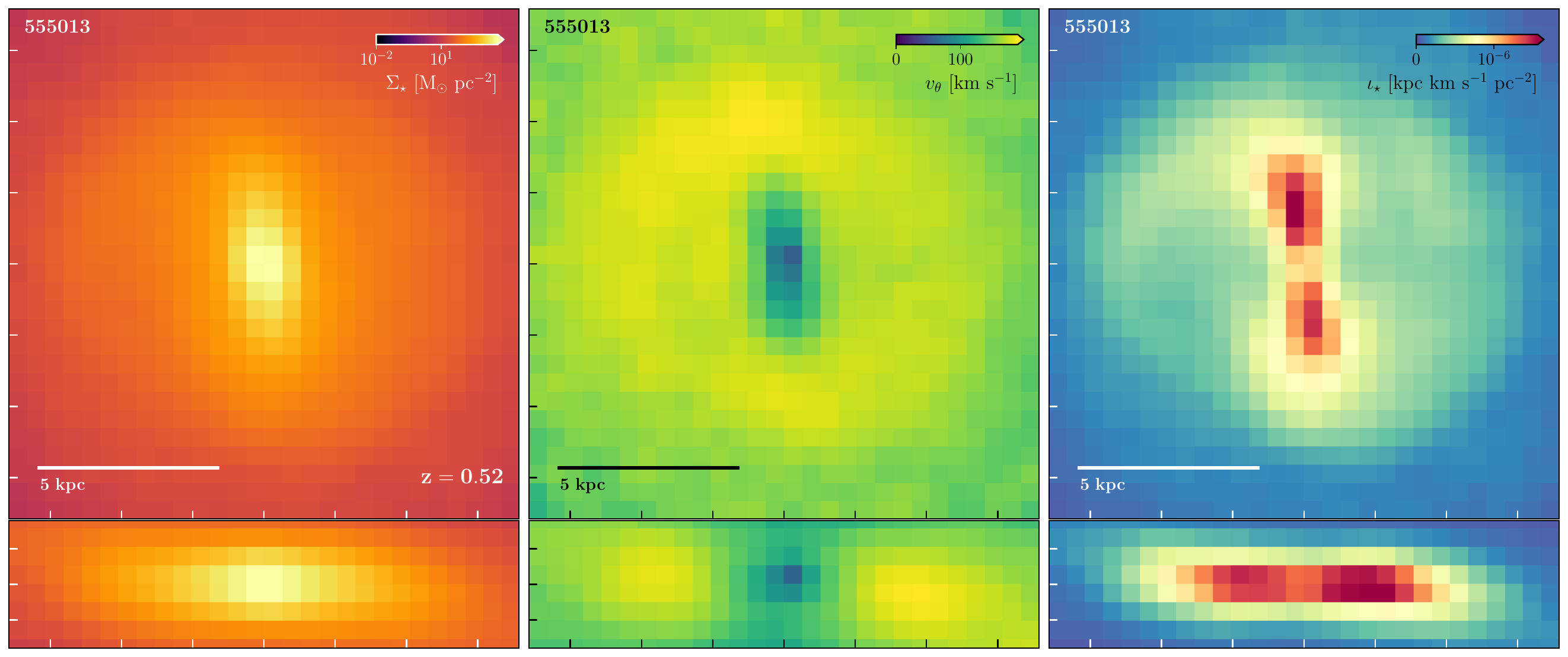}
    \end{subfigure}
    \\
    \begin{subfigure}[b]{0.49\textwidth}
        \includegraphics[width=\textwidth]{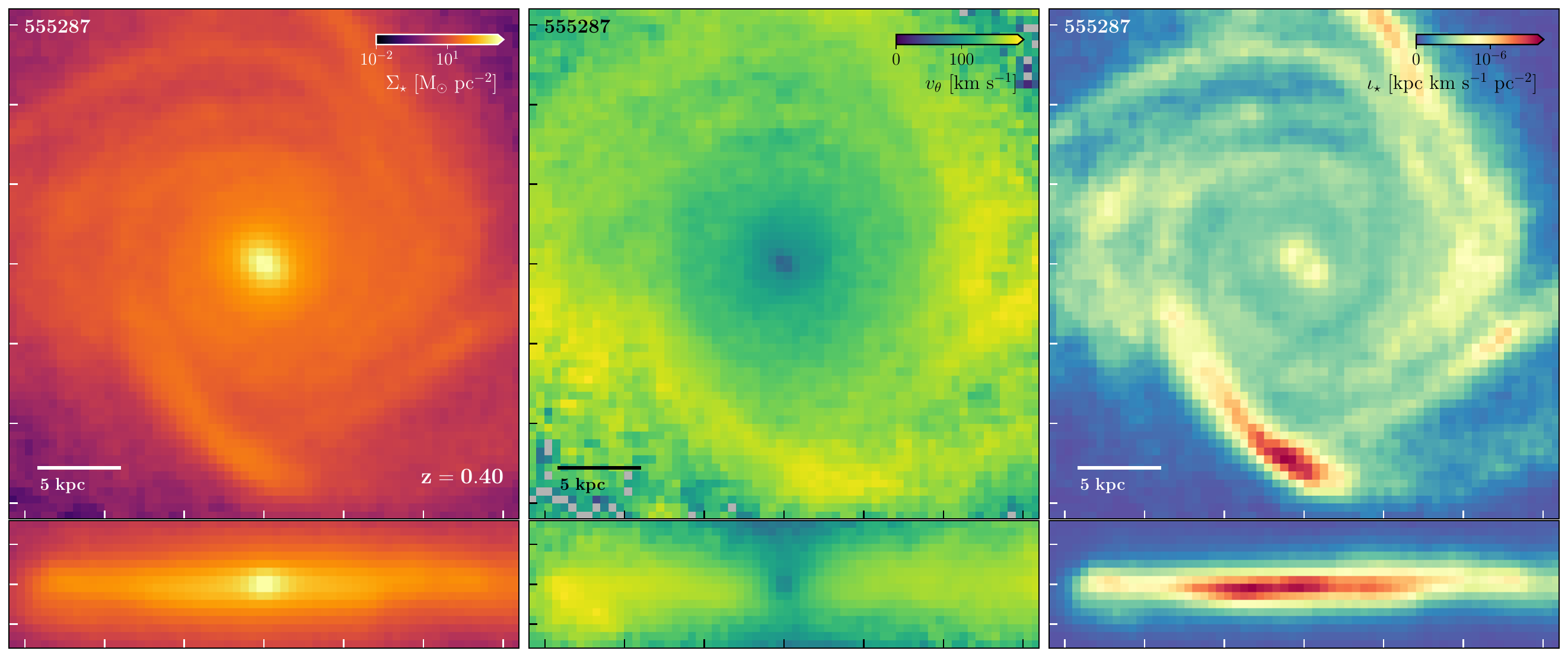}
    \end{subfigure}
    \begin{subfigure}[b]{0.49\textwidth}
        \includegraphics[width=\textwidth]{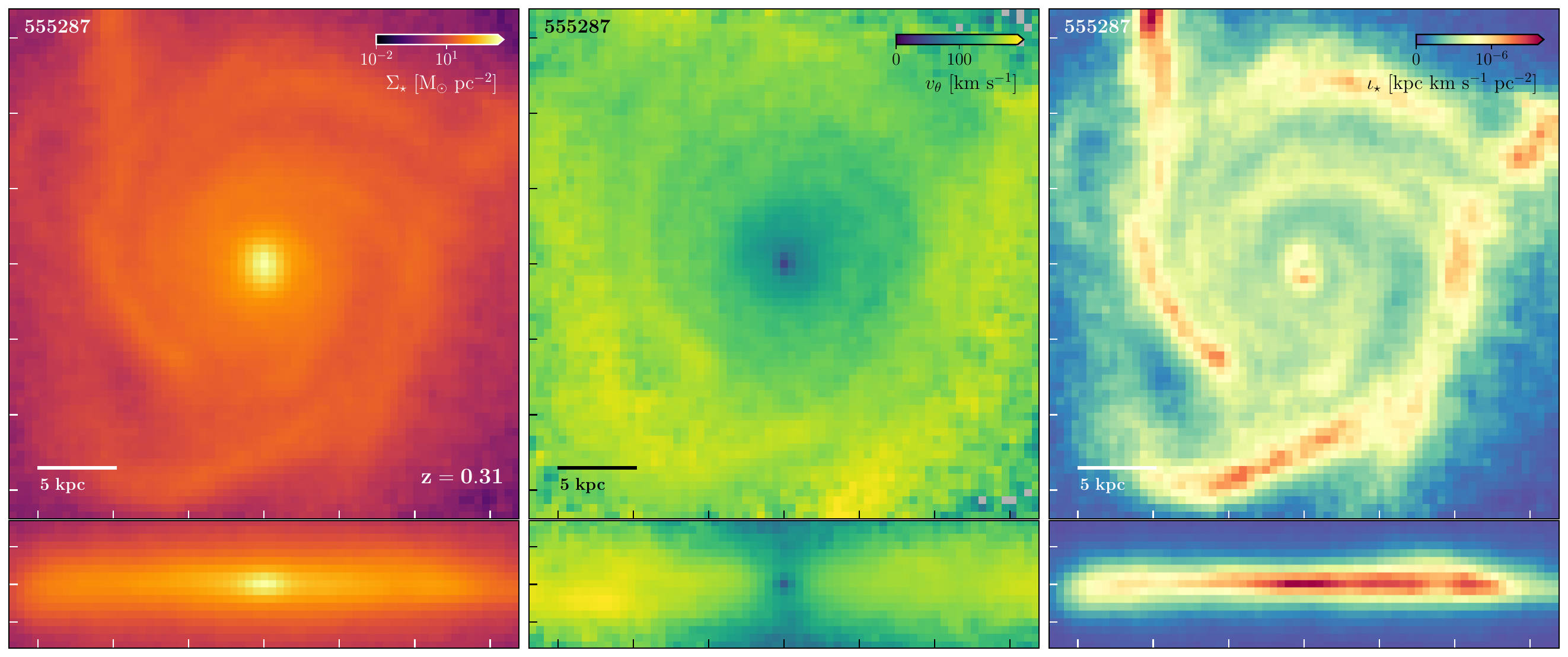}
    \end{subfigure}
    \\
    \begin{subfigure}[b]{0.49\textwidth}
        \includegraphics[width=\textwidth]{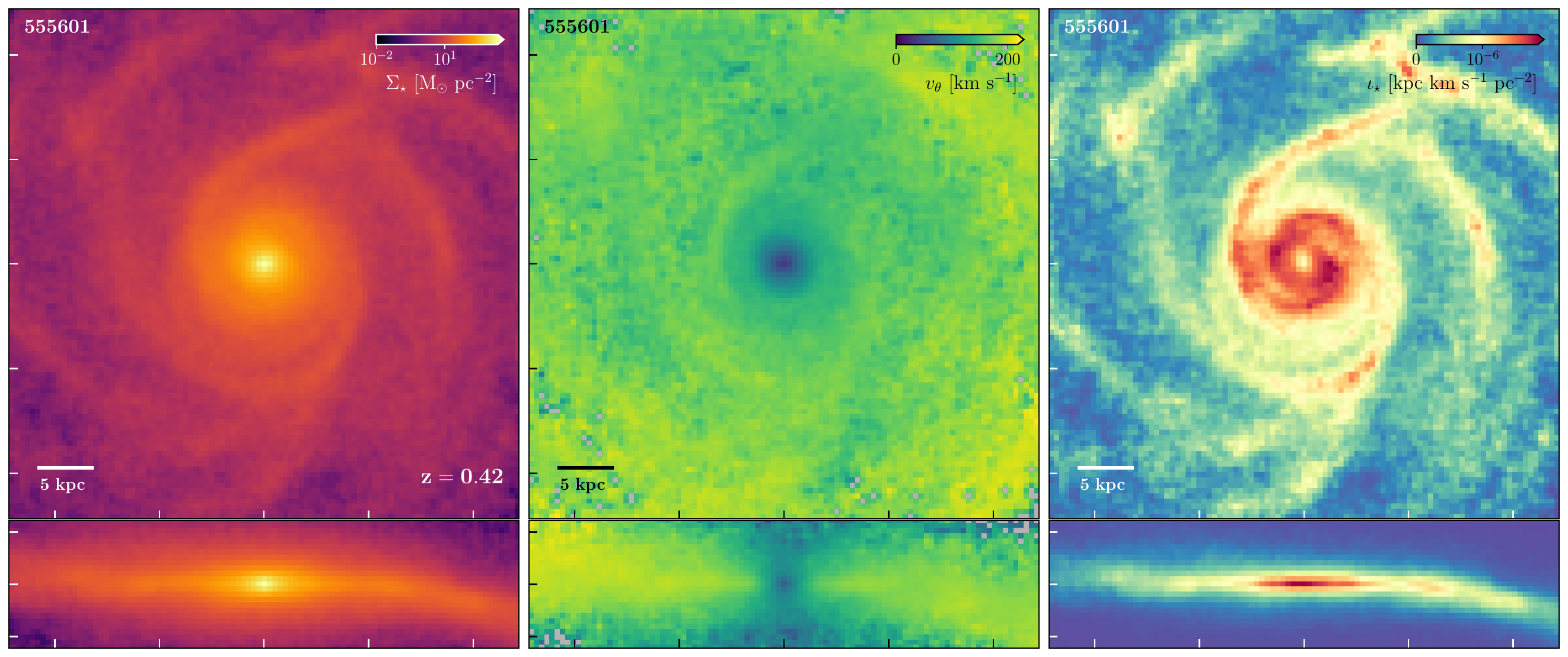}
    \end{subfigure}
    \begin{subfigure}[b]{0.49\textwidth}
        \includegraphics[width=\textwidth]{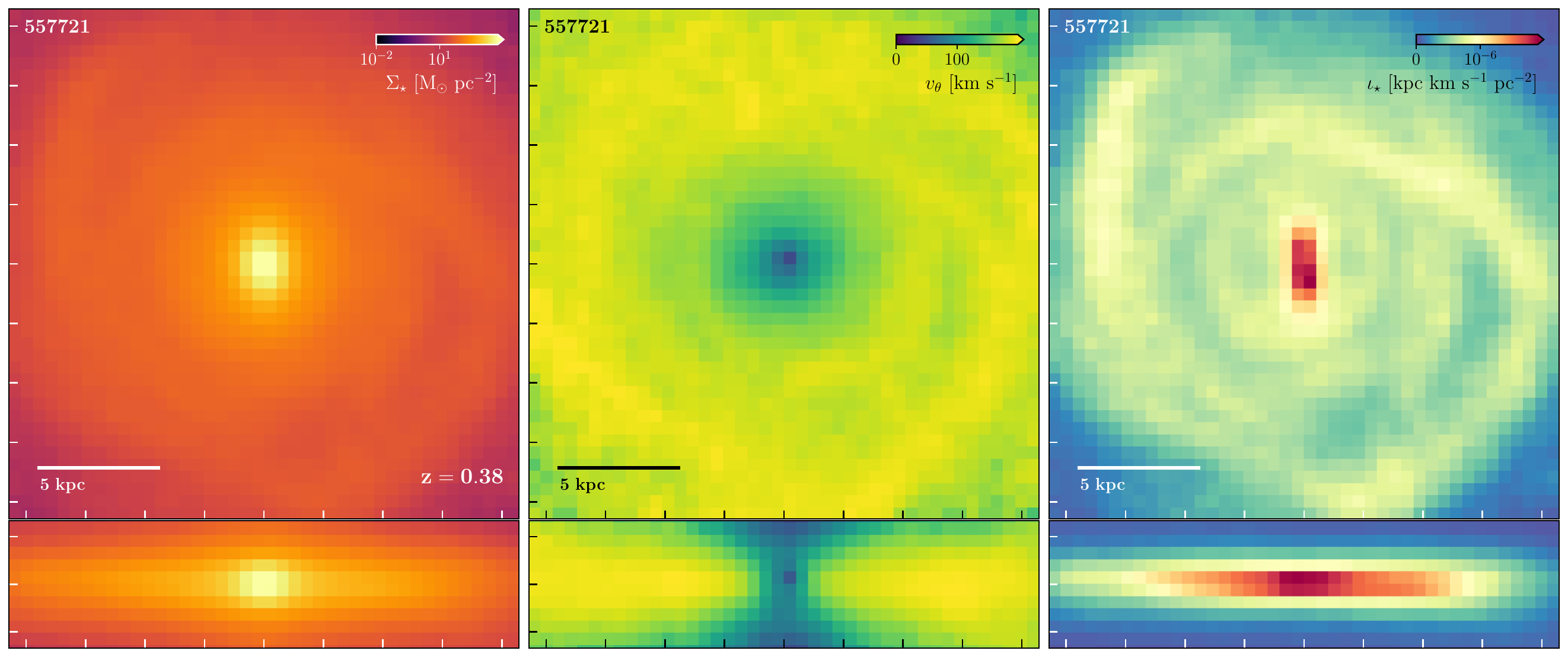}
    \end{subfigure}
    \\
    \begin{subfigure}[b]{0.49\textwidth}
        \includegraphics[width=\textwidth]{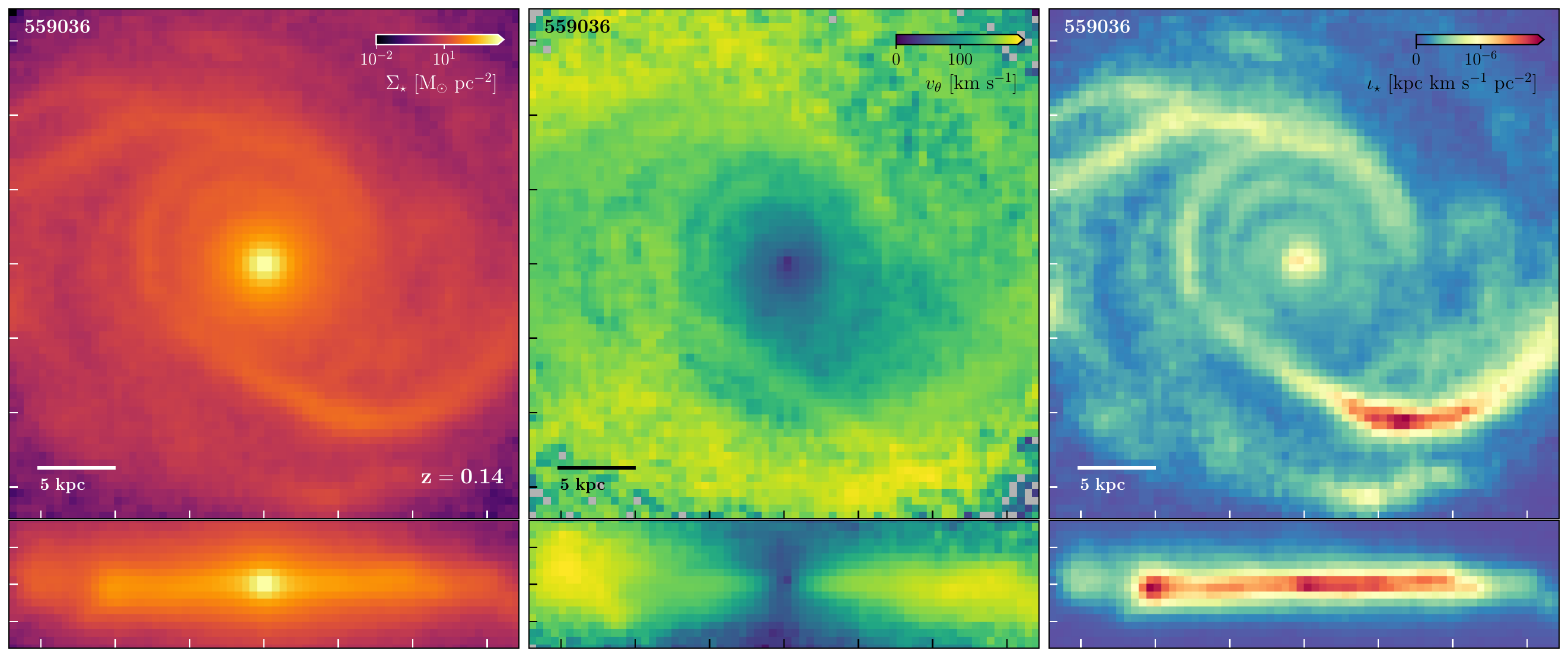}
    \end{subfigure}
    \begin{subfigure}[b]{0.49\textwidth}
        \includegraphics[width=\textwidth]{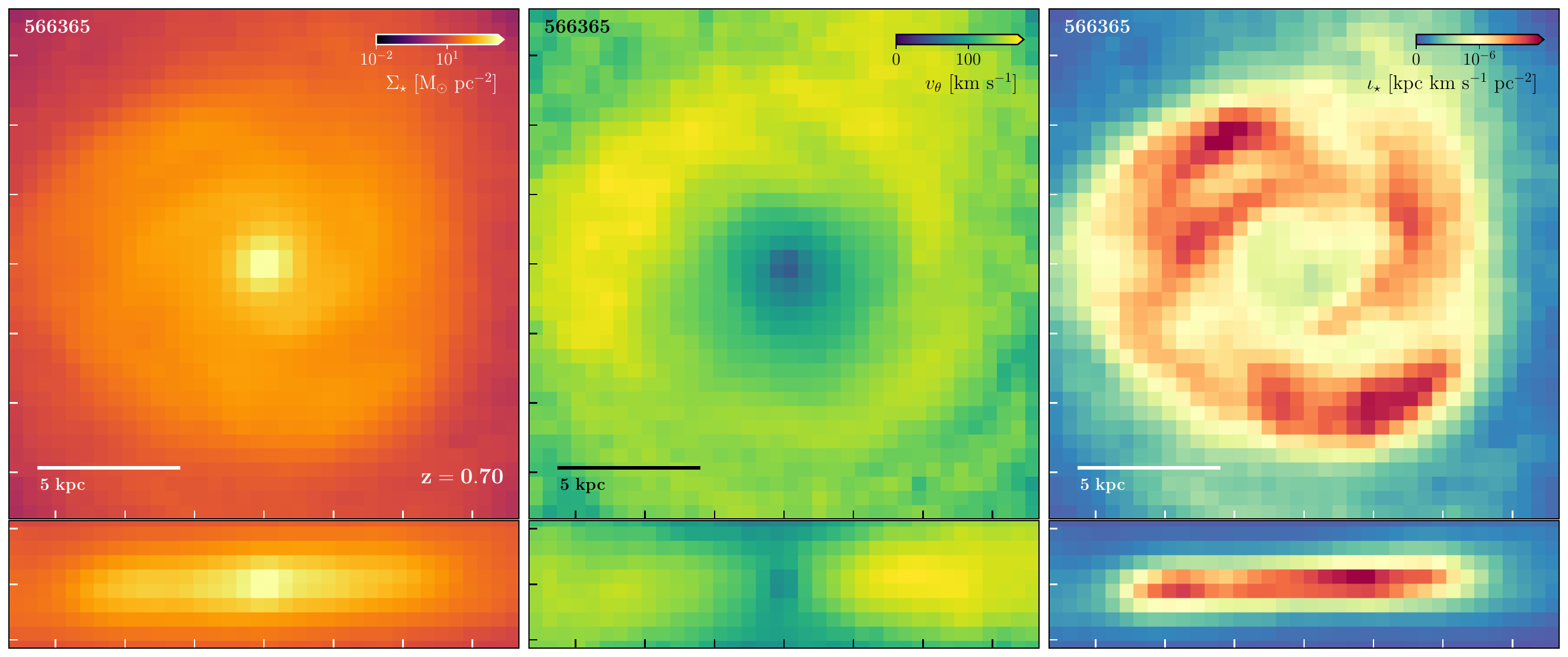}
    \end{subfigure}
    \\
    \begin{subfigure}[b]{0.49\textwidth}
        \includegraphics[width=\textwidth]{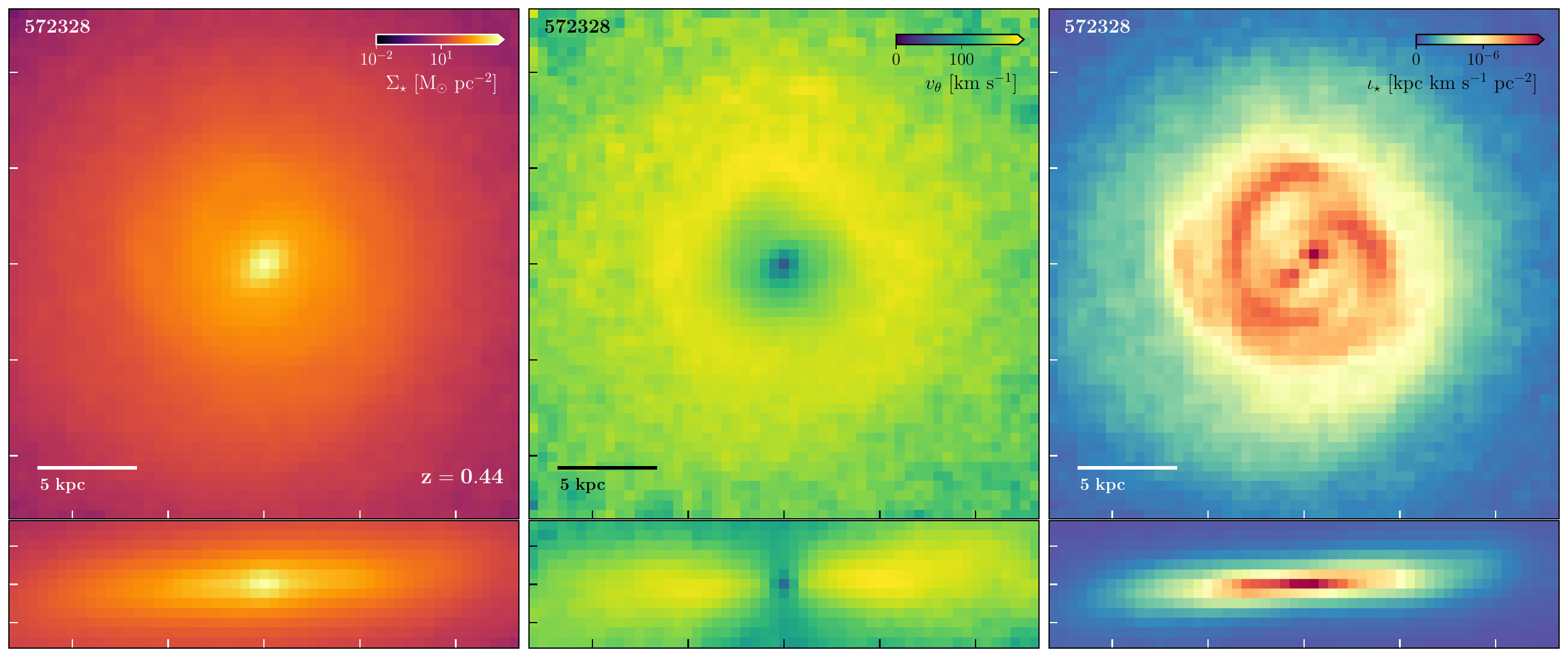}
    \end{subfigure}
    \begin{subfigure}[b]{0.49\textwidth}
        \includegraphics[width=\textwidth]{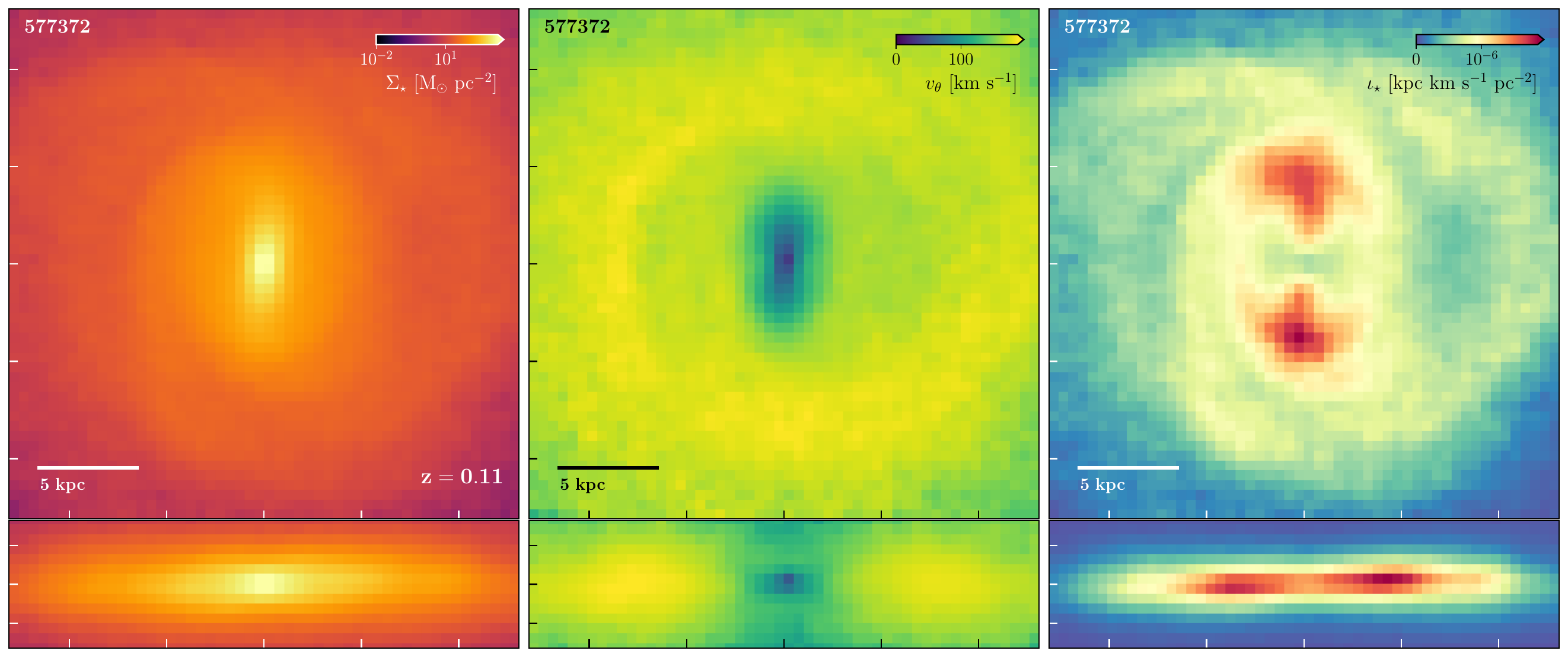}
    \end{subfigure}
    \caption{Continued.}
    \label{fig: zoo}
\end{figure*}

\FloatBarrier
\clearpage
\twocolumn

\section{Combined Fourier metrics}\label{app: Combined_Fourier}

Making use of the harmonic nature of the Fourier mode decomposition, and applying $m \in {0,1,2,3,4}$ to Eq. (\ref{eq: cm_discrete}), we define another set of morpho-kinematic metrics as follows:

\begin{equation}
a_{1+3}^{\max} = \max_{R_n \in [3\epsilon_\star,R_{\max}]} \frac{\left|c_1(R_n) + c_3(R_n)\right|}{\left|c_0(R_n)\right|},
\label{eq: a13}
\end{equation}

\begin{equation}
a_{2+4}^{\max} = \max_{R_n \in [3\epsilon_\star,R_{\max}]} \frac{\left|c_2(R_n) + c_4(R_n)\right|}{\left|c_0(R_n)\right|},
\label{eq: a24}
\end{equation}

\noindent and

\begin{equation}
R_{2+4}^{\max} = \frac{1}{R_{\max}} \left(\operatorname*{argmax}_{R_n \in [3\epsilon\star,,R_{\max}]} \frac{\left|c_2(R_n) + c_4(R_n)\right|}{\left|c_0(R_n)\right|}\right),
\label{eq: R_a24}
\end{equation}

\noindent The rapid decrease in Fourier amplitude with respect to $m$ ensures that all dominant substructures in the sAMSD space are captured in Eqs. (\ref{eq: a13}), (\ref{eq: a24}), and (\ref{eq: R_a24}). Grouping modes in $a_{1+3}^{\max}$, $a_{2+4}^{\max}$, and $R_{2+4}^{\max}$, reduces the individual harmonics fluctuations that can appear in $a_{1}^{\max}$, $a_{2}^{\max}$, and $R_{2}^{\max}$, as a consequence of finite particle sampling and numerical noise.

\begin{figure}[ht!]
\centering
\includegraphics[width=\hsize]{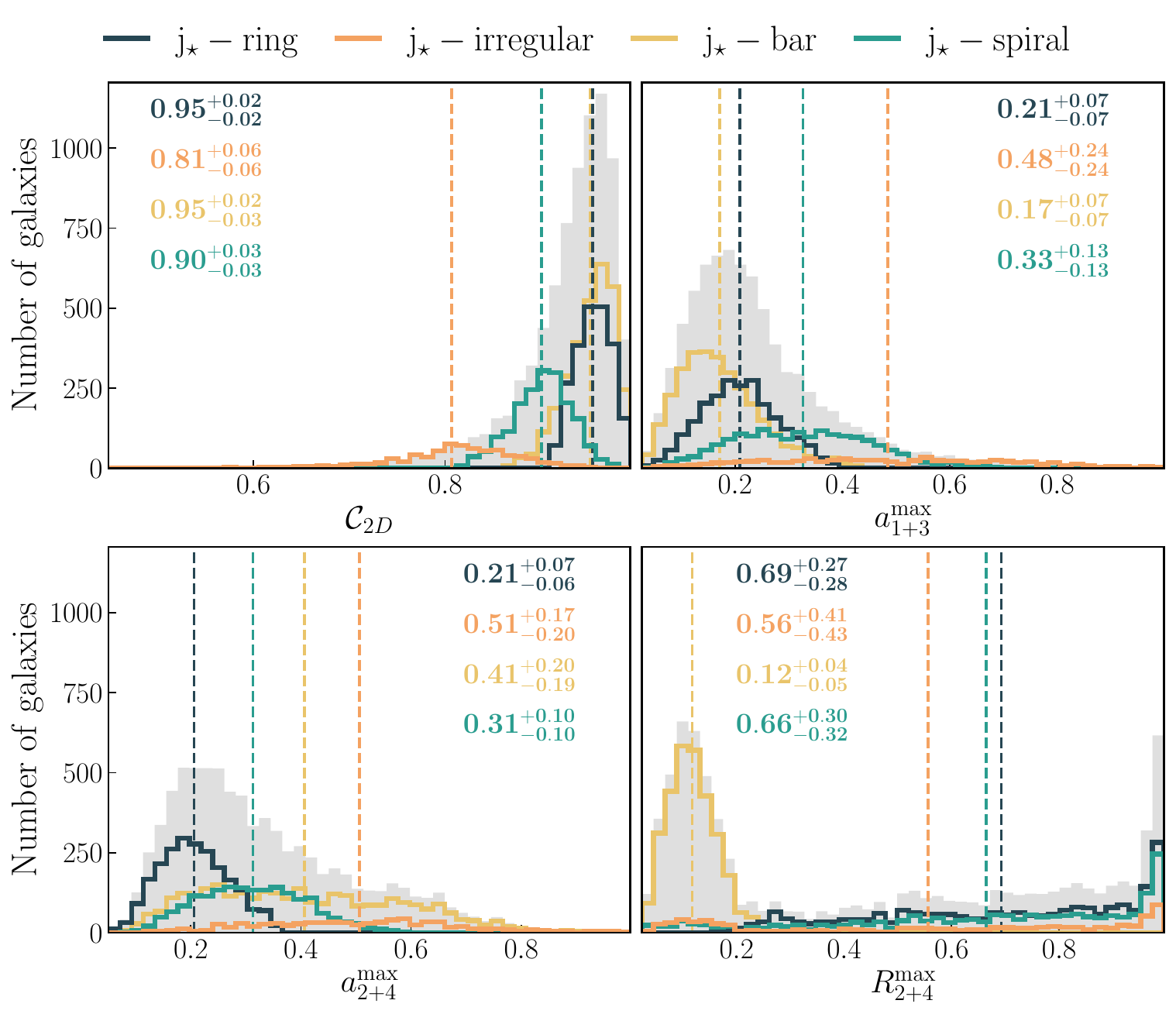}
\caption{Histograms of the Fourier combined morpho-kinematic metrics for each of the four components identified by the GMM. The panels show the number of galaxies per bin for $\mathcal{C}_{\rm 2D}$ (top-left), $a_{1+3}^{\max}$ (top-right), $a_{2+4}^{\max}$ (bottom-left), and $R_{2+4}^{\max}$ (bottom-right). The remaining description is the same as for Fig.~\ref{fig: GMM_distributions}.}
\label{fig: GMM_combined}
\end{figure}

To determine whether variations in the Fourier morpho-kinematic metrics affect our results, we applied our classification methodology, described in Sect.~\ref{Sec: GMM}, to the parameter space defined by $\mathcal{C}_{\rm 2D}$, $a_{1+3}^{\max}$, $a_{2+4}^{\max}$, and $R_{2+4}^{\max}$. Figure~\ref{fig: GMM_combined} shows the distributions of our four $j_\star$-types in this new space. When comparing these results with Fig.~\ref{fig: GMM_distributions}, we identify that there is no significant variation between the populations of $j_\star$-types defined by the simple or the combined Fourier metrics. The $j_\star$-irregularities, $j_\star$-bars and $j_\star$-spirals are so dominant that their strong imprint on the fundamental Fourier modes resists harmonic fluctuations.

\begin{figure}[h!]
\centering
\includegraphics[width=\hsize]{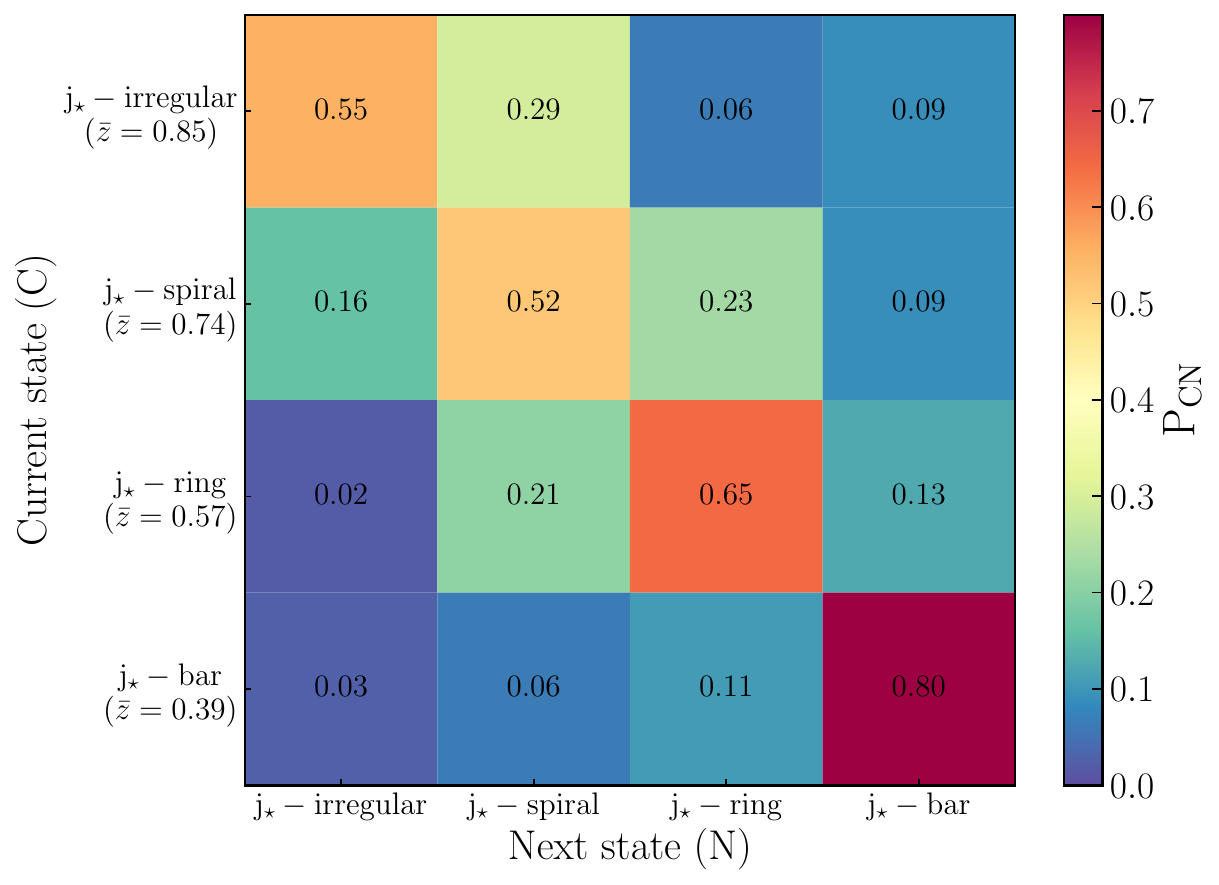}
\caption{Markov transition matrix for the different $j_\star$-types obtained from the posterior probabilities of the GMM classification applied to the combined Fourier metrics space. The remaining description is the same as for Fig.~\ref{fig: Markov_matrix}.}
\label{fig: Markov_matrix_combined}
\end{figure}

The morpho-kinematic evolution scenario for our galaxies, as defined by the simple Fourier metrics, also holds when the combined metrics are applied. The transition probabilities between the $j_\star$-types in Fig.~\ref{fig: Markov_matrix_combined}, determined by the GMM with basis $\mathcal{C}_{\rm 2D}$, $a_{1+3}^{\max}$, $a_{2+4}^{\max}$, and $R_{2+4}^{\max}$, are very similar to those already seen in the Markov transition matrix in Fig.~\ref{fig: Markov_matrix}. The only differences worth noting are that, compared to the simple metrics, the combined metrics establish that: the transition from $j_\star$-irregular is only significant towards $j_\star$-spiral, the transition from $j_\star$-spiral is no longer significant to $j_\star$-bar, there is one backward transition from $j_\star$-bar to $j_\star$-ring, the backward transition from $j_\star$-ring to $j_\star$-spiral is higher, the self-transition of $j_\star$-irregular is higher, and the self-transition of $j_\star$-rings is lower. These variations are all of the order of $\rm P_{CN} \leq 0.11$, and do not significantly alter the angular momentum redistribution scenario for secularly evolving discs described in Sect.~\ref{sec: secular_model}. As we have seen in this section, the simple Fourier morpho-kinematic metrics do not appear to be affected by noise or numerical artefacts in our simulations; for this reason, we have decided to retain them as the basis of our study, taking advantage of their simplicity and direct physical interpretation compared to their combined counterparts.

\clearpage

\section{Other \textsc{TNG50} properties explored}\label{app: correlation_properties}

\begin{figure}[h!]
\centering
\includegraphics[width=1\hsize]{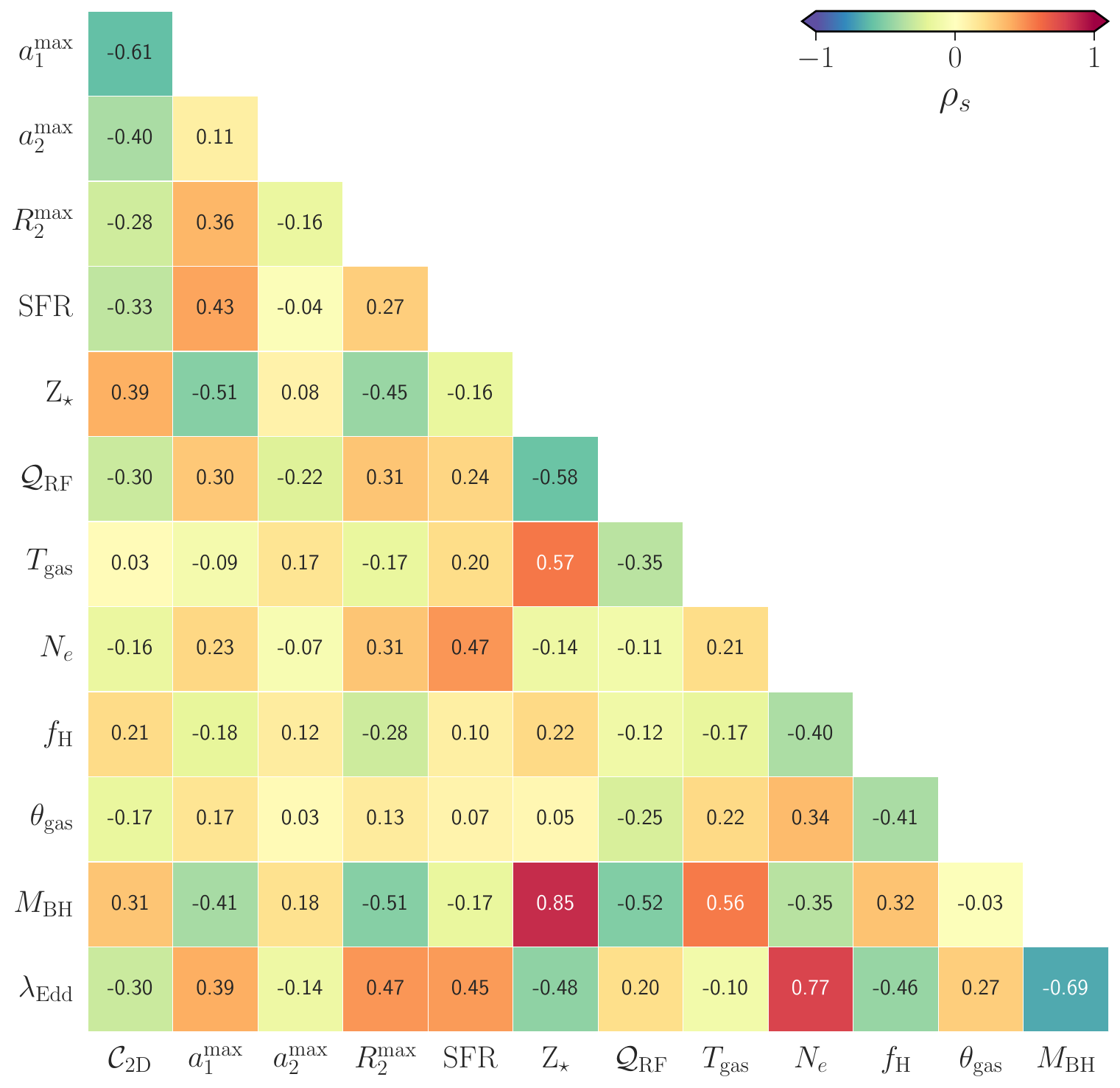}
\includegraphics[width=1\hsize]{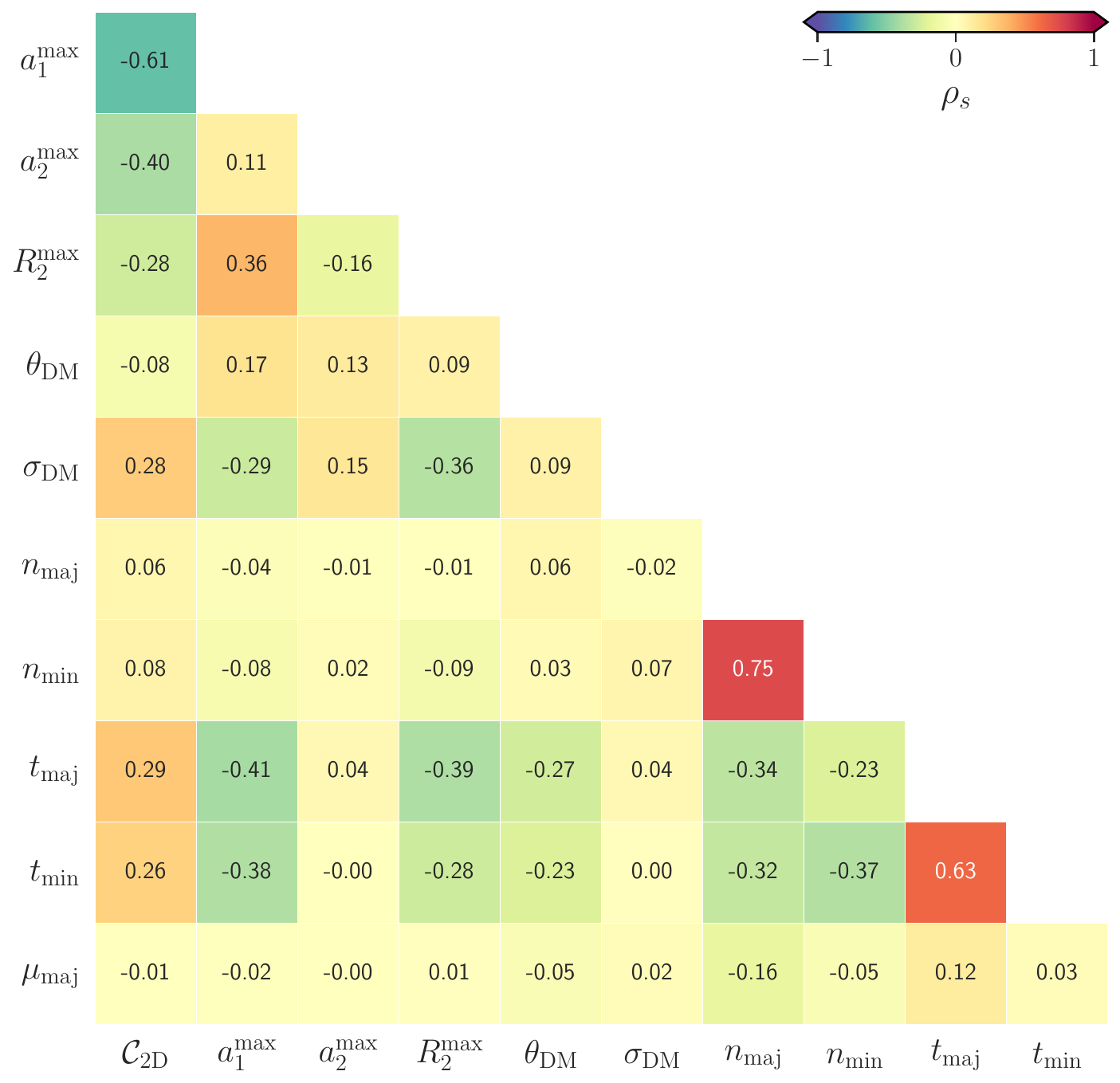}
\caption{$\rho_s$ matrices between the morpho-kinematic metrics and some \textsc{TNG50} galaxy properties.}
\label{fig: matrices_properties}
\end{figure}

In the top panel of Fig.~\ref{fig: matrices_properties}, we show the values of $\rho_s$ between our morpho-kinematic metrics and several key properties of the interstellar medium of our galaxies, including: the star formation rate ($\rm SFR$), the mass-weighted stellar metallicity ($\rm Z_\star$), the disc stability criterion proposed in \citet{romeo2013} ($\mathcal{Q}_{\rm RF}$), the gas temperature ($T_{\rm gas}$), the electron abundance ($N_{e}$), the hydrogen fraction ($f_{\rm H}$), the angle between the total angular momentum vector of the gas and the stars ($\theta_{\rm gas}$), the mass of the central supermassive black hole ($M_{\rm BH}$), and the Eddington ratio of the central supermassive black hole ($\lambda_{\rm Edd}$). All these properties have been extracted, or derived, directly from the TNG50 fiducial model while maintaining its unit system. As can be seen from the matrix, most of these properties have $|\rho_s| < 0.40$ across all our morpho-kinematic metrics, indicating that there is no significant (anti-)correlation between them and the shape of our galaxies in the sAMSD space. $\rm SFR$, $\rm Z_\star$, $M_{\rm BH}$ and $\lambda_{\rm Edd}$ are the only properties with $|\rho_s| \geq 0.40$  with respect to $a_1^{\max}$ and $R_2^{\max}$. However, these correlations arise as a consequence of the close relationship between $f_{\rm gas}$ and $\rm SFR$ ($\rho_s = 0.59$), and between $M_\star$ and $\rm Z_\star$ ($\rho_s = 0.81$), $M_{\rm BH}$ ($\rho_s = 0.89$) and $\lambda_{\rm Edd}$ ($\rho_s = -0.65$), and therefore do not contribute anything new to our discussion.

In the bottom panel of Fig.~\ref{fig: matrices_properties}, we show the values of $\rho_s$ between our morpho-kinematic metrics, some dark matter (DM) properties of our galaxies, and their merger history, including: the angle between the total angular momentum vector of the DM and the stars ($\theta_{\rm DM}$), the DM velocity dispersion ($\sigma_{\rm DM}$), the number of major mergers ($n_{\rm maj}$), the number of minor mergers ($n_{\rm min}$), the time since the last major merger ($t_{\rm maj}$), the time since the last minor merger ($t_{\rm min}$), and the mass ratio of the most recent major merger ($\mu_{\rm maj}$). The DM properties were extracted from the TNG50 fiducial model, keeping its unit system, while the merger history was recovered from the merger tree of our galaxies. All mergers with a mass ratio ($\mu$) bigger than or equal to 0.25 were classified as major, while those that satisfied $0.10 \leq \mu < 0.25$ were classified as minor. As can be seen from the matrix, all the properties, apart from $t_{\rm maj}$, have a $|\rho_s| < 0.40$ with respect to all our morpho-kinematic metrics, indicating that there is no significant (anti-)correlation between them and the shape of our galaxies in the sAMSD space. The $\rho = -0.41$ between $t_{\rm maj}$ and $a_1^{\max}$ can be explained by the close relationship between $f_{\rm gas}$ and $t_{\rm major}$ ($\rho_s = -0.45$).

\clearpage

\section{Bootstrapping sampling test}\label{app: bootstraping}

\begin{figure}[h!]
\centering
\includegraphics[width=.9\hsize]{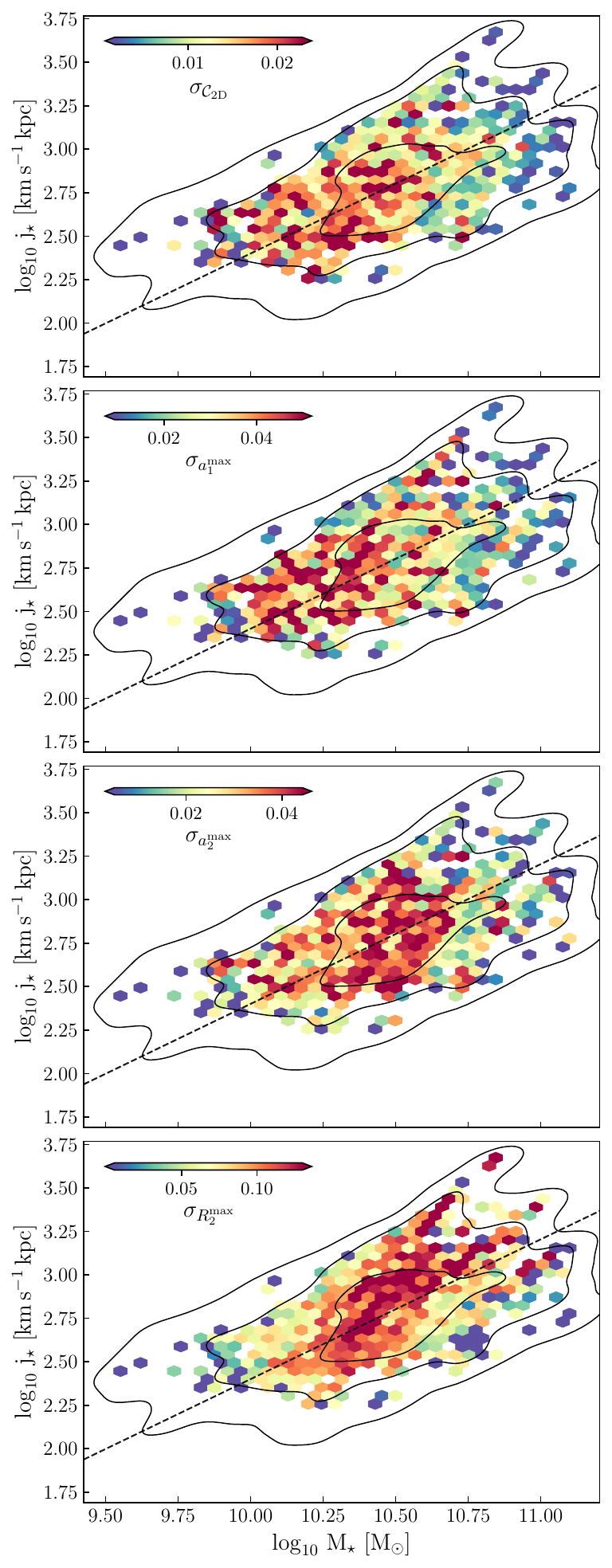}
\caption{Bootstrapping standard deviation for the morpho-kinematic metrics in the $j_\star-M_\star$ diagram. This process was carried out for a total of 1000 iterations, maintaining a fixed number of five galaxies per hexagonal bin. The axes, the minimum and maximum values of the coloured bars, the black dashed line, and the contours levels are described in Fig. \ref{fig: Fall_relation_TNG50}.}
\label{fig: bootstraping_std_TNG50}
\end{figure}

To assess the robustness of the mean trends identified in each hexagonal bin, we performed a bootstrap resampling analysis independently for every morpho-kinematic metric. For each bin containing at least five galaxies, we generated 1000 random realizations by selecting, without replacement, a subsample of five galaxies from the objects assigned to that bin. For every realization, we computed the mean value of the corresponding metric within each eligible bin. This yielded a distribution of mean values per bin, from which we derived the standard deviation across all bootstrap realizations. The resulting standard deviation provides an estimate of the uncertainty associated with finite sampling in each bin. These values were then mapped across the $j_\star-M_\star$ plane (Fig. \ref{fig: bootstraping_std_TNG50}) to test whether the trends discussed in the main analysis are sensitive to variations in the number or selection of galaxies within individual bins.

As shown in each panel of Fig. \ref{fig: bootstraping_std_TNG50}, the bootstrap standard deviations of all four metrics are consistently at least one order of magnitude smaller than the metric values themselves. This demonstrates that the distribution of the $j_\star$-structures across the $j_\star-M_\star$ plane is a robust result, largely unaffected by statistical fluctuations in the sample.

\section{Cosmic web reconstruction}\label{app: cosmic_web}

To identify all components of the cosmic web, namely voids, walls, filaments, and nodes, we use the publicly available and widely adopted structure finder \disperse \citep{sousbie2011a,sousbie2011}. We employ galaxies as tracers of the underlying large-scale structure. To handle discrete data sets, \disperse relies on discrete Morse theory \citep{forman2001}, building on the Delaunay tessellation. This framework provides a scale-free Delaunay Tessellation Field Estimator \citep[DTFE;][]{schaap2000} density field which is used to reconstruct the local topology. Furthermore, \disperse manages noisy data through topological persistence, filtering out structures that are topologically less robust. In practice, to reconstruct the cosmic web, we only include galaxies well above the resolution limit (stellar masses above $10^8$ M$_{\odot}$) and apply a standard persistence threshold $N_\sigma=3$.

\end{appendix}
\end{document}